\documentclass[11pt]{article}
\usepackage{mymacros}
\usepackage{tikz,pgfplots}
\usepackage{float}
\usepackage{appendix}
\usepackage{setspace}
\usepackage{caption}
\usepackage{subfiles}

\newtheorem{comment}[theorem]{Comment}
\renewcommand{\dag}{\dagger}

\newcommand{\good}{\textsc{good}}
\newcommand{\dnorm}[1]{\norm{#1}_{\diamond}}

\title{{\bf
Novel one-shot inner bounds for unassisted fully quantum channels
via rate splitting\footnote{A preliminary version of this work appeared in the proceedings of ISIT 2021 \cite{chakraborty:ratesplittingnew}.}
}}

\author{
Sayantan Chakraborty\thanks{Centre for Quantum Technologies, National University of Singapore}
\and
Aditya Nema\thanks{Institute for Quantum Information, RWTH Aachen}
\and
 Pranab Sen\textsuperscript{\textdagger}\thanks{
School of Technology and System Science, 
Tata Institute of Fundamental Research, Mumbai, India.\\
Email: {\sf
\{kingsbandz, aditya.nema30, pranab.sen.73\}@gmail.com
}
}
}

\date{}

\begin{document}
\maketitle

\begin{abstract}
We prove the first non-trivial one-shot inner bounds for sending quantum 
information over
an entanglement unassisted two-sender quantum multiple access channel 
(QMAC) and an unassisted two-sender two-receiver quantum interference 
channel (QIC). 
Previous works only studied the unassisted QMAC in the limit of many 
independent and identical uses of the channel also known as the 
asymptotic iid limit, and did not study the unassisted QIC at all. 
We employ two techniques, 
{\em rate splitting} and {\em successive
cancellation}, in order to obtain our inner bound.
Rate splitting was 
earlier used to obtain inner bounds, avoiding time sharing, 
for classical channels in the
asymptotic iid setting. Our main technical contribution is to extend
rate splitting from the classical asymptotic iid setting to the 
quantum one-shot setting. 
In the asymptotic iid limit our one-shot inner bound for QMAC approaches 
the rate region of Yard et al.~\cite{Yard_MAC}. For the QIC we
get novel non-trivial rate regions in the asymptotic iid setting.
All our results also
extend to the case where limited entanglement assistance is provided,
in both one-shot and asymptotic iid settings. The limited entanglement
results for one-setting for both QMAC and QIC are new. For the QIC
the limited entanglement results are new even in the asymptotic iid
setting.
\end{abstract}

\onehalfspacing

\section{Introduction}
The multiple access channel (MAC), where two independent senders Alice
(A) and Bob (B)
have to send their respective messages to a single receiver Charlie (C)
via a communication channel with two inputs and one output,
is arguably the simplest multiterminal channel. 
Yet, it abstracts out
important practical situations like several independent users 
transmitting their respective messages to a base station. 
Ahlswede~\cite{Ahlswede:mac}, and independently Liao~\cite{Liao:mac}, 
obtained the first optimal rate region for
the classical MAC in the asymptotic iid setting, using a powerful 
method called {\em simultaneous decoding}. 
Their region looks like the one in Figure~\ref{fig:MACregion}, 
where $I(A:B) := H(A) + H(B) - H(AB)$ denotes the mutual information
between two jointly distributed random variables $A$, $B$.
\begin{figure}[hhh]
\vspace*{-5mm}
\begin{center}
\begin{tikzpicture}
\draw [->, line width=1 pt] (1,-2) -- (1,2);
\draw [->, line width=1 pt] (1,-2) -- (5,-2);
\draw [line width=1 pt] (1,0) -- (2,0);
\draw [line width=1 pt] (3,-2) -- (3,-1);
\draw [line width=1 pt] (2,0) -- (3,-1);
\draw [line width=1 pt,dotted] (1,1) -- (4,-2);
\draw (1,2) node[anchor=east] {Bob};
\draw (5,-2) node[anchor=west] {Alice};
\draw (1.5,-1) node[anchor= west] {$n\rightarrow \infty$};
\draw (2.5,-0.5) node[anchor=west] {{\footnotesize $P$}};
\draw (1,0) node[anchor=north] {{\footnotesize $(0,I(B:AC))$}};
\draw (4,-2) node[anchor=north ] {{\footnotesize $(I(A:BC), 0)$}};
\draw (3,-1) node[anchor=west] 
{{\footnotesize $S = (I(A:BC), I(B:C))$}};
\draw (2,0) node[anchor=west] 
{{\footnotesize $T = (I(A:C),I(B:AC))$}};
\draw (1,1) node[anchor=west] {{\footnotesize $x+y= I(AB:C)$}};
\end{tikzpicture}
\end{center}
\vspace*{-5mm}
\caption{Achievable rate region per channel use for the classical MAC in 
the asymptotic iid setting.}
\label{fig:MACregion}
\end{figure}
Simultaneous decoding means that
Charlie is able to decode any point in the rate region, e.g. point P in
Figure~\ref{fig:MACregion}, by a one-step procedure.
Later on, other authors obtained the same 
rate region in a 
computationally less intensive fashion by using 
{\em successive cancellation} and {\em time sharing}.
In successive cancellation decoding Charlie first decodes Alice's message
and then uses it as an additional channel output in order to next decode
Bob's message, or vice versa.
In other words, Charlie can either decide
to decode point S or point T in Figure~\ref{fig:MACregion}. 
In order to decode
another point in the rate region, e.g. point P in 
Figure~\ref{fig:MACregion},
Charlie first figures out the convex combination $(\alpha, 1 - \alpha)$ 
of points S and T that would give point P. Out of $n$ iid channel uses, 
Charlie then decodes the first $\alpha n$ uses according to 
point S's decoding strategy and the remaining $(1-\alpha) n$ channel uses
according to point T's decoding strategy. This idea is called time sharing.

The interference channel is another important channel where sender
Alice wants to send her message to receiver Charlie and sender Bob,
whose message is independent of that of Alice, wants to send his message
to receiver Damru via a communication channel with two inputs and two
outputs. It abstracts out the
important practical situation where independent sender-receiver pairs
are communicating simultaneously via a noisy medium.
Han and Kobayashi \cite{HanKobayashi} (see also \cite{CMGElGamal})
obtained the best known inner bound for this channel in the classical
asymptotic iid setting.

The multiple access and interference channels can be defined in 
the context of quantum
information theory also. Early work studied the sending of classical 
information over a quantum MAC, without \cite{MAC_Winter} or with 
\cite{hsieh:qmac} entanglement assistance, in the asymptotic iid setting. 
These works obtained the natural quantum analogues of the optimal 
classical rate regions using 
successive cancellation and time sharing.
Later, Fawzi et al.~\cite{fawzi:interference} and 
Sen~\cite{sen:interference} studied the sending of classical 
information over a quantum interference channel in the 
asymptotic iid setting by first obtaining a simultaneous decoder for the
quantum MAC. The latter paper managed to obtain 
the natural quantum analogue
of the Han-Kobayashi inner bound.

For a variety of reasons recent research in Shannon theory has
studied in depth the {\em one-shot} setting where
the channel can be used only once. This is the most general setting and
subsumes the asymptotic iid, asymptotic non-iid aka information spectrum,
and finite block length settings.
Ideally, the one-shot inner bounds
should match or supersede the best inner bounds for the respective
channels in the asymptotic iid setting. 
Sen \cite{Sen18_1} obtained the natural one-shot quantum analogues
of best known classical rate regions for sending classical
information over entanglement unassisted and assisted quantum MACs and
quantum interference channels.
His one-shot inner bounds, obtained by simultaneous decoding, approach 
the optimal inner bounds known earlier
for the classical and quantum asymptotic iid settings. 

Note that presence of shared randomness does not affect the rates of
sending classical or quantum information over channels. Also the 
rates of sending quantum information and classical information
over entanglement assisted quantum channels are related by a factor of
two because of quantum teleportation. So the main setting left 
unstudied in the above works is the setting of sending quantum 
information over an entanglement unassisted quantum channel i.e. 
the senders and the receivers do not share 
any entanglement prior to the beginning of the protocol.
The first works to address this setting looked at a point-to-point
quantum channel in the asymptotic iid setting 
\cite{Lloyd,Shor_Notes}, culminating in the work of Devetak \cite{Devetak}
which showed with full rigour
that the {\em regularised coherent information} from sender $A$ to
receiver $B$ defined by
$I^*(A>B) := \lim_{k \rightarrow \infty} I(A^k>B^k) / k$,
$I(A^k > B^k) := H(B^k) - H(A^k B^k)$ where $A^k B^k$ is defined by
the channel action $(\mathcal{N}^{A' \to B})^{\otimes k}$ on an arbitrary
(in general entangled) pure state $\ket{\sigma}^{A^k (A')^k}$, is the
capacity of an unassisted quantum channel in the asymptotic iid limit. 
\begin{comment}
Devetak \cite{Devetak} provided the first fully rigorous proof
of this statement by proving an elegant connection between the
quantum capacity of an unassisted quantum channel and its classical
private capacity.
\end{comment}
Hayden et al.  \cite{Decoupling_Hayden_Horodecki}. 
showed that one can recover Devetak's 
result using a technique called {\em decoupling} 
\begin{comment}
Devetak's private coding technique was geared towards a specific
problem called {\em entanglement transmission} where the sender has
to transmit the quantum state of a system which might be entangled
with a second reference system, and the receiver must be able to decode
the channel output so that at the end of the protocol the joint quantum
state of the system plus reference is approximately preserved. 
Hayden et al.'s
decoupling technique was geared towards another problem called
{\em  entanglement generation} where the protocol aims to create EPR
pairs shared between the sender and the receiver. Once such EPR pairs
are created, it is easy to solve entanglement transmission by quantum
teleportation when an additional noiseless classical channel is provided.
Though entanglement generation looks to be weaker than entanglement
transmission and moreover requires an additional noiseless classical
channel, it was shown in \cite{Tema} that nevertheless both
these as well as several variant tasks are essentially equivalent.
\end{comment}

These works naturally lead one to consider unassisted multiterminal 
quantum channels. 
To the best of our knowledge, the only inner bound known
for the unassisted QIC is what one would obtain by treating the channel
as two independent unassisted point to point channels. For the unassisted
QMAC more is known.
Yard et al. \cite{Yard_MAC}
showed that the natural quantum analogue of the classical rate region,
with mutual information replaced by regularised coherent information
as in Figure~\ref{fig:QMACasymp},
is an inner bound for the unassisted quantum MAC (QMAC) in 
the asymptotic iid setting. 
They proved their inner bound by time sharing
and a suitable adaptation of successive cancellation.
\begin{figure}[hhh]
\vspace*{-5mm}
\begin{center}
\begin{tikzpicture}
\draw [->, line width=1 pt] (1,-2) -- (1,2);
\draw [->, line width=1 pt] (1,-2) -- (5,-2);
\draw [line width=1 pt] (1,0) -- (2,0);
\draw [line width=1 pt] (3,-2) -- (3,-1);
\draw [line width=1 pt] (2,0) -- (3,-1);
\draw [line width=1 pt,dotted] (1,1) -- (4,-2);
\draw (1,2) node[anchor=east] {Bob};
\draw (5,-2) node[anchor=west] {Alice};
\draw (1.5,-1) node[anchor= west] {$n\rightarrow \infty$};
\draw (2.5,-0.5) node[anchor=west] {{\footnotesize $P$}};
\draw (1,0) node[anchor=north] {{\footnotesize $(0,I^*(B>AC))$}};
\draw (4,-2) node[anchor=north ] {{\footnotesize $(I^*(A>BC), 0)$}};
\draw (3,-1) node[anchor=west] 
{{\footnotesize $S = (I^*(A>BC), I^*(B>C))$}};
\draw (2,0) node[anchor=west] 
{{\footnotesize $T = (I^*(A>C),I^*(B>AC))$}};
\draw (1,1) node[anchor=west] {{\footnotesize $x+y= I^*(AB>C)$}};
\end{tikzpicture}
\end{center}
\vspace*{-5mm}
\caption{Achievable rate region for the unassisted quantum MAC
per channel use in the asymptotic iid setting.}
\label{fig:QMACasymp}
\end{figure}

The above works behoove one to consider the problem of sending
quantum information over an unassisted quantum channel in the one-shot
setting. Buscemi and Datta \cite{Buscemi_Datta} proved the first
one-shot achievability result for the unassisted point-to-point channel
in terms of smooth modified R\'{e}nyi entropies. Their result was
generalised by Dupuis \cite{Dupuis_thesis} to the case where the receiver
has some side information about the sender's message. In the asymptotic 
iid limit,
these one-shot results approach the regularised coherent information
obtained in earlier works.

It is thus natural to study inner bounds for the unassisted QMAC in the 
one-shot setting. In this paper we take the first steps towards this
problem. Observe that successive cancellation can only
give the two endpoints $S$ and $T$ of the dominant line of the 
pentagonal rate region in Figure~\ref{fig:QMACasymp}.
Since time sharing cannot be used in the one-shot setting, it is not
clear how to obtain other rate tuples like the point $P$. An alternative
would be to develop a simultaneous decoder for the QMAC which can 
obtain a point like $P$ directly, but that is a major open problem
with connections to the notorious {\em simultaneous smoothing} open
problem \cite{chakraborty:simultaneousnew}.

Instead in this paper, we take inspiration from another powerful 
classical channel coding technique called {\em rate splitting}. Grant 
et al. \cite{Rate_Splitting_Urbanke} showed that 
it is possible to `split' Alice into two senders Alice\sub{0} and 
Alice\sub{1}, each sending disjoint parts of Alice's original message, 
such that any point in the pentagonal rate region of 
Figure~\ref{fig:QMACasymp} like P can be obtained without time sharing 
by a successive cancellation process where Charlie first decodes 
Alice\sub{0}'s message, then Bob's message using Alice\sub{0}'s message 
as side information and finally Alice\sub{1}'s message using Bob's and 
Alice\sub{0}'s messages as side information. Though Grant et al.'s rate 
splitting technique was developed for the classical MAC in the 
asymptotic iid setting, in this paper we show how it can be 
{\em adapted to the one-shot quantum setting}. This is a non-trivial 
task, which we tackle in two steps. In the first step we
use ideas from Yard et al. \cite{Yard_MAC} and Dupuis \cite{Dupuis_thesis} 
and suitably adapt
successive cancellation to the one-shot unassisted quantum 
setting. In the second step, we adapt the rate splitting function of 
Grant et al. \cite{Rate_Splitting_Urbanke} to the one-shot quantum 
setting. Our one-shot rates are in terms of the
smooth coherent R\'{e}nyi-2 information defined in
Section~\ref{sec:preliminaries}. Since the smooth coherent 
R\'{e}nyi-2 information
is not known to possess a chain rule with equality, we get an 
achievable rate region of the form in Figure~\ref{fig:QMAConeshot1Intro}. 
Our achievable rate region is a subset of the `ideal' 
pentagonal rate region shown by the dashed line. Nevertheless, 
using a quantum asymptotic 
equipartition result of Tomamichel et al. \cite{QAEP}, we 
show that this `subpentagonal' achievable rate region approaches the 
`pentagonal' region of Yard et al. \cite{Yard_MAC} (equal to the 
region demarcated by the dashed line) in the iid limit.
\begin{figure}[hhh]
\vspace*{-5mm}
\hspace*{-8mm}
\begin{tikzpicture}
\draw [->, line width=1 pt] (1,-2) -- (1,2);
\draw [->, line width=1 pt] (1,-2) -- (5,-2);
\draw [line width=1 pt] (1,0) -- (2,0);
\draw [line width=1 pt] (3,-2) -- (3,-1);
\draw [line width=1 pt,dash pattern=on 5pt off 5pt] (4.2,-2) -- (1,1.2);
\draw [shift={(3.04,0.0)},line width=1 pt]  
plot[domain=3.15:4.65,variable=\t]
({1*cos(\t r)},{1*sin(\t r)});
\draw (1,2) node[anchor=east] {Bob};
\draw (5,-2) node[anchor=west] {Alice};
\draw (2.25,-0.7) node[anchor= east] {{\footnotesize $n = 1$}};
\draw (2.75,-0.5) node[anchor= west] 
{{\footnotesize $n \rightarrow \infty$}};
\draw (3,-1) node[anchor=west] 
{{\footnotesize $S = (I^{\epsilon^2/800}_{2}(A>BC), 
		      I^{\epsilon^2/800}_{2}(B>C))$}};
\draw (2,0) node[anchor=west] 
{{\footnotesize $T = (I^{\epsilon^2/800}_{2}(A>C),
		      I^{\epsilon^2/800}_{2}(B>AC))$}};
\draw (4.2,-2) node[anchor=north] 
{{\footnotesize $(I^{\sqrt{\epsilon}}_{2}(AB>C), 0)$}};
\draw (1,1.2) node[anchor=east] 
{{\footnotesize $(0, I^{\sqrt{\epsilon}}_{2}(AB>C))$}};
\draw (2.5,-0.77) node[anchor=north] {{\footnotesize $P$}};
\end{tikzpicture}
\vspace*{-5mm}
\caption{One-shot achievable rate region for the unassisted QMAC 
(for single channel use only), contained
inside the `ideal' pentagonal region demarcated by the dashed line, and 
approaching it in the asymptotic iid limit. 
$O(\log \epsilon)$ additive factors have been ignored in the figure.}
\label{fig:QMAConeshot1Intro}
\end{figure}
The reason why splitting of Alice into Alice\sub{0} and Alice\sub{1}
allows one to obtain a `middle' rate point like $P$, in addition to the
`corner' points $S$ and $T$, is as follows. The rate point $P$ is
the projection onto the (Alice, Bob) plane of the `corner' rate point
$P'$ in the (Alice\sub{0}, Bob, Alice\sub{1}) space where the rates of
Alice\sub{0} and Alice\sub{1} are summed to obtain Alice's rate. The
point $P'$ can be obtained by a 3-step successive cancellation decoding. 
Note that the split of Alice depends on the rate point $P$ to be attained.
\begin{figure}[hhh]
\vspace*{-5mm}
\hspace*{-5mm}
\begin{tikzpicture}
\draw [->, line width=1 pt] (1,-2) -- (5,-2);
\draw [->, line width=1 pt] (1,-2) -- (0,-4);
\draw [->, line width=1 pt] (1,-2) -- (1,2);
\draw (5,-2) node[anchor=west] {Alice\sub{0}};
\draw (0,-4) node[anchor=east] {Bob};
\draw (1,2) node[anchor=east] {Alice\sub{1}};
\draw (4,1.75) node[anchor= west] {{\footnotesize $n = 1$}};
\draw [line width=1 pt] 
(3,1) -- (2,0.5) -- (1.75,-0.5) -- (2.3,-1) -- 
(3.7,-0.3) -- (4,0.5) -- (3,1);
\draw (1,-0.1) node[anchor=east] 
{{\footnotesize 
$
(I^{\epsilon^2/800}_{2}(A_0>C),
 I^{\epsilon^2/800}_{2}(B>A_0 C),
$
}};
\draw (1.8,-0.5) node[anchor=east] 
{{\footnotesize 
$
I^{\epsilon^2/800}_{2}(A_1>A_0 B C)) = P'
$
}};
\end{tikzpicture}
\vspace*{-5mm}
\caption{The `corner' point $P'$ can be obtained by successive
cancellation following the order Alice\sub{0} $\to$ Bob $\to$
Alice\sub{1} with splitting of Alice followed by one use of the unassisted
QMAC. Point $P'$ projects down to 
point $P$ in Figure~\ref{fig:QMAConeshot1Intro}. Only the `dominant face' of
the rate region is shown. Successive cancellation can only obtain the
corner points of the dominant face and all `sub-points' by `resource
wasting'. It cannot obtain `middle' points of the `dominant' face.
$O(\log \epsilon)$ additive factors have been ignored in the figure.}
\label{fig:QMAConeshot2}
\end{figure}

In fact, it turns out that our techniques are more general; they allow us to a obtain non-trivial achievable rate region for sending quantum information over a QMAC with \emph{rate limited entanglement assistance}. In the case of rate limited entanglement assistance, the amount of prior shared entanglement between the sender and the receiver is limited by a certain upper bound. If this upper bound is set to $0$, the situation reduces exactly to the unassisted case.  As the upper bound tends to infinity, the situation becomes the same as the QMAC with unlimited entanglement assistance.

We now state our result for the unassisted QIC. The trivial 
inner bound treats the QIC as two independent unassisted point to point
channels from Alice to Charlie and Bob to Damru. 
Rate splitting and successive cancellation can be similarly used to obtain
non-trivial rate regions for the unassisted QIC where one party,
say Alice, sacrifices her rate in order to boost Bob's rate with
respect to the trivial inner bound. The situation is summarised in
Figure~\ref{fig:QIConeshotIntro}.
\begin{figure}[hhh]
\vspace*{-5mm}
\begin{center}
\begin{tikzpicture}
\draw [->, line width=1 pt] (1,-2) -- (1,2);
\draw [->, line width=1 pt] (1,-2) -- (5,-2);
\draw (1,2) node[anchor=east] {Bob};
\draw (5,-2) node[anchor=west] {Alice};
\draw [line width=1 pt, dotted] (3,-2) -- (3,0);
\draw [line width=1 pt, dotted] (1,0) -- (3,0);
\draw (3,0) node[anchor=west] 
{{\footnotesize $(I^{\epsilon^2/800}_{2}(A>C),
		  I^{\epsilon^2/800}_{2}(B>D))$}};
\draw [line width=1 pt] (2,-2) -- (2,1);
\draw [line width=1 pt] (1,1) -- (2,1);
\draw (2,1) node[anchor=west] 
{{\footnotesize $(I^{\epsilon^2/800}_{2}(A_0>C), 
		  I^{\epsilon^2/800}_{2}(B>A_1 D))$}};
\draw [line width=1 pt, dash pattern=on 3pt off 2pt] (4,-2) -- (4,-1);
\draw [line width=1 pt, dash pattern=on 3pt off 2pt] (1,-1) -- (4,-1);
\draw (4,-1) node[anchor=west] 
{{\footnotesize $(I^{\epsilon^2/800}_{2}(A>B_1 C), 
		  I^{\epsilon^2/800}_{2}(B_0 > D))$}};
\draw (4.5,1.5) node[anchor= west] {{\footnotesize $n = 1$}};
\end{tikzpicture}
\end{center}
\vspace*{-5mm}
\caption{One-shot achievable rate region (for single channel use
only) for the unassisted QIC.
The trivial region is shown dotted. Alice can sacrifice her rate
in order to boost Bob's rate with respect to the trivial region, 
as shown by the solid rectangle. The dashed rectangle can be similarly
obtained by Bob sacrificing his rate in order to boost Alice's.
$O(\log \epsilon)$ additive factors have been ignored in the figure.}
\label{fig:QIConeshotIntro}
\end{figure}
Though the discussion above only involved unassisted QMAC and QIC, our
actual results also hold for the QMAC and QIC with limited entanglement 
assistance. However they seem to be inferior to the known results 
when entanglement assistance is unlimited \cite{Sen18_2}.

\paragraph{
Subsequent Works:
} After the arXiv and conference versions of this work were published \cite{chakraborty:ratesplittingnew}, Saus and Winter  \cite{colomer2023decoupling} obtained a partially smooth one shot simultaneous coding strategy for sending quantum information across the QMAC. They proved their nice result by proving a partially smoothed generalisation of the (non smooth) multi sender decoupling theorem given by Chakraborty et al. \cite{chakraborty:simultaneousnew}. As a result they obtain the natural smooth one shot analog of Figure \ref{fig:QMACasymp}. Hence they have an alternate derivation of the asymptotic iid rate region shown in Figure \ref{fig:QMACasymp} without appealing to rate splitting. However, their methods don't seem to be generalisable to the case of the QMAC with limited entanglement assistance. This is because their methods cannot smooth over the Choi state of the channel, which seems crucial for obtaining any non-trivial inner bounds in the limited entanglement assisted setting. Thus to the best of our knowledge, the present work is the only one providing a non-trivial smooth one shot achievable rate region for the QMAC with limited entanglement assistance. Besides, rate splitting has proved to be a powerful technique in classical network information theory, and so its generalisation to the most general one shot quantum setting should be of independent interest.

\section{Organisation of the Paper}
The paper is organised as follows. In Section \ref{sec:preliminaries} we present the definitions and facts regarding one-shot entropic quantities and other necessary mathematical tools that we will need throughout the paper. In Section \ref{chap:RateSplit1} we introduce the concept of quantum rate splitting and demonstrate it in the case of entanglement transmission across the point to point channel. We also develop the technique of successive cancellation decoding for entanglement transmission codes in this section. In Section \ref{chap:RateSplit2} we use the ideas introduced in Section \ref{chap:RateSplit1} to derive inner bounds for entanglement transmission over the QMAC and the QIC. Finally, in Section \ref{sec:IID} we present the asymptotic IID versions  of the one-shot inner bounds presented in  paper.

\section{Preliminaries}
\label{sec:preliminaries}
\subsection{Notation}

We will use the following conventions throughout the rest of the paper : 
\begin{enumerate}
\item We use the shorthand $M \cdot N := M N M^\dag$ for operators $M$ and $N$.
    \item Suppose that $\ket{\omega(U)}^{XA'B'}$ be a generic intermediary state (defined in \cref{sec:almostcptp}), where $X$ is a placeholder for other systems involved in the protocol. Suppose we are given a channel $\mathcal{N}^{A'B' \rightarrow C}$ and its corresponding Stinespring dilation $\mathcal{U}_{\mathcal{N}}^{A'B'\rightarrow CE}$. Then, we denote the state $\mathcal{U}_{\mathcal{N}}\ket{\omega(U)}^{XA'B'}$ by the symbol $\ket{\omega(U)}^{XCE}$. Although the two states are denoted using the same greek letter, we differentiate them by the systems on which they are defined. These systems will always be explicitly mentioned whenever we make use of this convention.
    \item We will use the same rule for control states, For example, suppose $\ket{\sigma}^{A"A'B"B'}$ is a control state for some channel coding protocol. Suppose we are given the channel $\mathcal{N}^{A'B'\rightarrow C}$ Then we use the following convention
    \begin{align*}
        \sigma^{A"B"C}\coloneqq \mathcal{N}\cdot \sigma^{A"A'B"B'}
    \end{align*}
    We will use this convention while specifying entropic quantities. It will be clear from context which state we refer to. For example, consider the expressions $H_{\min}^{\epsilon}(A")_{\sigma}$  and $I_{\min}^{\epsilon}(A"|C)_{\sigma}$. It is clear from the arguments of the entropic expressions that in the first case $\sigma=\sigma^{A"A'B"B'}$ and in the second case $\sigma=\sigma^{A"B'C}$.
    \item We will, on several occasions use the operator $\textup{op}^{X\rightarrow YA'B'}(\ket{\omega(U)}^{XYA'B'})$. To lessen the burden on notation, whenever we use this operator, we will not mention the systems on which the argument of the $\textup{op}$ operator is defined. It will however always mention the domain and range of the $\textup{op}$ operator in these cases to avoid any confusion.
\end{enumerate}

\subsection{Smooth Entropies}
For a pair of subnormalised density matrices $\rho$ and $\sigma$ in 
the same Hilbert space their {\em purified distance} is denoted by
$
P(\rho,\sigma):=\sqrt{1-F(\rho,\sigma)^2}
$
where 
$
F(\rho,\sigma):=
\norm{\sqrt{\rho} \sqrt{\sigma}}_1
+\sqrt{(1-\Tr[\rho])\cdot (1-\Tr[\sigma])}
$ 
is the {\em generalised fidelity} and $\norm{\cdot}_1$ is the Schatten
$1$-norm.
We use $\sigma \approx_\epsilon \rho$ as a shorthand for 
$P(\sigma, \rho) \leq \epsilon$. See \cite{Hmin_continuous} for more details.

The von Neumann entropy for a normalised quantum state
$\rho^A$ is defined by $H(A)_\rho := -\Tr [\rho \log \rho]$.
For a bipartite quantum state $\rho^{AB}$, the Coherent Information $I(A>B)$ is defined as $I(A>B)\coloneqq H(A|E)$ where the conditional entropy $H(A|E)$ is computed with respect to the purification $\ket{\rho}^{ABE}$ of the state $\rho^{AB}$.

\begin{definition}{\bf [{\em $\epsilon$-smooth sandwiched  R\'{e}nyi-2
conditional entropy}]}
    Given a bipartite state $\rho^{AB}$, the {\em $\epsilon$-smooth sandwiched  R\'{e}nyi-2
conditional entropy} is defined as
\[
\begin{array}{l}
H_2^\epsilon(A | E)_{\rho}  := \\
\displaystyle
-2 \log \min_{(\rho')^{AE} \approx_\epsilon \rho^{AE}} \min_{\sigma^E} 
\norm{(\one^A \otimes (\sigma^E)^{-1/4}) \cdot (\rho')^{AE}}_2,
\end{array}
\]
where $\sigma^E$ ranges over non-singular normalised states over $E$.
\end{definition}

\begin{definition}{\bf [$\epsilon$-smooth conditional min-entropy]}

The {\em $\epsilon$-smooth conditional min-entropy} is given by
\[
\begin{array}{l}
H_{\mathrm{min}}^\epsilon(A | E)_{\rho} := \\
\displaystyle
-\log \min_{(\rho')^{AE} \approx_\epsilon \rho^{AE}} 
\min_{\sigma^E: (\rho')^{AE} \leq \one^A \otimes \sigma^E} 
\Tr [\sigma^E],
\end{array}
\]
where $\sigma^E$ ranges over positive semidefinite operators on $E$.
    
\end{definition}

The unconditional smooth entropies are now defined from the conditional ones by taking the conditioning system to be one dimensional.

\begin{definition}{\bf [$\epsilon$-smooth coherent min-information]}\label{def:SmoothCoherentMinInformation}
Then the {\em $\epsilon$-smooth coherent min-information} aka
the negative of the {\em $\epsilon$-smooth conditional max-entropy} 
is given by
\[
I^\epsilon_{\mathrm{min}}(A \rangle  B)_\rho \coloneqq
-H_{\mathrm{max}}^\epsilon(A | B)_\rho \coloneqq
H_{\mathrm{min}}^\epsilon(A | E)_{\rho},
\]
where again $\ket{\rho}^{A B E}$ is a purification of $\rho^{AB}$. 
\end{definition}

As shown in \cite{Dupuis_thesis}, the smooth sandwiched R\'{e}nyi-$2$
conditional entropy upper bounds the 
smooth conditional min-entropy.
The smooth conditional min-entropy
is further lower bounded by the familiar conditional Shannon entropy
in the amortised sense in the asymptotic iid limit \cite{QAEP},
a result that is sometimes referred to as the fully quantum asymptotic
equipartition property.
To summarise, the smooth sandwiched R\'{e}nyi-$2$
coherent information  upper bounds the Shannon coherent information 
in the amortised sense in the asymptotic iid limit.

We will now state some properties on the smooth conditional min entropy that we will use throughout the rest of the paper.

\begin{fact}[Chaining for Smooth min-entropy \cite{chain_rule_hmin1,OneShotdecouplng_chainRuleHmin}]\label{chainingforminentropy}

Let $\epsilon>0$ and $\epsilon',\epsilon"\geq 0$ and let $\rho^{ABC}$ be a quantum state. Then
\[
H_{\min}^{\epsilon+2\epsilon'+\epsilon"}(AB|C)_{\rho}\geq H_{\min}^{\epsilon'}(A|BC)_{\rho}+H_{\min}^{\epsilon"}(B|C)_{\rho}-\log\frac{2}{\epsilon^2}
\]

\end{fact}

\begin{fact}[Unitary Invariance of Smooth min-entropy]\label{unitaryinvarforhmin}

Given $\epsilon\geq 0$, a quantum state $\rho^{AB}$ and isometries $U : \mathcal{H}_A\rightarrow \mathcal{H}_C$ and $V : \mathcal{H}_B\rightarrow\mathcal{H}_D$, define the state $\sigma^{CD}\coloneqq (U\otimes V)\rho^{AB}(U^{\dagger}\otimes V^{\dagger})$. Then

\[H_{\min}^{\epsilon}(A|B)_{\rho}=H_{\min}^{\epsilon}(C|D)_{\sigma}
\]
\end{fact}
\begin{fact}[Continuity of Smooth min-entropy]\label{continuityofhmin}
Given two quantum states $\rho^{AB}$ and $\sigma^{AB}$ such that $P(\rho{^{AB},\sigma^{AB}})\leq \delta$ and $\epsilon>0$, then 
\[
\abs{H_{\min}^{\epsilon}(A|B)_{\rho}-H_{\min}^{\epsilon}(A|B)_{\sigma}}\leq c\cdot \delta'
\]
where $c$ is an absolute constant and depends on the dimensions of system $A$ and $B$ and $\delta'=\sqrt{\delta^2+2\epsilon\delta}$
\end{fact}
The proofs of both \cref{unitaryinvarforhmin} and \cref{continuityofhmin} can be found in  \cite{Hmin_continuous}.

\begin{fact}[Quantum Asymptotic Equipartition Property \cite{QAEP}]\label{quantumaep}
Given a bipartite quantum state $\rho^{AB}$ on the system $\mathcal{H}_{A}\otimes \mathcal{H}_{B}$, $\epsilon>0$, an integer $n\in \mathbb{N}$ and the iid extension of the state $\rho^n_{AB}$ it holds that
\begin{align*}
    \lim\limits_{\epsilon\rightarrow 0}\lim\limits_{n\rightarrow \infty}\frac{1}{n}H_{\min}^{\epsilon}(A^n|B^n)_{\rho^n}=H(A|B)_{\rho}
\end{align*}
\end{fact}

\subsection{The \textbf{op} Operator}\label{subsec:opoperator}

One of the main technical tools we use in this paper, which is a workhorse in most of our proofs, is the notion of mapping a vector into an operator. This operation is denoted simply by '\textup{op}' and we compile some of its properties in this section for completeness. The interested reader is referred to \cite{Dupuis_thesis} for further details.

\begin{definition}{\bf[The \textup{op} operator]}
Given the systems $A$ and $B$, fix the standard bases $\ket{a_i}^A$ and 
$\ket{b_j}^B$. Then we define $\textup{op}^{A\rightarrow B} : A\otimes B \rightarrow L(A,B)$ as 
\begin{align*}
    \textup{op}^{A\rightarrow B}(\ket{a_i}\ket{b_j})\coloneqq \ket{b_j}\bra{a_i}~~~~~~~~~~~\forall i,j
\end{align*}
Notice that this definition is basis dependant and hence whenever we use this operator a choice of bases is implied, although not always explicitly mentioned.

\begin{fact}\label{swaplemma}
   Let $\ket{\psi}^{AB}$ and $\ket{\varphi}^{AC}$ be vectors on the systems $AB$ and $AC$ respectively. Then 
   \begin{align*}
       \textup{op}^{A\rightarrow C}(\ket{\varphi}^{AC})\ket{\psi}^{AB}=\textup{op}^{A\rightarrow B}(\ket{\psi}^{AB})\ket{\varphi}^{AC}
   \end{align*}
  \end{fact}

\begin{fact}\label{opepr}
Given a vector $\ket{\psi}^{AB}$, let $\ket{\Phi}^{AA'}$ be an EPR state, where $A\cong A'$. Then, 
\begin{align*}
    \sqrt{\abs{A}}\textup{op}^{A\rightarrow B}(\ket{\psi}^{AB})\ket{\Psi}^{AA'}=\ket{\psi}^{A'B}
\end{align*}
\end{fact}

\begin{fact}\label{opu}
For all vectors $\ket{\psi}^{AB}$ and any $M^{A\rightarrow C}$,
\begin{align*}
    \textup{op}^{C\rightarrow B}(M\ket{\psi})=\textup{op}^{A\rightarrow B}(\ket{\psi})M^T
\end{align*}
\end{fact}

\begin{fact}
For all $\ket{\psi}^{AB}$,
\begin{align*}
    \Tr_B[\psi^{AB}]=\textup{op}^{B\rightarrow A}(\ket{\psi})\textup{op}^{B\rightarrow A}(\ket{\psi})^{\dagger}
\end{align*}
\end{fact}
\end{definition}

\subsection{The Smooth Single Sender Decoupling Theorem}
\begin{fact}{\textup{\textbf{Smooth Decoupling Theorem \cite{Szehr_2013} }}}\label{fact:singledecoup}
Given $\epsilon>0$ a density matrix $\rho^{AE}$ and any completely positive operator $\mathcal{T}^{A\rightarrow R}$, define $\omega^{A'R}\coloneqq \big( \mathcal{T}\otimes \mathbb{I}^{A'}\big)\Phi^{AA'}$ such that $\Tr\left[\omega^{A'R}\right]=1$. Then
\begin{align*}
    \int\limits_{\mathbf{U(A)}}\norm{\mathcal{T}\big(U\cdot \rho\big)-\omega^E\otimes \rho^R}_1\leq {2}^{-\frac{1}{2}H^{\epsilon}_{2}(A'|R)_{\omega}-\frac{1}{2}H^{\epsilon}_{2}(A|E)_{\rho}}+8\epsilon
\end{align*}
where the integration is over the Haar measure on the set of all unitaries on the system $A$, denoted by $\mathbb{U(A)}$.
\end{fact} 

The single sender decoupling theorem implies the following channel coding theorem.

\begin{fact}{\mbox{\cite[Theorem~3.14]{Dupuis_thesis}}}
\label{fact:dupuis_pointtopoint}
Let  $\ket{\psi}^{ABR}$ be a pure state, $\mathcal{N}^{A' \to C}$ be any CPTP
superoperator with Stinespring dilation $U_{\mathcal{N}}^{A' \to CE}$, N and complementary channel
$\bar{\mathcal{N}}^{A' \to E}$, let $\omega^{A" CE} := U_{\mathcal{N}} \cdot \sigma^{A" A'}$, where $\sigma^{A" A'}$ is any pure state and $A" \cong A'$ , and let
$\epsilon > 0$. Then, there exists an encoding partial isometry $V^{A \to A'}$ and a decoding superoperator $\mathcal{D}^{CB \to AB}$ such that:
$$
\lVert \bar{\mathcal{N}}(V \cdot \psi^{AR}) - \omega^E \otimes \psi^R\rVert_1 \leq 2 \sqrt{2\delta_1}+\delta_2
$$
and
$$
\lVert (\mathcal{D} \circ \mathcal{N} \circ \mathcal{E})\psi^{ABR} - \psi^{ABR}\rVert_1 \leq 2 \sqrt{(4\sqrt{2\delta_1}+2\delta_2)}
$$
where
$\delta_1 := 3 \times 2^{\frac{1}{2}H_{\max}^\epsilon(A)_\psi-\frac{1}{2}H^\epsilon_{2}(A")_\omega} + 24 \epsilon$,
$\delta_2 := 3\cdot 2^{-\frac{1}{2}H_{2}^\epsilon(A"|E)_\omega-\frac{1}{2}H^\epsilon_{2}(A|R)_\psi} + 24 \epsilon$
\end{fact}

\subsection{Miscellaneous Useful Facts}

\begin{fact}\label{usefullemma}
 Given states $\rho^{ABC}, \sigma^A, \eta^C, \sigma^{AB}, \omega^{BC}$ such that
 \begin{align*}
     &\norm{\rho^{ABC}-\sigma^A\otimes \omega^{BC}}_1 \leq \epsilon_1 \label{ineq1}\\
     &\norm{\rho^{ABC}-\sigma^{AB}\otimes \eta^C}_1 \leq \epsilon_2 
 \end{align*}
 it holds that 
 \begin{align*}
     \norm{\rho^{ABC}-\sigma^A\otimes \sigma^B\otimes \eta^C}_1 \leq 2\epsilon_1+\epsilon_2
 \end{align*}
\end{fact}

\begin{fact}\label{normalisation}
For  any two density matrices $\rho$ and $\sigma$ and any real $c\in \mathbb{R}$, the following holds true:
\[ \norm{\rho-\sigma}_1\leq 2\norm{c\rho-\sigma}_1
\]
\end{fact}

\section{Quantum Channel Capacities: Definitions and Previous Work}\label{sec:EntanglementTransmission}

The capacity of a quantum channel can have many different and distinct interpretations, based on the information processing task being considered. The various definitions of the quantum capacity arise from considerations such as whether the information being sent through the channel is classical data or whether it consists of arbitrary quantum states.  Further, the definition of capacity changes whether the sender and the receiver can make use of pre-shared entanglement or EPR pairs, that they prepared before the protocol began. In this section, we will introduce the \emph{entanglement unassisted} and \emph{entanglement assisted} quantum capacities of a quantum channel. 

We will first define the capacities assuming that the sender Alice and the receiver Bob can utilise only one copy
of the channel i.e. the one-shot regime. We will then generalise to the case when many copies of the channel are available for use i.e. the iid regime.

The definition of the quantum capacity of a quantum channel stems from the following intuition: given a quantum channel $\mathcal{N}^{A'\to B}$, we want to exhibit a subspace $A_{\good}\subset A'$ such that channel acts approximately like the identity channel on this subspace. To make this precise, consider that the sender Alice has some system $A$ which holds her quantum message and an encoder $\mathcal{E}^{A\to A'}$. After receiving the quantum system $B$, Bob produces a guess for the contents of Alice's $A$ system, by using the decoding map $\mathcal{D}^{B\to A}$. Note that the state on system $A$ can be arbitrarily entangled with systems that are not accessible to the protocol. To capture this notion, we consider the purification $\ket{\psi}^{AR}$ of the state on the system $A$. Thus, the goal of the protocol is to fulfil the condition:
\[
\norm{\psi^{AR}-\mathcal{D}\circ \mathcal{N} \circ \mathcal{E} (\psi)}_1 \leq \eps
\]
for all pure states $\ket{\psi}^{AR}$. This is precisely equivalent to the condition that
\[
\dnorm{\I^{A}-\mathcal{D}\circ\mathcal{N}\circ\mathcal{E}}\leq \eps .
\]
It is difficult to show the existence of coding schemes using this definition of the quantum capacity, due to the maximisation over all pure states. However, in 2003, Werner and Kretschmann \cite{Tema} showed that there exist several other equivalent definitions of the quantum capacity that are operationally more useful. One such definition is the \emph{entanglement transmission capacity} of a quantum channel:
\begin{definition}{\bf [Entanglement Transmission Capacity]}
A $(Q,\epsilon)$ entanglement transmission code consists of an encoder $\mathcal{E}^{A\to A'}$ and a decoding CPTP $\mathcal{D}^{B\to A}$ such that
\begin{align*}
    \ket{\Phi}^{RA}=\frac{1}{2^{Q}}\sum\limits_{i=1}^{2^Q}\ket{i}^R\ket{i}^A, \\
    F\left( \ket{\Phi}^{RA},\mathbb{I}^R\otimes (\mathcal{D}\circ \mathcal{N}\circ \mathcal{E})(\Phi^{RA})\right) \geq 1-\epsilon .
\end{align*}

$Q$  is said to be an achievable rate for entanglement transmission if there exists a $(Q,\epsilon)$ entanglement transmission code. The supremum of the set of all achievable rates $Q$, where the supremum is taken over all encoding and decoding maps, is defined to be the entanglement transmission capacity of the channel.
\end{definition}
Werner and Kretschmann showed that given a $(Q,\eps)$ entanglement transmission code with the encoder decoder pair $(\mathcal{E},\mathcal{D})$, there exists another encoder decoder pair $(\mathcal{E}',\mathcal{D}')$ such that, for all pure states $\ket{\psi}^{AR}$ 
\[
\norm{\psi^{AR}-\mathcal{D}'\circ \mathcal{N} \circ \mathcal{E}' (\psi)}_1 \leq \eps,
\]
where
\[
\log~ \abs{A}\geq Q-1 .
\]
Refer to \cite{Tema, Buscemi_Datta} and \cite{BarnumKnillNielson1} for details. In this paper, we will only prove the existence of codes for entanglement transmission. 

We will now consider the case when Alice and Bob share EPR pairs to potentially boost the rate of entanglement transmission. In this setting, Alice and Bob share the EPR state $\ket{\Phi}^{\Tilde{A}B}$ where $\Tilde{A}$ is with Alice and $B$ lies with Bob. The two parties are allowed to use this state during the protocol, which aims to transmit the $M$ system of the maximally entangled state $\ket{\Phi}^{RM}$ from Alice to Bob. Thus Alice needs to possess an encoder $\mathcal{E}^{M\Tilde{A}\to A'}$ and Bob a decoder $\mathcal{D}^{B\to M}$ such that
\[
    F\left( \ket{\Phi}^{RM},\mathbb{I}^R\otimes (\mathcal{D}\circ \mathcal{N}\circ \mathcal{E})(\Phi^{RM}\otimes \Phi^{\Tilde{A}B})\right) \geq 1-\epsilon .
\]
This is known as entanglement transmission with \emph{entanglement assistance}. As before, let $Q$ denote the rank of the EPR state to be transmitted (in this case $\log~\abs{M}$) and $E$ denote the rate at which pre-shared entanglement is available for use during the protocol (in this case $\log ~\abs{\Tilde{A}}$). Then, the rate $(Q, E)$ is said to be $\eps$-achievable for entanglement transmission with entanglement assistance if there exists an encoder and decoder pair for which the above fidelity condition holds. $Q$ is said to be achievable for \emph{unassisted transmission} if no pre-shared entanglement is used during the protocol i.e. if $(Q,0)$ is $\eps$-achievable.

Now suppose that instead of constraining Alice and Bob to code for only one copy of the channel, we allow them to code for $n$ tensor copies i.e. the channel $\mathcal{N}^{\otimes n}$, where $n$ can be arbitrarily large. In this case, we define the capacity of the channel as the maximum rate at which qubits can be transmitted across the channel per channel use. We state the formal definition below:
\begin{definition}{\bf [Quantum Capacity in the iid Regime]} An $(n,Q)$ code for a quantum $\mathcal{N}^{A'\to B}$ consists of an encoding map $\mathcal{E}_n\coloneqq\mathcal{E}^{A\to A^{'n}}$ and a decoding map $\mathcal{D}_n\coloneqq\mathcal{D}^{C^n\to A}$ such that
\[
\dnorm{\I^{A}-\mathcal{D}_n\circ\mathcal{N}^{\otimes n}\circ\mathcal{E}_n}\leq \eps .
\]
The rate $Q=\frac{1}{n}\log \abs{A}$ is said to be an achievable rate for a the channel $\mathcal{N}^{A'\to B}$ if there exists a sequence of $(n,Q)$ codes $(\mathcal{E}_n,\mathcal{D}_n)$ such that
\[
\lim\limits_{n\to \infty} \dnorm{\I^{A}-\mathcal{D}_n\circ\mathcal{N}^{\otimes n}\circ\mathcal{E}_n} =0 .
\]
The quantum capacity of $\mathcal{N}^{A'\to B}$ is the supremum of all achievable rates for this channel.
\end{definition}

One can similarly generalise the above definition to the case when entanglement assistance is available. 

\subsection{Entanglement Transmission over the QMAC}

\begin{definition}{\bf [One-Shot Entanglement Transmission over the QMAC]}
Given the QMAC $\mathcal{N}^{A'B'\to C}$, with senders Alice and Bob and receiver Charlie, suppose that Alice and Bob are given the $A$ and $B$ parts of the maximally entangled states $\ket{\Phi_1}^{R_1A}$ and $\ket{\Phi_2}^{R_2B}$. Alice and Bob want to send the systems $A$ and $B$ to Charlie via the QMAC with  high fidelity. An entanglement transmission code for the QMAC then consists of the encoding maps $\mathcal{E}_1^{A\to A'}$ and $\mathcal{E}_2^{B\to B'}$ belonging to Alice and Bob respectively, and the decoding map $\mathcal{D}^{C\to AB}$ such that
\[
F(\ket{\Phi_1}\ket{\Phi_2},\mathcal{D}\circ\mathcal{N}\circ \mathcal{E}_1\otimes\mathcal{E}_2 \left(\Phi_1\otimes \Phi_2\right) ) \geq 1-\eps.
\]
The rate of the code is defined as
\[
\begin{aligned}
R_A &\coloneqq \log \abs{A} \\
R_B &\coloneqq \log \abs{B}.
\end{aligned}
\]
\end{definition}
\begin{comment}
Note that the above definition is easily generalized to the asymptotic iid case, as well as the case when Alice and Bob have access to pre-shared entanglement with Charlie.
\end{comment}

\begin{definition}{\bf [Entanglement Transmission Capacity Region of the QMAC]}
Any rate pair $(R_A,R_B)$ for which there exists a corresponding entanglement transmission code is called $\epsilon$-achievable. The union of all $\epsilon$-achievable rate pairs is defined as the achievable rate region for entanglement transmission over the QMAC.
\end{definition}

A natural question is whether we can strengthen the definition of the achievable region for the QMAC to include all states that lie in the spaces corresponding to the systems $A$ and $B$. To that end, we define the task of \emph{strong subspace transmission}
 \cite{Yard_MAC}:
 \begin{definition}{\bf [Strong Subspace Transmission]}
 Suppose Alice and Bob posses some pure quantum states $\ket{\psi}^{R_1A}$ and $\ket{\varphi}^{R_2B}$, where we place no restrictions on the systems $R_1$ and $R_2$ other than that they be finite dimensional. A strong subspace transmission code then consists of encoding maps $\left(\mathcal{E}_1^{A\to A'},\mathcal{E}_2^{B\to B'}\right)$ and a decoding map $\mathcal{D}^{C\to AB}$ such that, for all $\ket{\psi}^{R_1A}$ and $\ket{\varphi}^{R_2B}$
 \[
 F(\ket{\psi}\ket{\varphi},\mathcal{D}\circ \mathcal{N}\circ \mathcal{E}_1\otimes \mathcal{E}_2 \left(\psi\otimes\varphi\right))\geq 1-\eps.
 \]
 The rate pair
 \[
 (R_A,R_B)\coloneqq (\log \abs{A},\log \abs{B})
 \]
 are said to be achievable for strong subspace transmission if there exists a corresponding strong subspace transmission code. The union of all achievable rates gives the achievable region for this task.
 \end{definition}
 
 In \cite[Section 5]{Yard_MAC}, the authors showed that given that Alice and Bob have access to independent public coins with Charlie, the rate regions for entanglement transmission and strong subspace transmission over the QMAC are equivalent. Thus, in this paper, we will design all our protocols for entanglement transmission.
 
 The authors of the paper \cite{Yard_MAC} also provide the best known achievable bounds for this task in the asymptotic iid setting. We state their result below:
 \begin{theorem}
 Given the QMAC $\mathcal{N}^{A'B'\to C}$ its capacity region is given by the closure of 
 \[
 \bigcup\limits_{k=1}^{\infty}\frac{1}{k}\mathcal{Q}(\mathcal{N}^{\otimes k}),
 \]
 where the region $\mathcal{Q}(\mathcal{M})$ equals the pairs of non-negative rates $(R_A,R_B)$ satisfying
 \begin{align*}
     R_A <& I(A\rangle BC)_{\sigma} \\
     R_B <& I(B\rangle AC)_{\sigma} \\
     R_A+R_B <& I(AB\rangle C)_{\sigma},
 \end{align*}
 where all the entropic quantities are computed with respect to the control state
 \[
 \sigma^{ABC}\coloneqq \left(\I^{AB}\otimes \mathcal{M}\right)\left(\Omega^{AA'}\otimes \Delta^{BB'}\right)
 \]
 for a pair of pure states $\ket{\Omega}^{AA'}$ and $\ket{\Delta}^{BB'}$.
 \end{theorem}
 The achievable region shown in the theorem can be picturized by the rate region given in \cref{fig:QMACasymp3}. In the figure, we use the shorthand
\[
I^{*}(A\rangle B)\coloneqq \lim\limits_{k\to \infty}\frac{1}{k}I(A^k\rangle B^k).
\]
Please note that the above shorthand is informal since the quantity on the right hand side is computed with respect to a state on the systems $A^kB^k$. Thus, the precise description of the rate region actually requires a \emph{union} over all such states, over all values of $k$. We use this informal notation to emphasise the shape of the rate region and the fact that the rate expressions are regularised.
\begin{figure}[h]
\vspace*{-5mm}
\begin{center}
\begin{tikzpicture}
\draw [->, line width=1 pt] (1,-2) -- (1,2);
\draw [->, line width=1 pt] (1,-2) -- (5,-2);
\draw [line width=1 pt] (1,0) -- (2,0);
\draw [line width=1 pt] (3,-2) -- (3,-1);
\draw [line width=1 pt] (2,0) -- (3,-1);
\draw [line width=1 pt,dotted] (1,1) -- (4,-2);
\draw (1,2) node[anchor=east] {Bob};
\draw (5,-2) node[anchor=west] {Alice};
\draw (1.5,-1) node[anchor= west] {$n\rightarrow \infty$};
\draw (2.5,-0.5) node[anchor=west] {{\footnotesize $P$}};
\draw (1,0) node[anchor=north] {{\footnotesize $(0,I^*(B\rangle AC))$}};
\draw (4,-2) node[anchor=north ] {{\footnotesize $(I^*(A\rangle BC), 0)$}};
\draw (3,-1) node[anchor=west] 
{{\footnotesize $S = (I^*(A\rangle BC), I^*(B\rangle C))$}};
\draw (2,0) node[anchor=west] 
{{\footnotesize $T = (I^*(A\rangle C),I^*(B\rangle AC))$}};
\draw (1,1) node[anchor=west] {{\footnotesize $x+y= I^*(AB\rangle C)$}};
\end{tikzpicture}
\end{center}
\vspace*{-5mm}
\caption[Achievable rate region for the unassisted quantum MAC
per channel use in the asymptotic iid setting]{Achievable rate region for the unassisted quantum MAC
per channel use in the asymptotic iid setting for a fixed control state. The full region is the convex closure of all such pentagonal regions corresponding to all bipartite input control states.}
\label{fig:QMACasymp3}
\end{figure}

\subsection{Unassisted vs. Rate Limited Assistance}

As mentioned previously, a rate pair $(Q, E)$ is said to be $\eps$-achievable for entanglement assisted entanglement transmission across a point-to-point quantum channel $\mathcal{N}^{A'\to B}$ if there exists an encoder and decoder pair which consume pre-shared entanglement at a rate $E$ to faithfully transmit one half of a maximally entangled state at rate $Q$. The two extreme cases are when $E=0$ (the unassisted case) and when $E$ can be arbitrarily large. Recall that Lloyd, Shor and Devetak \cite{Lloyd,Shor_Notes,Devetak} showed that an achievable rate for the unassisted transmission of entanglement across the point-to-point channel is given by the maximum of the coherent information
\[
I(A\rangle B)
\]
over all control states of the form $\Omega^{AB}$. For the case of entanglement assisted transmission, Bennet, Shor, Smolin and Thapliyal \cite{EntanglementAssistedBennett} showed that the rate 
\[
\frac{1}{2} I(A:B)
\]
is achievable, whenever entanglement assistance is available at the rate $\frac{1}{2}\left(H(A)+H(A|B)\right)$. We mention that the one-shot analogue of this result was proved by Anshu, Jain and Warsi \cite{AnshuJainHypothesistesting}. In this paper, we will be interested in proving theorems which interpolate between these two cases. To be precise, we will prove achievable bounds for entanglement transmission of the following sort:
\[
\begin{aligned}
Q+E <& I_1\\
Q-E <& I_2.
\end{aligned}
\]

Note that to recover the unassisted achievable bounds, one simply sets $E=0$. On the other hand, to recover the entanglement assisted bounds, can simply saturate the first condition. Thus this situation is more general, where we can limit the rate of entanglement assistance. We therefore call this case entanglement transmission with \emph{rate limited entanglement assistance} .

It is not hard to prove a coding theorem in the case of rate limited entanglement assistance in the asymptotic iid setting, simply by time sharing between the two types of protocols. However, the situation is not so easy in the one-shot setting, since in that regime time sharing is impossible. To the best of our knowledge, bounds of this kind appeared first in the work of Dupuis \cite{Dupuis_thesis} . All the theorems in this paper are written with this more general setting in mind.

\section{Quantum Rate Splitting I}\label{chap:RateSplit1}

In this section we introduce the tools required to do quantum rate splitting. We demonstrate the technique for the point-to-point quantum channel. We will apply the tools introduced in this section to the problem of one-shot entanglement transmission over the QMAC and QIC in later sections. Another key element in our proof, which we discuss in this section, is a way to do successive cancellation decoding for entanglement transmission codes when the receiver has some side information. This allows us to generalise our bounds to the case when the sender and the receiver may share some limited number of EPR states before the protocol starts.

\subsection{Rate Splitting for point-to-point Channels}
\label{ratesplitting} 
\subsubsection{Rate Splitting in the Classical Regime}
\label{ieeesec:ratesplitclassical} 
In this section, we briefly review the idea of rate splitting, as detailed in
\cite{Rate_Splitting_Urbanke}. Consider the classical point-to-point 
channel $(\mathcal{A},P_{B|A},\mathcal{B})$ between Alice and Bob and 
let $P_A$ be the input distribution that maximises $I(A:B)$. The idea is 
to split Alice into two independent senders, Alice\sub{0} and 
Alice\sub{1} and then have Bob decode their messages via a successive 
cancellation strategy. To do this, we create two new distributions 
\[
P_U^{\theta}\textup{ and } P_V^{\theta}
\]
 with respect to some parameter $\theta\in [0,1]$, where the random variables $U$ and $V$ both range over the same classical alphabet $\mathcal{A}$. These distributions are meant to be the input distributions of Alice\sub{0} and Alice\sub{1} respectively. However, we must maintain the invariant that the distribution at the input to the channel must be $P_A$. To do this, we define a deterministic function $f$ which has the following properties:
\begin{align*}
    &f: \mathcal{A}\times \mathcal{A}\to \mathcal{A} \\
    &f(U,V)\sim P_A, \textup{ where } U\sim P_U^{\theta}~\textup{ and }~V\sim P_V^{\theta}.
\end{align*}
Moreover, when $\theta = 0$, 
$V$ is distributed exactly like $A$ and $U$ is a single point distribution,
and when $\theta = 1$, 
$U$ is distributed exactly like $A$ and $V$ is a single point 
distribution. Furthermore, appealing to the 
properties of the mutual information one can show that 
\[
I(A:B) = I(U^{\theta} V^{\theta}:B) 
       = I(U^{\theta}:B)+I(V^{\theta}:B U^{\theta}) .
\]
From the 
above discussion it is clear that a simple encoding-decoding strategy is 
as follows:
\begin{enumerate}
    \item Alice is split into Alice\sub{0} and Alice\sub{1}.
    \item Alice\sub{0} uses a code of rate $I(U^{\theta}:B)$ 
regarding Alice\sub{1} as noise and Alice\sub{1} uses a code of rate  $I(V^{\theta}:B U^{\theta})$ regarding Alice\sub{0} as side information at the receiver.
\item  Bob decodes via successive cancellation.
\item 
 Finally, one 
can show that $(I(U^{\theta}:B),I(V^{\theta}:B U^{\theta}))$ is a 
continuous function in $\theta\in [0,1]$ and so the ordered pair 
traces out the straight 
line joining the points $(0,I(A:B))$ and $(I(A:B),0)$ due to the chain
rule of mutual information with equality.
\end{enumerate} With this 
construction in hand, one can design an encoding and decoding scheme for 
the classical MAC without appealing to time sharing or jointly typical 
simultaneous decoding. 
Firstly, split Alice into the two users Alice\sub{0} and 
Alice\sub{1} by the construction above. Then Charlie does a successive 
cancellation decoding for this $3$ sender MAC with senders Alice\sub{0}, Bob and Alice\sub{1}: first decode 
Alice\sub{0}'s message treating the other senders as noise, then 
decode Bob's message regarding 
Alice\sub{0}'s message as side information and Alice\sub{1} as noise, 
and finally, decode 
Alice\sub{1}'s message regarding Bob's and Alice\sub{0}'s message as 
side information. Thus three point-to-point channel decodings are done
by Charlie in order to decode the sent messages at the rate triple 
$(I(U^{\theta}:C),I(C:B U^{\theta}),I(V^{\theta}:C B U^{\theta}))$. 
Notice that, all points in the dominant face of the achievable region 
in Figure~\ref{fig:MACregion}
can be achieved in this way due to continuity as $\theta$ varies from
$0$ to $1$. Also, observe that the split of Alice depends on $\theta$.

The triple $(f,P^{\theta}_U,P^{\theta}_V)$ with respect to the distribution $P_A$ is called a split of $P_A$. That such a triple exists is  given by the following fact:

\begin{fact}\label{split}
Given a distribution $P_A$ on the set $\mathcal{A}$, there exist two distributions $P_{U}^{\theta}$ and $P_V^{\theta}$ (both defined on $\mathcal{A}$), parameter $\theta\in [0,1]$ and a function $f:\mathcal{A}\times \mathcal{A}\rightarrow \mathcal{A}$ such that the following hold true:
\begin{enumerate}
    \item $f(U,V)\sim P_A$
    \item For fixed values of $x$ and $u$, $P^{\theta}_{f(U,V)|U}(a|u)$ is a continuous function of $\theta$.
    \item For $\theta=0$, $P^{\theta}_{f(U,V)|U}(a|u)=P_A(a)$.
    \item For $\theta=1$, and all $u\in \mathcal{A}$, $P^{\theta}_{f(U,V)|U}(a|u)$ puts all its mass on one element.
\end{enumerate}
\end{fact}
\begin{proof}
We demonstrate an explicit construction, as shown in \cite{Rate_Splitting_Urbanke}.  Assume that $\mathcal{A}$ is an ordered set. We describe the distribution in terms of distribution  functions, for which we use the letter $F$ along with the appropriate subscript. Then, define, for all $i\in \mathcal{A}$:
\begin{align*}
    &F_U^{\theta}(i)\coloneqq \theta F_A(i)+1-\theta \\
   & F_V^\theta(i)\coloneqq \frac{F_A(i)}{F_U^{\theta}(i)}\\
   & f(u,v)\coloneqq \max\brak{u,v} ~~\forall u,v\in \mathcal{A}.
\end{align*}
It is easy to check the triple defined above satisfies all the properties in \cref{split}. The interested reader may look at \cite{Rate_Splitting_Urbanke} for details.
\end{proof}

\subsubsection{Rate Splitting in the Quantum Case}\label{sec:ratesplitquantum}
 
To describe rate splitting in the entanglement transmission scenario, we will define an abstract \emph{splitting scheme} with some properties of interest:
\begin{definition}{\textbf{Splitting Scheme}}\label{splittingschme}
Given a control state $\ket{\Omega}^{A''A'}$ and systems $A''_0$ and $A''_1$ such that $A''\cong A''_1 \cong A''_0$, we define a splitting scheme to be a family of isometric embeddings $\brak{U_{\theta}^{A''\rightarrow A''_0A''_1}}$  parametrized by a variable $\theta\in [0,1]$, such that:
\begin{enumerate}
    \item For all $\theta, \theta' \in [0,1]$ and $\epsilon>0$ there exists $\delta>0$ such that whenever $\abs{\theta-\theta'}\leq \delta$,  $\norm{U_{\theta}\cdot \Omega-U_{\theta'}\cdot\Omega}_1\leq \epsilon$.
    \item Given any channel $\mathcal{N}^{A'\rightarrow C}$ and its Stinespring dilation $\mathcal{U}_{\mathcal{N}}^{A'\rightarrow CE}$, 
    \begin{align*}
        I(A''_0\rangle  C)_{\mathcal{U}_{\mathcal{N}}\cdot \Omega_0}= I(A''_1\rangle  C)_{\mathcal{U}_{\mathcal{N}}\cdot \Omega_1}=
        I(A''\rangle  C)_{\mathcal{U}_{\mathcal{N}}\cdot \Omega}
    \end{align*}
    where $\Omega_0\coloneqq U_0\cdot \Omega$ and $\Omega_1\coloneqq U_1\cdot \Omega$.
\end{enumerate}
\end{definition}

The splitting scheme defined above can be defined with respect to the more general control state \[\ket{\sigma}^{A''A'B''B'}\coloneqq\ket{\Omega}^{A''A'}\ket{\Delta}^{B''B'}.\] This will be useful when we describe the splitting protocol for more general multi-terminal channels, viz. $\mathcal{N}^{A'B'\to C}$. In that case, the second condition in Definition \ref{splittingschme} can be stated as
\begin{align*}
        I(A''_0B''\rangle  C)_{\mathcal{U}_{\mathcal{N}}\cdot\sigma_0}= I(A''_1B''\rangle  C)_{\mathcal{U}_{\mathcal{N}}\cdot \sigma_1}=
        I(A''B''\rangle  C)_{ \sigma}
    \end{align*}
where $\sigma_0\coloneqq U_0\cdot \sigma$ and $\sigma_1\coloneqq U_1\cdot \sigma$. For the purposes of this section, where we only demonstrate splitting for the point-to-point channel, one may simply ignore the state $\ket{\Delta}$. Also, note that the invariants in the splitting scheme are specified in terms of the coherent information. A more general definition would be to specify the invariants in terms of the smooth min-entropy. We work with this more general definition.

We will first give an overview of the strategy for the \emph{unassisted} case. We will then
state and prove the main technical lemma of this section, \cref{achievability}. The ideas used in proving this lemma will generalise easily to the setting of the multi-terminal channels such as the QMAC and the QIC.

We will 
emulate the strategy outlined in \cref{ieeesec:ratesplitclassical} for a bipartite pure quantum state
\[
\ket{\Omega}^{A'' A'} \coloneqq
\sum_{a''\in \mathcal{A''}}\sqrt{P_{A''}(a)}\ket{a}^{A''}\ket{\zeta_a}^{A'} ,
\]
where $\ket{a''}$ runs over the computational basis of $A''$ and
$P_{A''}$ is a probability distribution on the basis set $\mathcal{A''}$.
We split the system $A''$ into two registers $A''_0$ and $A''_1$
corresponding to the two senders Alice\sub{0} and Alice\sub{1}. Let the
split $(P_{U}^{\theta},P_{V}^{\theta},f)$ be as in 
the previous subsection. Define the isometric embedding
$U_{\textsc{split}}(\theta)^{A''\rightarrow A''_0 A''_1}$ as follows:
\[
\sqrt{P_A(a)}\ket{a}^{A''} 
\overset{U_{\textsc{split}}(\theta)}{\longmapsto} 
\sum_{(u,v)\in f^{-1}(a'')}
\sqrt{P_{U}^{\theta}(u)P_V^{\theta}(v)}\ket{u}^{A''_0}\ket{v}^{A''_1} ,
\]
and
$
\ket{\Omega(\theta)}^{A''_0 A''_1 A'} \coloneqq
U_{\textsc{split}}(\theta)\ket{\Omega}^{A''A'}.
$

We now pass the system $A'$ through a point-to-point channel
$\mathcal{N}^{A' \to B}$ and obtain the quantum state
$\ket{\Omega(\theta)}^{A''_0 A''_1 B}$.
By unitary invariance,
$
I^\epsilon_{\mathrm{min}}(A'' \rangle  B)_\Omega =
I^\epsilon_{\mathrm{min}}(A''_0 A''_1 \rangle  B)_{\Omega(\theta)}.
$
From the works of \cite{Dupuis_thesis, Szehr_2013} applied to
transmission of quantum information over one-shot unassisted point to 
point quantum channels, we
first realise that Bob can decode Alice\sub{0}'s quantum message
at the rate of 
$
I^{O(\epsilon^2)}_{\mathrm{min}}(A''_0 \rangle  B)_{\Omega(\theta)} 
- O(\log \epsilon^{-1})
$ 
with error at most $O(\sqrt{\eps})$. 
Then, employing the successive cancellation methods of Yard et al.
\cite{Yard_MAC} 
Bob can decode Alice\sub{2}'s quantum message
at the rate of 
$
I^{O(\epsilon^2)}_{\mathrm{min}}(A''_1 \rangle  B A''_0)_{\Omega(\theta)} 
- O(\log \epsilon^{-1})
$ 
with error at most $O(\sqrt{\eps})$.

Doing both the steps above requires us to overcome a few technical 
challenges. We
do this by defining a notion of {\em almost CPTP} maps (see \cref{sec:almostcptp}) and combining it with another proof technique
by Dupuis for the unassisted quantum broadcast channel 
\cite{Dupuis_thesis}. We believe that the notion of almost CPTP maps should be useful
in other situations as well.

We have thus operationally shown the chain rule inequality
$
I^{O(\epsilon^2)}_{\mathrm{min}}(A''_1 \rangle  B A''_0)_{\Omega(\theta)} + 
I^{O(\epsilon^2)}_{\mathrm{min}}(A''_0 \rangle  B)_{\Omega(\theta)}  \leq  
I^{\epsilon}_{\mathrm{min}}(A''_0 A''_1 \rangle  B)_{\Omega(\theta)}
$ (suppressing the log factors). One can prove this fact independently using the chain rule for the smooth min-entropy \cref{chainingforminentropy}.
We now see that as $\theta$ varies from $0$ to $1$, the point
$
(R_0(\theta),R_1(\theta)) = 
(I^{O(\epsilon^2)}_{\mathrm{min}}(A''_0 \rangle  B)_{\Omega(\theta)}, 
 I^{O(\epsilon^2)}_{\mathrm{min}}(A''_1 \rangle  B A''_0)_{\Omega(\theta)}) 
$ 
traces out a continuous curve 
that lies on or below the line segment joining the point
$(I^\epsilon_{\mathrm{min}}(A \rangle  B), 0)$ to the point
$(0, I^\epsilon_{\mathrm{min}}(A \rangle  B))$ and meets it at its endpoints.
The continuity of the curve follows from the continuity of the states 
and the functionals involved. Continuity of the functionals is implied by \cref{continuityofhmin}, whereas continuity of the states is implied by \cref{contomega}. This rate splitting and successive cancellation idea can now be easily generalised to the case of entanglement transmission over QMAC with limited entanglement assistance.

We will now consider the general case when Bob has side information available at the decoder. Recall that the users Alice\sub{0} and Alice\sub{1} obtained from splitting the sender Alice are treated as \emph{independent} senders. Hence, suppose Alice\sub{0} and Alice\sub{1} wish to transmit the systems $A_0$ and $A_1$ of the states $\ket{\eta}^{A_0B_0R_0}$ and $\ket{\psi}^{A_1B_1R_1}$ to Bob. We wish to prove there exists an encoder $\mathcal{E}^{A_0A_1\rightarrow A'}$ and a decoder $\mathcal{C}^{BC_0C_1\rightarrow A_0A_1}$ such that 
 
 \[
F\big(\mathcal{C}\circ\mathcal{N}\circ \mathcal{E} (\eta\otimes \psi),\eta\otimes \psi\big)\geq 1-\epsilon.
\]
 Given that such an encoder decoder pair exist, set $\ket{\eta}^{A_0B_0R_0}\gets \ket{\Phi}^{M_0R_0}\ket{\Phi}^{\tilde{A}_0B_0}$ and $\ket{\psi}^{A_1B_1R_1}\gets \ket{\Phi}^{A_1M_1}\ket{\Phi}^{\Tilde{A}_1B_1}$. Let $Q_{A_0}=\log \abs{M_0}, Q_{A_1}=\log \abs{M_2}$, and $E_{A_0}=\log\abs{B_0}, E_{A_1}=\log \abs{B_1}$. The rates $Q_{A_0}, Q_{A_1}$ are the entanglement transmission rates of Alice\sub{0} and Alice\sub{1} and $E_{A_0}$ and $E_{A_1}$ quantify the amount of pre-shared entanglement available to them before the protocol begins.
 
 We will consider the simpler case, when Alice\sub{0} does not share any entanglement with Bob, but Alice\sub{1} does, i.e. the register $C_0$ is trivial. We quantify the rates in the following lemmas:

  \begin{proposition}\label{achievability} Given the control state $\ket{\Omega}^{A''A'}$, the point-to-point quantum channel $\mathcal{N}^{A'\rightarrow C}$ and the splitting scheme $\brak{U_{\theta}^{A''}}$, suppose Alice has to send states $\ket{\eta}^{A_0R_0}\otimes \ket{\psi}^{A_1BR_1}$ to Bob, where $A_0$ and $A_1$ are the message registers and $B$ models the side information Bob has about the $A_1$. $R_0$ and $R_1$ are reference systems. We define $\ket{\Omega'(\theta)}^{A''_0A''_1A'}\coloneqq U_{\theta}^{A''}\ket{\Omega}^{A''A'}$ and 
  \begin{align*}
      \ket{\Omega(\theta)}^{A''_0A''_1CE}\coloneqq \mathcal{U}_{\mathcal{N}}^{A'\rightarrow CE}\ket{\Omega(\theta)}^{A''_0A''_1A'}.
  \end{align*}
  Then there exist an encoder  $\mathcal{E}^{A_0A_1\rightarrow A'}$ and a decoder $\mathcal{C}^{BC\rightarrow A_0A_1}$ such that
  \begin{align*}
    \norm{  \mathcal{C}\circ\mathcal{N}\circ \mathcal{E} (\eta^{A_0R_0}\otimes \psi^{A_1B_1R_1})-\eta^{A_0R_0}\otimes \psi^{A_1B_1R_1}}_1\leq \delta,
  \end{align*}
  where $\delta= 4\sqrt{2\delta_{\textup{dec}}(0)} +2\sqrt{2\delta_{\textup{dec}}(1)}+2\sqrt{2\delta_{\textup{enc(0)}}+2\delta_{\textup{enc}}(1)}$ \\ \vspace{2mm} \\
  and
  \begin{align*}
      \delta_{\textup{dec}}(0) &= 20\cdot 2^{-\frac{1}{2} H_{2}^{\epsilon}(A_0|R_0)_{\eta}-\frac{1}{2} H_{\min}^{\epsilon_0}(A''_0|A''_1E)_{\Omega(\theta)}}+160\epsilon \\
      \delta_{\textup{dec}}(1) &=  20\cdot 2^{-\frac{1}{2} H_{2}^{\epsilon}(A_1|R_1)_{\psi}-\frac{1}{2} H_{\min}^{\epsilon_0}(A''_1|E)_{\Omega(\theta)}}+160\epsilon \\
      \delta_{\textup{enc}}(0) &= 20\cdot 2^{\frac{1}{2} H_{\max}^{\epsilon}(A_0)_{\eta}-\frac{1}{2} H_{\min}^{\epsilon_0}(A''_0|A''_1)_{\Omega(\theta)}}+160\epsilon \\
 \delta_{\textup{enc}}(1) &= 2^{\frac{1}{2} H_{\max}^{\epsilon}(A_1)_{\psi}-\frac{1}{2} H_{\min}^{\epsilon}(A''_1)_{\Omega(\theta)}}+12\epsilon,
  \end{align*}
  
  where $\epsilon_0=O(\epsilon^2)$ for some positive $\epsilon$.
  \end{proposition}
  
    An easy corollary of \cref{achievability} is the following:
  \begin{corollary}\label{corol:achievability}
  Given the control state $\ket{\Omega}^{A''A'}$, the parameter $\theta\in [0,1]$ and the point-to-point channel $\mathcal{N}^{A'\rightarrow B}$, and the splitting scheme $\brak{U^{A''\rightarrow A''_0A''_1}_{\theta}}$, Alice can transmit EPR states to Bob at the rate $Q_{A_0}+Q_{A_1}$ given $E_{A_1}$ bits of pre-shared entanglement, with error at most $240\sqrt{\epsilon}$ whenever
  \begin{align*}
  Q_{A_0} &< H_{\max}^{\epsilon_0}(A''_0|A''_1)_{\Omega'(\theta)}+\log4\epsilon^2 \\
  Q_{A_0} &< I_{\min}^{\epsilon_0}(A''_0\rangle B)_{\mathcal{U}_{\mathcal{N}}\cdot\Omega'(\theta)}+\log4\eps^2 \\
  Q_{A_1}+E_{A_1} &< H_{\max}^{\epsilon}(A''_1)_{\Omega'(\theta)}+\log4\epsilon^2 \\
  Q_{A_1}-E_{A_1} &<I_{\min}^{\epsilon_0}(A''_1\rangle  A''_0B)_{\mathcal{U}_{\mathcal{N}}\cdot\Omega'(\theta)}+\log4\epsilon^2,
  \end{align*}
  where $\epsilon_0=O(\epsilon^2)$ and $\ket{\Omega'(\theta)}^{A''_0A''_1A'}=U_{\theta}^{A''}\ket{\Omega}^{A''A'}$.
  
  \end{corollary}

\begin{proof}
We initialise the states $\ket{\eta}^{A_0R_0} and \ket{\psi}^{A_1R_1B_1}$ as follows 
  \begin{align*}
      &\ket{\eta}^{A_0R_0}\gets \ket{\Phi}^{A_0R_0} \\
      &\ket{\psi}^{A_1R_1B_1}\gets \ket{\Phi}^{R_1M_1}\ket{\Phi}^{\tilde{A}_1B_1} .
  \end{align*}
  Here, the registers $M_1\tilde{A}_1$ play the roles of $A_1$, and the notation $\Phi$ is used generically to mean an EPR state. Let 
  \begin{align*}
      &\abs{R_0}=2^{Q_{A_0}} \\
      &\abs{R_1}=2^{Q_{A_1}} \textup{ and } \abs{B_1}=2^{E_{A_1}}.
  \end{align*}
  Note that Alice's actual rate $Q_A$ is $Q_{A_0}+Q_{A_1}$. The following relations are easy to check:
  \begin{align*}
      &H_{\max}(A_0)_{\eta}=Q_{A_0} \implies H_{\max}^{\epsilon}(A_0)_{\eta}\leq Q_{A_0} \\
      &H_{\max}(M_1\tilde{A}_1)_{\psi}=Q_{A_1}+E_{A_1} \implies H_{\max}^{\epsilon}(M_1\tilde{A}_1)_{\psi}\leq Q_{A_1}+E_{A_1} \\
      & H_{\min}(A_0|R_0)_{\eta}=Q_{A_0} \implies H_{\min}^{\epsilon}(A_0)_{\eta}\geq Q_{A_0} \\
      &H_{\min}(M_1\tilde{A}_1|R_1)_{\psi}=E_{A_1}-Q_{A_1} \implies H_{\min}^{\epsilon}(M_1\tilde{A}_1|R_1)_{\psi}\geq E_{A_1}-Q_{A_1} .
  \end{align*}
  Then, from \cref{achievability}, we set 
  \begin{align*}
      &\delta_{\textup{dec}}(0)< 200\epsilon \\
      &\delta_{\textup{dec}}(1)< 200\epsilon \\
      &\delta_{\textup{enc}}(0)<200\epsilon \\
      & \delta_{\textup{enc}}(1)< 16\epsilon .
  \end{align*}
  
  Plugging in these numbers in the bounds shown in \cref{achievability} completes the proof.
  
\end{proof}

\subsection{Tools for Successive Cancellation: Intermediate States and Almost CPTP Maps}\label{sec:almostcptp}

As mentioned in section \cref{sec:ratesplitquantum}, we will require the notion of almost CPTP maps to be able to successive cancellation decoding for entanglement transmission codes. An upshot of this technique is that it allows the decoder to use side information that the receiver may have, to boost the sender's entanglement transmission rate. This is what essentially allows us to provide the bounds in the general case when only limited entanglement is available.

The problem that we consider is as follows: we are given the channel $\mathcal{N}^{A'\to C}$ and the split control state $\ket{\Omega}^{A''_0A''_1A'}$. The two split senders Alice\sub{0} and Alice\sub{1} wish to send the $A_0$ and $A_1$ parts of the states $\ket{\eta}^{R_0A_0}$ and $\ket{\psi}^{A_1BR_1}$ to the receiver Charlie. Additionally, Charlie also holds the system $B$ as side information, which he can potentially use to boost Alice\sub{1}'s rate. As a first step, we embed the systems $A_0$ and $A_1$ into the systems $A''_0$ and $A''_1$ via the action of the isometries $W_0^{A_0\to A''_0}$ and $W_1^{A_1\to A''_1}$:
\begin{align*}
    \ket{\eta}^{R_0A''_0}&\coloneqq W_0\ket{\eta}^{R_0A_0} \\
    \ket{\psi}^{R_1BA''_1}&\coloneqq W_1\ket{\psi}^{R_1BA''_1}.
\end{align*}

 Our encoder will be of the form
\begin{align*}
    \mathcal{E}^{A''_0A''_1\to A'}(\cdot) \equiv \abs{A''_0A''_1}\op^{A''_0A''_1\to A'}(\Omega^{A''_0A''_1A'})U^{A''_0}\otimes U^{A''_1}(\cdot),
\end{align*}
where $U^{A''_0}$ and $U^{A''_1}$ are random unitaries, picked independently from the Haar measure. The above map is not trace preserving in general, and is only CPTP on average over the choices of the two random unitaries. One of our main aims will be to show that, with respect to the states $\eta$ and $\psi$, there exist fixed instantiations of $U^{A''_0}$ and $U^{A''_1}$ such that 
\[
\mathcal{E}^{A''_0A''_1\to A'}\circ W_0\otimes W_1\cdot (\eta\otimes \psi) \equiv V^{A_0A_1\to A'}\cdot(\eta \otimes \psi)
\]
where $V^{A_0A_1\to A'}$ is an isometry. This isometry should also have a corresponding decoding map $\mathcal{D}^{BB\to A_0A_1}$ such that 
\[
F\left(\ket{\eta}^{A_0R_0}\ket{\psi}^{A_1BR_1},\mathcal{D}\circ\mathcal{N}\circ\mathcal{E}(W_0\cdot \eta\otimes W_1\cdot \psi)\right)\geq 1-\eps .
\]

To show the existence of the encoder $V$ and its corresponding decoder $\mathcal{D}$, we first need a good way to manipulate the quantity 
$\mathcal{E}^{A''_0A''_1\to A'}\circ W_0\otimes W_1\cdot (\eta \otimes \psi)$, which we henceforth abbreviate as $\mathcal{E}(\eta\otimes \psi)$. To that end, we define \emph{intermediate states}.

\subsubsection{Intermediate States}

\begin{definition}{{\bf Intermediate State}}
Given the control state $\ket{\Omega}^{A''_0A''_1A'}$ and the state $\ket{\psi}^{A''_1BR_1}$, we define the intermediate state 
\[
    \ket{\omega}^{A''_0BR_1A'}\coloneqq \sqrt{\abs{A''_1}}~\op^{A''_1\to A''_0A'}(\ket{\Omega}^{A''_0A''_1A'})U^{A''_1}\ket{\psi}^{A''_1BR_1} .
\]
\end{definition}

\noindent The following lemma will enable us to write the encoded state in terms of the intermediate state.

\begin{lemma}{\bf Intermediate State Lemma}\label{lem:IntermediateStateLemma}
Given the intermediate state $\ket{\omega}^{A''_0BR_1A'}$, the following holds
\[
\mathcal{E}(\eta\otimes \psi) = \abs{A''_0}\op^{A''_0\to BR_1A'}\left( \omega^{A''_0BR_1A'}\right) \circ U^{A''_0}\cdot (\eta^{A''_0R_0}) .
\]
 
\end{lemma}
\begin{proof}
Consider the following series of equalities:
\begin{align*}
     &\sqrt{A''_0A''_1}~\op^{A''_0A''_1\to A'}(\ket{\Omega}^{A''_0A''_1A'})~(U^{A''_0}\otimes U^{A''_1})~\ket{\eta}^{A''_0R_0}\ket{\psi}^{A''_1BR_1} \\
    =& \sqrt{A''_0A''_1}~\op^{A''_0A''_1\to R_0BR_1}\left((U^{A''_0}\otimes U^{A''_1})~\ket{\eta}^{A''_0R_0}\ket{\psi}^{A''_1BR_1}\right) \ket{\Omega}^{A''_0A''_1A'} \\
    =& \left(\sqrt{A''_0}~\op^{A''_0\to R_0}(U^{A''_0}\ket{\eta}^{A''_0R_0})\otimes \sqrt{A''_1}~\op^{A''_1\to BR_1}(U^{A''_1}\ket{\psi}^{A''_1BR_1})\right)\ket{\Omega}^{A''_0A''_1A'} \\
    =& \sqrt{A''_0}~\op^{A''_0\to R_0}(U^{A''_0}\ket{\eta}^{A''_0R_0}) \left(\sqrt{A''_1}~\op^{A''_1\to BR_1}(U^{A''_1}\ket{\psi}^{A''_1BR_1})\ket{\Omega}^{A''_0A''_1A'}\right) \\
    =& \sqrt{A''_0}~\op^{A''_0\to BR_1A'} \left(\sqrt{A''_1}~\op^{A''_1\to BR_1}(U^{A''_1}\ket{\psi}^{A''_1BR_1})\ket{\Omega}^{A''_0A''_1A'}\right) U^{A''_0}\ket{\eta}^{A''_0R_0} \\
    =& \sqrt{A''_0}~\op^{A''_0\to BR_1A'} \left(\sqrt{A''_1}~\op^{A''_1\to A''_0A'}(\ket{\Omega}^{A''_0A''_1A'})U^{A''_1}\ket{\psi}^{A''_1BR_1}\right)U^{A''_0}\ket{\eta}^{A''_0R_0}
\end{align*}
The above derivation uses the properties of the op operator proved in \cref{subsec:opoperator} of \cref{sec:preliminaries}. Writing the first and the last terms in the state notation gives us the required result.
\end{proof}

\noindent Note that the intermediate states may not be quantum states in the sense that they may not have trace $1$. In the following lemma, we prove that intermediate states have trace $1$ on average over the choice of random unitaries, assuming some entropic inequalities are satisfied.
\begin{lemma}{\bf Trace of Intermediary States}\label{lem:TraceIntermediateStateLemma}
Given the intermediary state
\[
\ket{\omega}^{A''_0BR_1A'}=\sqrt{\abs{A''_1}}\op^{A''_1\to A''_0A'}(\ket{\Omega^{A''_0A''_1A'}})U^{A''_1}\ket{\psi}^{A''_1R_1B}
\]
where $U^{A''_1}$ is a random unitary sampled from the Haar measure and given that
\[
H_{\max}^{\eps}(A)_{\psi}\leq H_{\min}^{\eps}(A''_1)_{\Omega}+O(\log\eps),
\]
the following holds
\[
\E_{U^{A''_1}}\left[\abs{\Tr[\omega]-1}\right]\leq O(\eps).
\]
\end{lemma}
\begin{proof}
Using the single sender decoupling theorem~\ref{fact:singledecoup}, we see that
\begin{align*}
    &\E_{U^{A''_1}}\norm{\abs{A''_1}\Tr_{A''_0A'}\op^{A''_1\to A''_0A'}(\ket{\Omega^{A''_0A''_1A'}})U^{A''_1}\cdot\psi^{A''_1R_1B}-\psi^{R_1B}}_1 \\
    \leq~ &2^{-\frac{1}{2}H_{\min}^{\eps}(A_1|R_1B)_{\psi}-\frac{1}{2}H_{\min}^{\eps}(A''_1)_{\Omega}}+12\epsilon \\
    =~ &2^{\frac{1}{2}H_{\max}^{\eps}(A_1)_{\psi}-\frac{1}{2}H_{\min}^{\eps}(A''_1)_{\Omega}}+12\epsilon.
\end{align*}
We can replace $A''_1$ with the system $A_1$ in the entropic quantity corresponding to $\ket{\psi}^{A''_1BR_1}$ since $\ket{\psi}^{A''_1BR_1}$ is an isometric embedding of $\ket{\psi}^{A_1R_1B}$. The last equality follows from the duality of the smooth min- and max- entropies for pure states. This concludes the proof. 
\end{proof}

We will now show the following approximate data processing type inequality:\linebreak Suppose we are given the intermediary state 
\begin{align*}
\ket{\omega}^{A''_0BR_1A'}=\sqrt{\abs{A''_1}}\op^{A''_1\to A''_0A'}(\ket{\Omega^{A''_0A''_1A'}})U^{A''_1}\ket{\psi}^{A''_1R_1B}
\end{align*}
where $U^{A''_1}$ is a Haar random unitary. Then, with constant probability over the choice of $U^{A''_1}$ 
\[
H_{\min}^{\epsilon}(A''_0|BR_1)_{\omega}\geq H_{\min}^{f(\epsilon)}(A''_0|A''_1)_{\Omega}-O(1),
\]
where $f(\epsilon)$ is some function of $\epsilon$. There are several technical issues that one should note here. For example, the expression $H_{\min}^{\epsilon}(A''_0|BR_1)_{\omega}$ is, strictly speaking, not defined, since $\omega$ is not really a normalised state. Hence, what we actually want to show is that the above data processing like inequality holds for the quantity $H_{\min}^{\epsilon}(A''_0|BR_1)_{\tilde{\omega}}$, which is defined with respect to he normalised version of $\omega$. To that end, we first define almost CPTP maps in the following section.

\subsubsection{Almost CPTP Maps}
\begin{definition}[Almost CPTP]\label{def:almostcptp}
We call a linear map $\mathcal{T}^{A''_1\rightarrow BR_1}$ as an almost CPTP if $\mathcal{T}$ has the following properties:
\begin{enumerate}
\item $\mathcal{T}$ is CP.
\item $\Tr[\mathcal{T}(\pi^{A''_1})]\in [1-\delta,1+\delta]$ for some small $\delta\geq 0$.
\item $\int\mathcal{T}(U^{A''_1}\cdot \xi)d\mu=\Tr[\xi]\mathcal{T}(\pi^{A''_1})$.
\end{enumerate}
\end{definition}

\begin{lemma}{\bf [Approximate Data Processing Inequality for Almost CPTP Maps]}\label{almostcptp}
When the measure $\mu$ is set to be the Haar measure on the unitary group on $A''_1$, and given the condition that 
\[
H_{\max}^{f(\epsilon)}(A_1)_{\psi}\leq H_{\min}^{f(\epsilon)}(A''_1)_{\Omega}+O(\log f(\epsilon)),
\]
then the following holds with constant probability over the choice of $U^{A''_1}$,
\begin{align*}
    H_{\min}^{\epsilon}(A''_0|BR_1)_{\tilde{\omega}}\geq H_{\min}^{O(\epsilon^2)}(A''_0|A''_1)_{\Omega}-O(1)
\end{align*}
where $\tilde{\omega}\coloneqq \frac{\omega}{\Tr[\omega]}$ and $f(\epsilon)=O(\epsilon^2)$.
\end{lemma}
   
   \begin{proof}
First, given the condition that $
H_{\max}^{f(\epsilon)}(A_1)_{\psi}\leq H_{\min}^{f(\epsilon)}(A''_1)+2\log f(\epsilon) $, from \cref{lem:TraceIntermediateStateLemma} we see that 
\begin{align*}
    \E_{U^{A''_1}} \abs{\Tr[\omega]-1} \leq 13f(\epsilon).
\end{align*}
Next, define
\begin{align*}
    \mathcal{T}^{A''_1\rightarrow BR_1}(\xi)\coloneqq \abs{A''_1}\big(\textup{op}^{A''_1\rightarrow BR_1}(\psi)\cdot \xi\big).
\end{align*}

\noindent\textbf{Checking $\mathcal{T}$ is almost CPTP}

Firstly, it is clear that $\mathcal{T}$ is CP. Next, we see that
\begin{align*}
    \Tr[\mathcal{T}(\pi^{A''_1})] &= \Tr[\textup{op}^{A''_1\rightarrow BR_1}(\psi)\textup{op}^{A''_1\rightarrow BR_1}(\psi)^{\dagger}] \\
    &= \Tr[\Tr_{A''_1}(\psi)] \\
    &= 1.
\end{align*}
It is another easy verification, using the properties of Haar integrals, that 
\begin{align*}
    \int\mathcal{T}(U^{A''_1}\cdot\xi)d\mu=\Tr[\xi]\mathcal{T}(\pi^{A''_1}).
\end{align*}
This shows that $\mathcal{T}$ is indeed an almost CPTP. 
\\ \vspace{1mm} \\
\noindent \textbf{Applying $\mathcal{T}$ to the operator inequality} \\
\vspace{1mm}\\

\noindent Again, using the properties of the $\textup{op}$ operator we see that 
\begin{align*}
    \mathcal{T}\big((U^{A''_1})^{T}\cdot\Omega\big) &=\abs{A''_1}\Big( \textup{op}^{A''_1\rightarrow BR_1}(\psi)\cdot \big((U^{A''_1})^{T}\cdot\Omega^{A''_0A''_1A'} \big)\Big) \\
    &= \abs{A''_1}\Big( \textup{op}^{A''_1\rightarrow BR_1}(U^{A''_1}\psi)\cdot \Omega^{A''_0A''_1A'} \Big) \\
    &= \abs{A''_1}\Big( \textup{op}^{A''_1\rightarrow A''_0A'}(\Omega)\cdot \big(U^{A''_1}\cdot\psi^{A''_1BR_1}\big) \Big) \\
    &= \omega.
\end{align*}
Now, suppose that $\Tilde{\Omega}$ is the optimiser in the definition of $H_{\min}^{f(\epsilon)}(A''_0|A''_1)_{\Omega}$  and that $\norm{\Tilde{\Omega}-\Omega}_1\leq 2f(\epsilon)$. Suppose also that $\lambda^{A''_1}$ be a positive semidefinite matrix such that $\Tr[\lambda^{A''_1}]=2^{-H_{\min}^{f(\epsilon)}(A''_0|A''_1)_{\Omega}}$ and 
\begin{align*}
    \Tilde{\Omega}^{A''_0A''_1}\leq\mathbb{I}^{A''_0}\otimes \lambda^{A''_1}.
\end{align*}
Then, using the fact that $\mathcal{T}$ is a CP map, we see that
\begin{align*}
   \mathcal{T}\Big(\big(U^{A''_1}\big)^T\cdot \Tilde{\Omega}^{A''_0A''_1}\Big)&\leq\mathbb{I}^{A''_0}\otimes\mathcal{T}\Big(\big(U^{A''_1}\big)^T\cdot\lambda^{A''_1}\Big)^{BR_1}.
\end{align*}
First notice that, by properties $2$ and $3$ of almost CPTP maps (\cref{def:almostcptp}),
\begin{align*}
    \int\mathcal{T}\Big(\big(U^{A''_1}\big)^T\cdot\lambda^{A''_1}\Big)dU^{A''_1} &= \int\mathcal{T}\Big(U^{A''_1}\cdot\lambda^{A''_1}\Big)dU^{A''_1} \\
    &= \Tr[\lambda^{A''_1}]\mathcal{T}(\pi^{A''_1})\\
    \intertext{Taking trace on both sides} 
    \Tr\left[\int\mathcal{T}\Big(\big(U^{A''_1}\big)^T\cdot\lambda^{A''_1}\Big)dU^{A''_1}\right]&=2^{-H_{\min}^{f(\epsilon)}(A''_0|A''_1)_{\Omega}},
\end{align*}
where the last equality stems from the fact that for $\mathcal{T}$, property 2 holds with $\delta=0$.
Next, from the fact that $\tilde{\Omega}-\Omega$ is Hermitian, we can write $\tilde{\Omega}-\Omega=\Delta_{+}-\Delta_{-}$ where $\Delta_{\pm}$ are positive semidefinite matrices with disjoint support. This implies that
\begin{align*}
    \norm{\tilde{\Omega}-\Omega}_1 &=\Tr[\Delta_+]+\Tr[\Delta_-] \\
    &\leq 2f(\epsilon), \\
    \intertext{then}
    \int\norm{\mathcal{T}\Big(\big(U^{A''_1}\big)^T\cdot\tilde{\Omega}\Big)-\mathcal{T}\Big(\big(U^{A''_1}\big)^T\cdot\Omega\Big)}_1dU^{A''_1} &=\int\norm{\mathcal{T}\Big(\big(U^{A''_1}\big)^T\cdot\big(\Delta_+-\Delta_- \big) \Big)}_1 dU^{A''_1} \\
    &\leq \int \Tr\Big[\mathcal{T}\Big(\big(U^{A''_1}\big)^T\cdot\big(\Delta_+\Big)\Big]dU^{A''_1}+\int\Tr\Big[\mathcal{T}\Big(\big(U^{A''_1}\big)^T\cdot\big(\Delta_-)\Big)\Big] dU^{A''_1} \\
    &= \big( \Tr[\Delta_+] + \Tr[\Delta_-]\big)\Tr[\mathcal{T}(\pi^{A''_1})] \\
    &\leq 4f(\epsilon).
\end{align*}

\noindent\textbf{Derandomization}

Consider the following random variables:
\begin{enumerate}
    \item $X_1\coloneqq \abs{\Tr[\omega]-1}$ . 
    \item $X_2\coloneqq \Tr[\mathcal{T}\Big(\big(U^{A''_1}\big)^T\cdot\lambda^{A''_1}\Big)]$ .
    \item $X_3\coloneqq \norm{\mathcal{T}\Big(\big(U^{A''_1}\big)^T\cdot\tilde{\Omega}\Big)-\omega}_1$ .
\end{enumerate}

We know from the previous arguments that 
\begin{enumerate}
    \item $\E[X_1]\leq 13f(\epsilon)\eqqcolon \mu_1$ .
    \item $\E[X_2]=2^{-H_{\min}^{f(\epsilon)}(A''_0|A''_1)_{\Omega}}\eqqcolon \mu_2$ .
    \item $\E[X_3]\leq 4f(\epsilon) \eqqcolon \mu_3$ .
\end{enumerate}

Then, by Markov's inequality and a union bound, for some integer $k\geq 4$, we see that 
\begin{align*}
    \Pr[\prod\limits_{i\in [3]}\brak{X_i\leq k\cdot \mu_i}] \geq 1-\frac{3}{k}.
\end{align*}
This implies that there exists, with at least constant probability, a fixed value of $U^{A''_1}$ such that 
\begin{enumerate}
    \item $\norm{\Tr[\omega]-1}_1 \leq k\cdot 13f(\epsilon)$ .
    \item $\Tr[\mathcal{T}\Big(\big(U^{A''_1}\big)^T\cdot\lambda^{A''_1}\Big)]\leq k\cdot 2^{-H_{\min}^{f(\epsilon)}(A''_0|A''_1)_{\Omega}}$ . \label{item:two}
    \item $\norm{\mathcal{T}\Big(\big(U^{A''_1}\big)^T\cdot\tilde{\Omega}\Big)-\omega}_1 \leq k\cdot 4f(\epsilon)$ .
\end{enumerate}
Consider now 
\begin{align*}
    \norm{\frac{\mathcal{T}\Big(\big(U^{A''_1}\big)^T\cdot\tilde{\Omega}\Big)}{\Tr[\mathcal{T}\Big(\big(U^{A''_1}\big)^T\cdot\tilde{\Omega}\Big)]}-\frac{\omega}{\Tr[\omega]}}_1 &\leq \norm{\frac{\mathcal{T}\Big(\big(U^{A''_1}\big)^T\cdot\tilde{\Omega}\Big)}{\Tr[\mathcal{T}\Big(\big(U^{A''_1}\big)^T\cdot\tilde{\Omega}\Big)]}-\mathcal{T}\Big(\big(U^{A''_1}\big)^T\cdot\tilde{\Omega}\Big)}_1 \\
    & + \norm{\mathcal{T}\Big(\big(U^{A''_1}\big)^T\cdot\tilde{\Omega}\Big)-\omega}_1 \\
    & + \norm{\omega-\frac{\omega}{\Tr[\omega]}}_1 \\
    &= \abs{\Tr[\mathcal{T}\Big(\big(U^{A''_1}\big)^T\cdot\tilde{\Omega}\Big)]-1} \\
    &+ \norm{\mathcal{T}\Big(\big(U^{A''_1}\big)^T\cdot\tilde{\Omega}\Big)-\omega}_1 + \abs{\Tr[\omega]-1} \\
    & \leq k\cdot 17f(\epsilon)+ k\cdot 4f(\epsilon) + k\cdot 13f(\epsilon) =k\cdot 34f(\epsilon) \coloneqq k'f(\epsilon).
\end{align*}

Notice that $\frac{\mathcal{T}\Big(\big(U^{A''_1}\big)^T\cdot\tilde{\Omega}\Big)}{\Tr[\mathcal{T}\Big(\big(U^{A''_1}\big)^T\cdot\tilde{\Omega}\Big)]}$ is a normalized state in the $\sqrt{k'f(\epsilon)}$ ball (w.r.t the purified distance) around the state $\tilde{\omega}$. This means that $\frac{\mathcal{T}\Big(\big(U^{A''_1}\big)^T\cdot\tilde{\Omega}\Big)}{\Tr[\mathcal{T}\Big(\big(U^{A''_1}\big)^T\cdot\tilde{\Omega}\Big)]}$ is a candidate optimiser for $H_{\min}^{\sqrt{k'\cdot f(\epsilon)}}(A''_0|BR_1)_{\tilde{\omega}}$. To be precise, using Item~\ref{item:two}, we see that 
\begin{align*}
    H_{\min}^{\sqrt{k'\cdot f(\epsilon)}}(A''_0|BR_1)_{\tilde{\omega}} \geq H_{\min}^{f(\epsilon)}(A''_0|A''_1)_{\Omega}-\log k +\log (1-17kf(\epsilon)).
\end{align*}
We now set the function $f(\epsilon)$ as $\frac{\epsilon^2}{k'}$. Then, substituting we get
\begin{align*}
    H_{\min}^{\epsilon}(A''_0|BR_1)_{\tilde{\omega}} \geq H_{\min}^{O(\epsilon^2)}(A''_0|A''_1)_{\Omega}-O(1),
\end{align*}
which concludes the proof.
\end{proof}

\noindent The following lemma demonstrates the use of almost CPTP maps with the decoupling theorem in a channel coding scenario.

\begin{lemma}\label{onehaar}
Given the intermediary state $\ket{\omega}^{A''_0A'BR_1}$ and the channel $\mathcal{N}^{A'\rightarrow C}$ with Stinespring dilation $\mathcal{U}^{A'\rightarrow CE}$, let the measure $\mu$ to be the Haar measure over the unitary group corresponding to the system $A''_1$. Let $\ket{\tilde{\omega}}$ we the normalised unit vector obtained from $\ket{\omega}$. Suppose that we are given the condition
\[
H_{\max}^{f(\epsilon)}(A_1)_{\psi}\leq H_{\min}^{f(\epsilon)}(A''_1)_{\Omega}+O(\log f(\epsilon)).
\]
Then following holds true with constant probability over the choices of $U^{A''_0}$ and $U^{A''_1}$
\begin{align*}
    &\norm{\Big(\abs{A''_0}~\Tr_C~\mathcal{U}_{\mathcal{N}} ~\textup{op}^{A''_0\rightarrow A'BR_1}(\tilde{\omega})~U^{A''_0}\cdot \eta^{A''_0R_0}\Big)^{R_0BR_1E}-\eta^{R_0}\otimes \tilde{\omega}^{EBR_1}}_1 \\
    &\leq 2^{-\frac{1}{2}H_{\min}^{O(\epsilon^2)}(A''_0|A''_1E)_{\mathcal{U}_{\mathcal{N}}\cdot \Omega}-\frac{1}{2}H_{\min}^{\epsilon}(A_0|R_0)_{\eta}+2\log k}+12k\epsilon,
\end{align*}
where $k$ is a constant positive integer and $f(\epsilon)=O(\epsilon^2)$.
\end{lemma}
 \begin{proof}
First, we define the intermediary state
\begin{align*}
    \ket{\omega}^{A''_0BR_1CE} \coloneqq \mathcal{U}_{\mathcal{N}}^{A'\to CE} \ket{\omega}^{A''_0BR_1A'}.
\end{align*}
It is not hard to see from the properties of the $\textup{op}$ operator that 
\begin{align*}
    \ket{\omega}^{A''_0BR_1CE} &= \sqrt{\abs{A''_1}} \Big(\textup{op}^{A''_1\to A''_0CE}(\mathcal{U}_{\mathcal{N}}\ket{\Omega}) U^{A''_1}\ket{\psi}^{A''_1BR_1}\Big).
\end{align*}

Note that since $\mathcal{U}_{\mathcal{N}}$ is trace preserving, the traces of $\omega^{A''_0BR_1CE}$ and $\omega^{A''_0BR_1A'}$ are the same. We refer to this trace quantity as $\Tr[\omega]$ throughout the proof.

Recall that, the condition $
H_{\max}^{f(\epsilon)}(A)_{\psi}\leq H_{\min}^{f(\epsilon)}(A''_1)_{\Omega}+2\log f(\epsilon)
$ along with \cref{lem:TraceIntermediateStateLemma} implies that 
\[
    \E_{A''_1}\abs{\Tr[\omega]-1}\leq 13f(\epsilon).
\]
 We work with the same almost CPTP map $\mathcal{T}^{A''_1\rightarrow BR_1}$ as in \cref{almostcptp}. Suppose $\tilde{\Omega}^{A''_0A''_1E}$ is the optimiser in the definition of $H_{\min}^{f(\epsilon)}(A''_0|A''_1E)_{\mathcal{U}_{\mathcal{N}}\cdot\Omega}$ and 
 $\norm{\mathcal{U}_{\mathcal{N}}\cdot\Omega-\tilde{\Omega}}\leq 2f(\epsilon)$. Let $\lambda^{A''_1E}$ be a positive semidefinite matrix such that
 \begin{align*}
     \Tr[\lambda^{A''_1E}] = &2^{-H_{\min}^{f(\epsilon)}(A''_0|A''_1E)_{\mathcal{U}_{\mathcal{N}}\cdot \Omega}} \\
     \intertext{and}
     & \tilde{\Omega} \leq \mathbb{I}^{A''_0}\otimes \lambda^{A''_1E} \\
     \implies & \mathcal{T}\Big(\big(U^{A''_1}\big)^T\cdot\Tilde{\Omega}\Big) \leq \mathbb{I}^{A''_0}\otimes \mathcal{T}\Big(\big(U^{A''_1}\big)^T\cdot\lambda^{A''_1E}\Big).
 \end{align*}
 As before, we note that the action of the random map $\mathcal{T}\Big(\big(U^{A''_1}\big)^T(\cdot)\Big)$ does not change the trace of $\lambda^{A''_1E}$ on average:
\begin{align*}
    \int\mathcal{T}\Big(\big(U^{A''_1}\big)^T\cdot\lambda^{A''_1E}\Big)dU^{A''_1} &= \int\mathcal{T}\Big(U^{A''_1}\cdot\lambda^{A''_1E}\Big)dU^{A''_1} \\
    &= \mathcal{T}(\pi^{A''_1})\otimes \lambda^{E}.\\
    \intertext{Taking trace on both sides} 
    \Tr[\int\mathcal{T}\Big(\big(U^{A''_1}\big)^T\cdot\lambda^{A''_1E}\Big)dU^{A''_1}]&=\Tr[\lambda^E] \\
    &=\Tr[\lambda^{A''_1E}]\\
    &=2^{-H_{\min}^{f(\epsilon)}(A''_0|A''_1E)_{\mathcal{U}_{\mathcal{N}}\Omega}}.
\end{align*}
It is also not hard to see via the definition of $\ket{\omega}^{A''_0BR_1CE}$ that 
\begin{align*}
    \mathcal{T}\Big(\big(U^{A''_1}\big)^T\mathcal{U}_{\mathcal{N}}\cdot \Omega\Big)^{A''_0BR_1CE} = \omega^{A''_0BR_1CE}.
\end{align*}

We will now apply the smooth single sender decoupling theorem to the quantity on the left in the theorem statement, after appropriate normalisation:
\begin{align*}
    &\int\frac{1}{\Tr[\omega]}\norm{\Big(\abs{A''_0}~\Tr_C~\mathcal{U}_{\mathcal{N}} ~\textup{op}^{A''_0\rightarrow A'BR_1}(\omega)~U^{A''_0}\cdot \eta^{A''_0R_0}\Big)^{R_0BR_1E}-\eta^{R_0}\otimes \omega^{BR_1E}}_1dU^{A''_0} \\
    &\leq 2^{-\frac{1}{2}H_{\min}^{\epsilon}(A''_0|BR_1E)_{\tilde{\omega}}-\frac{1}{2}H_{\min}^{\epsilon}(A_0|R_0)_{\eta}}+12\epsilon.
\end{align*}

\textbf{Derandomization}

Next, define the random variables
\begin{enumerate}
\item $X_1\coloneqq \norm{\Tr[\omega]-1}_1 $
    \item $X_2\coloneqq \Tr[\mathcal{T}\Big(\big(U^{A''_1}\big)^T\cdot\lambda^{A''_1E}\Big)]$
    \item $X_3 \coloneqq \norm{\Big(\mathcal{T}\Big(\big(U^{A''_1}\big)^T\cdot\Tilde{\Omega}\Big)^{A''_0BR_1E}-\omega^{A''_0BR_1E}}_1$
    \item $X_4\coloneqq \frac{1}{\Tr[\omega]}\norm{\Big(\abs{A''_0}~\Tr_C~\mathcal{U}_{\mathcal{N}} ~\textup{op}^{A''_0\rightarrow A'BR_1}(\omega)~U^{A''_0}\cdot \eta^{A''_0R_0}\Big)^{R_0BR_1E}-\eta^{R_0}\otimes \omega^{BR_1E}}_1$
    \item $\mu_4(U^{A''_1})\coloneqq 2^{-\frac{1}{2}H_{\min}^{\epsilon}(A''_0|BR_1E)_{\tilde{\omega}}-\frac{1}{2}H_{\min}^{\epsilon}(A_0|R_0)_{\eta}}+12\epsilon$
\end{enumerate}

We define $\mu_1, \mu_2$ and $\mu_3$ analogously as in \cref{almostcptp}. We already know from the single sender decoupling theorem that 
\begin{align*}
    \E_{U^{A''_0}}[X_4~|~U^{A''_1}] \leq \mu_4(U^{A''_1}).
\end{align*}

Let $k\in \mathbb{N}$ be some positive integer $\geq 5$. Then, via the conditional Markov inequality we see that 
\begin{align*}
    \Pr\limits_{U^{A''_0}, U^{A''_1}}[X_4 \geq k\cdot \mu_4(A''_1)] &= \sum\limits_{U^{A''_1}} \Pr\limits_{U^{A''_0}}[X_4 \geq k\cdot \mu_4(A''_1)~|~U^{A''_1}]\cdot\Pr[U^{A''_1}] \\
    &\leq \sum\limits_{U^{A''_1}}\frac{1}{k}\Pr[U^{A''_1}] \\
    &= \frac{1}{k}.
\end{align*}
Since $X_1, X_2$ and $X_3$ are only functions of $U^{A''_1}$, Markov's inequality along with a union bound imply that 
\begin{align*}
    \Pr\limits_{U^{A''_0}, U^{A''_1}}[\prod\limits_{i\in [4]}\brak{X_i\leq k\cdot \mu_i}] \geq 1-\frac{4}{k}.
\end{align*}
 Then, repeating the arguments in \cref{almostcptp}, one can see that there exists, with probability at least $1-\frac{4}{k}$, fixed unitaries  $U^{A''_0}$ and $U^{A''_1}$ such that the following holds:
\begin{align*}
    & H_{\min}^{\sqrt{34k f(\epsilon)}}(A''_0| BR_1E)_{\tilde{\omega}} \geq H_{\min}^{f(\epsilon)}(A''_0|A''_1E)_{\mathcal{U}_{\mathcal{N}}\cdot \Omega}-\log k + \log (1-17k f(\epsilon)) \\
    \intertext{and}
    &\frac{1}{\Tr[\omega]}\norm{\Big(\abs{A''_0}~\Tr_C~\mathcal{U}_{\mathcal{N}} ~\textup{op}^{A''_0\rightarrow A'BR_1}(\omega)~U^{A''_0}\cdot \eta^{A''_0R_0}\Big)^{R_0BR_1E}-\eta^{R_0}\otimes \omega^{BR_1E}}_1 \\
    &\leq k\cdot2^{-\frac{1}{2}H_{\min}^{\epsilon}(A''_0|BR_1E)_{ \tilde{\omega}}-\frac{1}{2}H_{\min}^{\epsilon}(A_0|R_0)_{\eta}}+12k\epsilon.
\end{align*}
Setting $f(\epsilon)=\frac{\epsilon^2}{34k}$ and making the appropriate substitutions, we see that 
\begin{align*}
    &\frac{1}{\Tr[\omega]}\norm{\Big(\abs{A''_0}~\Tr_C~\mathcal{U}_{\mathcal{N}} ~\textup{op}^{A''_0\rightarrow A'BR_1}(\omega)~U^{A''_0}\cdot \eta^{A''_0R_0}\Big)^{BR_0R_1E}-\eta^{R_0}\otimes \omega^{BR_1E}}_1 \\
    &\leq k\cdot \frac{1}{(1-\frac{\epsilon^2}{2})}\cdot 2^{-\frac{1}{2}H_{\min}^{\frac{\epsilon^2}{34k}}(A''_0|A''_1E)_{ \mathcal{U}_{\mathcal{N}}\cdot\Omega}-\frac{1}{2}H_{\min}^{\epsilon}(A_0|R_0)_{\eta}+\log k }+12k\epsilon \\
    &\leq 2^{-\frac{1}{2}H_{\min}^{\frac{\epsilon^2}{34k}}(A''_0|A''_1E)_{ \mathcal{U}_{\mathcal{N}}\cdot\Omega}-\frac{1}{2}H_{\min}^{\epsilon}(A_0|R_0)_{\eta}+O(\log k) }+12k\epsilon.
\end{align*}
This concludes the proof.
\end{proof}

\subsection{Proof of Proposition \ref{achievability}}
We are now ready to prove our main theorem in this section, which is \cref{achievability}.
  \begin{proof}
  At the outset we assume that
  \[
  H_{\max}^{f(\eps)}(A_1)_{\psi}\leq H_{\min}^{f(\eps)}(A''_1)_{\Omega} + O(\log f(\eps)),
  \]
  where $f(\epsilon)=O(\epsilon^2)$. Consider the randomised encoder
  \begin{align*}
      \mathcal{E}^{A_0A_1\rightarrow A'}_{\textsc{rand}}\equiv\sqrt{\abs{A''_0}\abs{A''_1}}\textup{op}^{A''_0A''_1\rightarrow A'}(\Omega'(\theta))~\big(U^{A''_0}W_0^{A_1\rightarrow A''_0}\otimes U_1^{A''_1}W_1^{A_1\rightarrow A''_1}\big) .
  \end{align*}
  From the \cref{lem:IntermediateStateLemma} we know that
  \[
  \mathcal{E}(\eta\otimes \psi)=\abs{A''_0}\op^{A''_0\to BR_1A'}(\ket{\omega}^{A''_0BR_1A'})U^{A''_0}\cdot \eta^{A''_0R_0},
  \]
  where $\ket{\omega}^{A''_0BR_1A'}$ is the intermediate state defined as 
  \[
  \ket{\omega}^{A''_0BR_1A'}=\sqrt{\abs{A''_1}}\op^{A''_1\to A''_0A'}(\ket{\Omega^{A''_0A''_1A'}})U^{A''_1}\ket{\psi}^{A''_1R_1B}.
  \]
  We also use the convention that $\Tilde{\omega}$ is the normalised version of $\omega$.
  \\ \vspace{1mm} \\ {\large{\bf The Decoupling Step}} \\
  We consider the four decoupling equations corresponding to the encoding and decoding steps for Alice\sub{0} and Alice\sub{1}.
  \\ \vspace{2mm} \\ {\bf The Encoding Equations} 
    \begin{align*}\label{eq:encAliceZero}
      \E_{U^{A''_0}}\Big[ \left\lVert\abs{A''_0}\Tr_{A'}~\textup{op}^{A''_0\rightarrow A'BR_1}(\Tilde{\omega})~U^{A''_0}W_0\cdot \eta^{R_0A_0}\right.-&\left.\eta^{R_0}\otimes \Tilde{\omega}^{BR_1}\right\rVert_1\Big] \\
      &\leq 2^{\frac{1}{2} H_{\max}^{\epsilon}(A_0)_{\eta}-\frac{1}{2} H_{\min}^{\epsilon}(A''_0|B_1R_1)_{\Tilde{\omega}}}+12\epsilon . \tag{enc\_Alice\sub{0}}
  \end{align*}
  
  \begin{align*}\label{eq:encAliceOne}
      \E_{U^{A''_1}}\Big[ \left\lVert\abs{A''_1}\Tr_{A''_0A'}~\textup{op}^{A''_1\rightarrow A''_0A'}(\Omega(\theta))~U^{A''_1}W_1\cdot \psi^{R_1BA_1}\right.-&\left.\psi^{R_1B}\right\rVert_1\Big] \\
      &\leq 2^{\frac{1}{2} H_{\max}^{f(\epsilon)}(A_1)_{\psi}-\frac{1}{2} H_{\min}^{f(\epsilon)}(A''_1)_{\Omega}}+12f(\epsilon) . \tag{enc\_Alice\sub{1}}
  \end{align*}
  \vspace{2mm} \\ {\bf The Decoding Equations} 
    \begin{align*}\label{eq:decAliceZero}
      \E_{U^{A''_0}}\Big[ \left\lVert\abs{A''_0}\Tr_{C}\mathcal{U}_{\mathcal{N}}~\textup{op}^{A''_0\rightarrow A'BR_1}(\Tilde{\omega})~U^{A''_0}W_0\cdot \eta^{R_0A_0}\right.-&\left.\eta^{R_0}\otimes \Tilde{\omega}^{BR_1E}\right\rVert_1\Big] \\
      &\leq 2^{-\frac{1}{2} H_{2}^{\epsilon}(A_0|R_0)_{\eta}-\frac{1}{2} H_{\min}^{\epsilon}(A''_0|BR_1E)_{\Tilde{\omega}}}+12\epsilon. \tag{dec\_Alice\sub{0}}
  \end{align*}
  \begin{align*}\label{eq:decAliceOne}
      \E_{U^{A''_1}}\Big[ \left\lVert\abs{A''_1}\Tr_{CA''_0}\mathcal{U}_{\mathcal{N}}~\textup{op}^{A''_1\rightarrow A''_0A'}(\Omega)~U^{A''_1}W_1\cdot \psi^{R_1A_1}\right.-&\left.\psi^{R_1}\otimes \Omega^{E}\right\rVert_1\Big] \\
      &\leq 2^{-\frac{1}{2} H_{2}^{\epsilon}(A_1|R_1)_{\psi}-\frac{1}{2} H_{\min}^{\epsilon}(A''_1|E)_{\mathcal{U}_{\mathcal{N}}\cdot\Omega}}+12\epsilon . \tag{dec\_Alice\sub{1}}
  \end{align*}
  \vspace{2mm} {\large{\bf Derandomisation}} \\
  Note that under the assumption that
  \[
    H_{\max}^{f(\eps)}(A_1)_{\psi}\leq H_{\min}^{f(\eps)}(A''_1)_{\Omega} + O(\log f(\eps)),
  \]
  the upper bound in \cref{eq:encAliceOne} is at most $13f(\eps)$. Then, following steps that are similar to the arguments in \cref{onehaar}, we see that for a large enough but constant integer $k$, there exist unitaries $U^{A''_0}$ and $U^{A''_1}$ such that
  
\[
  \begin{aligned}
  &\begin{aligned}
      \left\lVert\abs{A''_0}\Tr_{A'}~\textup{op}^{A''_0\rightarrow A'BR_1}(\Tilde{\omega})~U^{A''_0}W_0\cdot \eta^{R_0A_0}\right.-&\left.\eta^{R_0}\otimes \Tilde{\omega}^{BR_1}\right\rVert_1 \\
      &\leq k\cdot 2^{\frac{1}{2} H_{\max}^{\epsilon}(A_0)_{\eta}-\frac{1}{2} H_{\min}^{\epsilon}(A''_0|B_1R_1)_{\Tilde{\omega}}}+12k\epsilon,
  \end{aligned} \\
  &\begin{aligned}
       \left\lVert\abs{A''_1}\Tr_{A''_0A'}~\textup{op}^{A''_1\rightarrow A''_0A'}(\Omega(\theta))~U^{A''_1}W_1\cdot \psi^{R_1BA_1}\right.-&\left.\psi^{R_1B}\right\rVert_1 \\
      &\leq 13kf(\epsilon),
  \end{aligned}\\
  &\begin{aligned}
  \left\lVert\abs{A''_0}\Tr_{C}\mathcal{U}_{\mathcal{N}}~\textup{op}^{A''_0\rightarrow A'BR_1}(\Tilde{\omega})~U^{A''_0}W_0\cdot \eta^{R_0A_0}\right.-&\left.\eta^{R_0}\otimes \Tilde{\omega}^{BR_1E}\right\rVert_1 \\
      &\leq k\cdot 2^{-\frac{1}{2} H_{2}^{\epsilon}(A_0|R_0)_{\eta}-\frac{1}{2} H_{\min}^{\epsilon}(A''_0|BR_1E)_{\Tilde{\omega}}}+12k\epsilon,
  \end{aligned} \\
  &\begin{aligned}
  \left\lVert\abs{A''_1}\Tr_{CA''_0}\mathcal{U}_{\mathcal{N}}~\textup{op}^{A''_1\rightarrow A''_0A'}(\Omega)~U^{A''_1}W_1\cdot \psi^{R_1A_1}\right.-&\left.\psi^{R_1}\otimes \Omega^{E}\right\rVert_1 \\
      &\leq k\cdot 2^{-\frac{1}{2} H_{2}^{\epsilon}(A_1|R_1)_{\psi}-\frac{1}{2} H_{\min}^{\epsilon}(A''_1|E)_{\mathcal{U}_{\mathcal{N}}\cdot\Omega}}+12k\epsilon,
  \end{aligned} \\
  &H_{\min}^{\epsilon}(A''_0|BR_1)_{\Tilde{\omega}}\geq H_{\min}^{\frac{\eps^2}{34k}}(A''_0|A''_1)_{\Omega}-\log k + \log(1-\frac{\epsilon^2}{34}), \\
  &H_{\min}^{\epsilon}(A''_0|BR_1E)_{\Tilde{\omega}}\geq H_{\min}^{\frac{\eps^2}{34k}}(A''_0|A''_1E)_{\mathcal{U}_{\mathcal{N}}\cdot \Omega}-\log k + \log(1-\frac{\epsilon^2}{34}),
  \end{aligned}
  \]
  where we get the data processing inequalities  by setting $f(\eps)=\frac{\eps^2}{34k}$. Simplifying the above and using the definition of $\omega$ we see that the above inequalities imply that
  \[
  \begin{aligned}
  &\begin{aligned}
  \left\lVert\abs{A''_0}\Tr_{A'}~\textup{op}^{A''_0\rightarrow A'BR_1}(\Tilde{\omega})~U^{A''_0}W_0\cdot \eta^{R_0A_0}\right.-&\left.\eta^{R_0}\otimes \Tilde{\omega}^{BR_1}\right\rVert_1 \\
      &\leq 2^{\frac{1}{2} H_{\max}^{\epsilon}(A_0)_{\eta}-\frac{1}{2} H_{\min}^{O(\epsilon^2)}(A''_0|A''_1)_{\Omega}+O(\log k)}+12k\epsilon \\
      &\coloneqq \delta_{\textup{enc}}(1),
  \end{aligned}\\
  &\begin{aligned}
  \norm{\omega^{R_1B}-\psi^{R_1B}}_1 \leq 13kf(\eps) \coloneqq \delta_{\textup{enc}}(2),
  \end{aligned} \\
  &\begin{aligned}
  \left\lVert\abs{A''_0}\Tr_{C}\mathcal{U}_{\mathcal{N}}~\textup{op}^{A''_0\rightarrow A'BR_1}(\Tilde{\omega})~U^{A''_0}W_0\cdot \eta^{R_0A_0}\right.-&\left.\eta^{R_0}\otimes \Tilde{\omega}^{BR_1E}\right\rVert_1 \\
      &\leq 2^{-\frac{1}{2} H_{2}^{\epsilon}(A_0|R_0)_{\eta}-\frac{1}{2} H_{\min}^{O(\epsilon^2)}(A''_0|A''_1E)_{\mathcal{U}_{\mathcal{N}}\cdot\Omega}+O(\log k)}+12k\epsilon \\
      &\coloneqq \delta_{\textup{dec}}(1),
  \end{aligned} \\
  &\begin{aligned}
  \norm{\omega^{R_1E}-\psi^{R_1}\otimes \Omega^E}_1 \leq 2^{-\frac{1}{2} H_{2}^{\epsilon}(A_1|R_1)_{\psi}-\frac{1}{2} H_{\min}^{\epsilon}(A''_1|E)_{\mathcal{U}_{\mathcal{N}}\cdot\Omega}+O(\log k)}+12k\epsilon \coloneqq \delta_{\textup{dec}}(2).
  \end{aligned}
  \end{aligned}
  \]
  \vspace{2mm} {\large{\bf Normalisation and Uhlmann's Theorem}} \\
  
  Note that, the matrices on the left inside each $1$-norm expression is unnormalised. We use \cref{lem:NormalisationLemma} to replace each of these with their normalised counterparts, which increases each of the upper bounds by a multiplicative factor of $2$. Also note that, by \cref{cor:purificationLemma}, 
  \[
  \norm{\frac{\abs{A''_0}\Tr_{A'}~\textup{op}^{A''_0\rightarrow A'BR_1}(\omega)~U^{A''_0}W_0\cdot \eta^{R_0A_0}}{\Tr\left[ \abs{A''_0}\Tr_{A'}~\textup{op}^{A''_0\rightarrow A'BR_1}(\omega)~U^{A''_0}W_0\cdot \eta^{R_0A_0}\right]}-\eta^{R_0}\otimes \Tilde{\omega}^{BR_1}}_1\leq 2\delta_{\textup{enc}}(1)
  \]
  which, by the definition of $\mathcal{E}(\eta\otimes \psi)$ is equivalent to
  \[
  \norm{\Tr_{A'}\frac{\mathcal{E}(\eta\otimes \psi)}{\Tr\left[\mathcal{E}(\eta\otimes \psi)\right]}-\eta^{R_0}\otimes \Tilde{\omega}^{BR_1}}_1 \leq 2\delta_{\textup{enc}}(1).
  \]
  Then, combining the first two inequalities (after appropriately appending $\eta^{R_0}$ to the second inequality) we see that
  \[
    \norm{\Tr_{A'}\frac{\mathcal{E}(\eta\otimes \psi)}{\Tr\left[\mathcal{E}(\eta\otimes \psi)\right]}-\eta^{R_0}\otimes \psi^{R_1B}}_1 \leq 2\delta_{\textup{enc}}(1)+2\delta_{\textup{enc}}(2).
  \]
  Thus, applying Uhlmann's theorem, we see that there exists an encoding isometry $V_{\textup{enc}}^{A_0A_1\to A'}$ such that
  \[
  \norm{\frac{\mathcal{E}(\eta\otimes \psi)}{\Tr\left[\mathcal{E}(\eta\otimes \psi)\right]}-V^{A_0A_1\to A'}_{\textup{enc}}\cdot \eta^{A_0R_0}\otimes \psi^{A_1BR_1}}_1 \leq 2\sqrt{2\delta_{\textup{enc}}(1)+2\delta_{\textup{enc}}(2)}.
  \]
  Next, note that $\frac{1}{\sqrt{\Tr[\omega]}}\ket{\omega}^{A''_0BR_1CE}$ is a valid purification of the state $\Tilde{\omega}^{BR_1E}$ appearing in the inequality corresponding to $\delta_{\textup{dec}}(1)$ and also of the state $\Tilde{\omega}^{R_1E}$ appearing in the normalised version of the inequality corresponding to $\delta_{\textup{dec}}(2)$. Then, via Uhlmann's theorem we see that there exist isometries 
  \begin{align*}
&V_1^{C\to A_0\gol{C}\gol{A''}_0} \\    
&V_2^{\gol{A}''_0\gol{C}B\to A_1BF},
\end{align*}
such that 
\begin{align*}
    &\norm{V_1^{C\to A_0\gol{C}\gol{A''}_0}\cdot \mathcal{U}_{\mathcal{N}}\left(\frac{\mathcal{E}(\eta\otimes \psi)}{\Tr\left[\mathcal{E}(\eta\otimes \psi)\right]}\right)-\eta^{A_0R_0}\otimes \Tilde{\omega}^{\gol{A}''_0\gol{C}R_1BE} }_1 \leq 2\sqrt{2\delta_{\textup{dec}}(1)} \\
    \intertext{and}
    &\norm{V_2^{\gol{A}''_0\gol{C}B\to A_1BF}\cdot \Tilde{\omega}^{\gol{A}''_0\gol{C}R_1BE}-\psi^{A_1BR_1}\otimes \Omega^{EF}}_1 \leq 2\sqrt{2\delta_{\textup{dec}}(2)}.
\end{align*}
Then, by using the triangle inequality after appending the state $\eta^{A_0R_0}$ to the second inequality and applying the isometry $V_2$ to the first inequality, we see that
\[
\begin{aligned}
\norm{V_2\circ V_1 \circ \mathcal{U}_{\mathcal{N}}\left(\frac{\mathcal{E}(\eta\otimes \psi)}{\Tr\left[\mathcal{E}(\eta\otimes \psi)\right]}\right)-\eta^{A_0R_0}\otimes \psi^{A_1BR_1}\otimes \Omega^{EF}}_1 \leq&~2\sqrt{2\delta_{\textup{dec}}(1)}+2\sqrt{2\delta_{\textup{dec}}(2)} \\
\coloneqq & ~\delta'.
\end{aligned}
\]
Defining 
\[
\mathcal{C}^{BC\to A_0A_1} \coloneqq \Tr_F V_2\circ V_1
\]
and discarding the $E$ system, we see that
\[
\norm{\mathcal{C}\circ \mathcal{N} \left(\frac{\mathcal{E}(\eta\otimes \psi)}{\Tr\left[\mathcal{E}(\eta\otimes \psi)\right]}\right)-\eta^{A_0R_0}\otimes \psi^{A_1BR_1}}_1 \leq \delta'.
\]
A further application of the triangle inequality with the expression which bounds the encoding error (after acting the operator $\mathcal{C}\circ \mathcal{N} $ on it) shows that
\[
\begin{aligned}
\norm{\mathcal{C}\circ\mathcal{N}\circ V_{\textup{enc}}(\eta\otimes \psi)-\eta^{A_0R_0}\otimes \psi^{A_1BR_1}}_1 \leq&~ \delta'+  2\sqrt{2\delta_{\textup{enc}}(1)+2\delta_{\textup{enc}}(2)}\\
\coloneqq&~ \delta.
\end{aligned}
\]
\\ \vspace{2mm} {\large{\bf Successive Cancellation}} \\
The decoding algorithm is now clear. 
\begin{enumerate}
    \item Alice creates a state close to $\left(\frac{\mathcal{E}(\eta\otimes \psi)}{\Tr\left[\mathcal{E}(\eta\otimes \psi)\right]}\right)$ by using the encoding isometry $V_{\textup{enc}}$ on $\ket{\eta}^{A_0R_0}\ket{\psi}^{A_1BR_1}$.
    \item Charlie first recovers the $A_0$ system of $\ket{\eta}^{A_0R_0}$ by applying the isometry $V_1$. This isometry also does the job of locally simulating the system $A''_0$ of the state $\omega$ at Charlie's end.
    \item Using the locally created $A''_0$ system and the pre-shared $B$ system as 'side information', Bob decodes the $A_1$ register of the state $\ket{\psi}^{A_1BR_1}$.
    \item The entire decoding procedure, after discarding the purifying system $F$ is encapsulated by the operator $\mathcal{C}$.
\end{enumerate}

This completes the proof of the theorem.

\end{proof}

\section{Quantum Rate Splitting II}\label{chap:RateSplit2}

In this section, we apply the quantum rate splitting and successive cancellation decoding techniques developed in \cref{chap:RateSplit1} to the problem of entanglement transmission across the QMAC and the QIC in the one-shot setting. This allows us to prove non-trivial achievability results in the one-shot setting in terms of smoothed one-shot entropic quantities, without appealing to a simultaneous decoder. We have already mentioned that Yard et al. \cite{Yard_MAC} showed that the natural quantum analogue of the pentagonal rate region, with the mutual information replaced by the regularised coherent information, is achievable for the QMAC. To the best of our knowledge, the only inner bound known for the QIC (for both the unassisted and entanglement assisted regimes) is what one would obtain by treating the
channel as two independent unassisted point-to-point channels.

Recall that the idea in rate splitting is to `split' Alice into two senders Alice\sub{0} and 
Alice\sub{1}, each sending disjoint parts of Alice's original message, 
such that any point in the pentagonal rate region of 
Figure~\ref{fig:QMACasymp2} like P can be obtained without time sharing 
by a successive cancellation process where Charlie first decodes 
Alice\sub{0}'s message, then Bob's message using Alice\sub{0}'s message
as side information and finally Alice\sub{1}'s message using Bob's and 
Alice\sub{0}'s messages as side information. \\
\vspace{2mm} \\

\begin{figure}[hhh]
\vspace*{-5mm}
\begin{center}
\begin{tikzpicture}
\draw [->, line width=1 pt] (1,-2) -- (1,2);
\draw [->, line width=1 pt] (1,-2) -- (5,-2);
\draw [line width=1 pt] (1,0) -- (2,0);
\draw [line width=1 pt] (3,-2) -- (3,-1);
\draw [line width=1 pt] (2,0) -- (3,-1);
\draw [line width=1 pt,dotted] (1,1) -- (4,-2);
\draw (1,2) node[anchor=east] {Bob};
\draw (5,-2) node[anchor=west] {Alice};
\draw (1.5,-1) node[anchor= west] {$n\rightarrow \infty$};
\draw (2.5,-0.5) node[anchor=west] {{\footnotesize $P$}};
\draw (1,0) node[anchor=north] {{\footnotesize $(0,I(B\rangle AC))$}};
\draw (4,-2) node[anchor=north ] {{\footnotesize $(I(A\rangle BC), 0)$}};
\draw (3,-1) node[anchor=west] 
{{\footnotesize $S = (I(A\rangle BC), I(B\rangle C))$}};
\draw (2,0) node[anchor=west] 
{{\footnotesize $T = (I(A\rangle C),I(B\rangle AC))$}};
\draw (1,1) node[anchor=west] {{\footnotesize $x+y= I(AB\rangle C)$}};
\end{tikzpicture}
\end{center}
\vspace*{-5mm}
\caption[Achievable rate region for the unassisted quantum MAC
per channel use in the asymptotic iid setting.]{Achievable rate region for the unassisted quantum MAC
per channel use in the asymptotic iid setting with respect to a fixed bipartite input control state. The actual rate region is the convex closure of all such pentagonal regions.}
\label{fig:QMACasymp2}
\end{figure}
Our one-shot rates are in terms of the
smooth coherent min-information defined in
\cref{def:SmoothCoherentMinInformation}. Since the smooth coherent 
min-information
is not known to possess a chain rule with equality, we get an 
achievable rate region of the form in Figure~\ref{fig:QMAConeshot1}. 
Our achievable rate region is a subset of the `ideal' 
pentagonal rate region shown by the dashed line.  Nevertheless, 
using a quantum asymptotic 
equipartition result of Tomamichel et al. \cite{QAEP}, we 
show that this `subpentagonal' achievable rate region approaches the 
`pentagonal' region of Yard et al. \cite{Yard_MAC} (equal to the 
region demarcated by the dashed line) in the iid limit.
\begin{figure}[hhh]
\centering
\vspace*{-5mm}
\begin{tikzpicture}
\draw [->, line width=1 pt] (1,-2) -- (1,2);
\draw [->, line width=1 pt] (1,-2) -- (5,-2);
\draw [line width=1 pt] (1,0) -- (2,0);
\draw [line width=1 pt] (3,-2) -- (3,-1);
\draw [line width=1 pt,dash pattern=on 5pt off 5pt] (4.2,-2) -- (1,1.2);
\draw [shift={(3.04,0.0)},line width=1 pt]  
plot[domain=3.15:4.65,variable=\t]
({1*cos(\t r)},{1*sin(\t r)});
\draw [shift={(3.1,0.1)},dotted]  
plot[domain=3.15:4.65,variable=\t]
({1*cos(\t r)},{1*sin(\t r)});
\draw (1,2) node[anchor=east] {Bob};
\draw (5,-2) node[anchor=west] {Alice};
\draw (2.25,-0.7) node[anchor= east] {{\footnotesize $n = 1$}};
\draw (2.75,-0.5) node[anchor= west] 
{{\footnotesize $n \rightarrow \infty$}};
\draw (3,-1) node[anchor=west] 
{{\footnotesize $(I^{O(\eps^2)}_{\mathrm{min}}(A\rangle BC), 
		      I^{O(\eps^2)}_{\mathrm{min}}(B\rangle C))$}};
\draw (2,0) node[anchor=west] 
	{{\footnotesize $(I^{O(\eps^2)}_{\mathrm{min}}(A\rangle C),
		      I^{O(\eps^2)}_{\mathrm{min}}(B\rangle AC))$}};
\draw (4.2,-2) node[anchor=north] 
{{\footnotesize $(I^{\sqrt{\epsilon}}_{{\mathrm{min}}}(AB\rangle C), 0)$}};
\draw (1,1.2) node[anchor=east] 
{{\footnotesize $(0, I^{\sqrt{\epsilon}}_{{\mathrm{min}}}(AB\rangle C))$}};
\draw (2.5,-0.77) node[anchor=north] {{\footnotesize $P$}};
\draw (1,1) node[anchor=west] 
{{\footnotesize $x+y= I^{\sqrt{\epsilon}}_{\mathrm{min}}(AB\rangle C)$}};
\end{tikzpicture}
\vspace*{-5mm}
\caption[One-shot achievable rate region for the unassisted QMAC.]{One-shot achievable rate region for the unassisted QMAC 
(for single channel use only), contained
inside the `ideal' pentagonal region demarcated by the dashed line, and 
approaching it in the asymptotic iid limit. 
$O(\log \epsilon)$ additive factors have been ignored in the figure.
The dotted curve shows the situation for some large finite $n$,
intuitively indicating that the region is approaching the dashed 
line when $n \rightarrow \infty$. Note that the above region corresponds to a fixed bipartite input control state. The actual region is a union over all such regions.}
\label{fig:QMAConeshot1}
\end{figure}

\subsection{The QMAC}
In this section, we will use the techniques developed till now to show the existence of encoders and a decoder for entanglement transmission over the QMAC $\mathcal{N}^{A'B'\to C}$ with senders Alice and Bob and receiver Charlie. As usual, we consider the Stinespring dilation of this operator $\mathcal{U}_{\mathcal{N}}^{A'B'\to CE}$. We will first specify the control state
\[
\ket{\sigma}^{A''A'B''B'}\coloneqq \ket{\Omega}^{A''A'}\ket{\Delta}^{B''B'}.
\]
The state $\ket{\Omega}^{A''A'}$ is associated with Alice and $\ket{\Delta}^{B''B'}$ is associated with Bob. As before we consider the action of a splitting scheme on this state. In particular, we will again split Alice into the two senders Alice\sub{0} and Alice\sub{1}, by acting a unitary $U_{\theta}$ on the system $A''$. Thus we will deal with the split control state 
\[
\ket{\sigma(\theta)}^{A''_0A''_1A'B''B'}\coloneqq \ket{\Omega(\theta)}^{A''_0A''_1A'}\ket{\Delta}^{B''B'}.
\]
As before we will often omit the $\theta$ in the notation for the state. The states to be transmitted are denoted as $\ket{\eta}^{A_0C_0R_0},~\ket{\psi}^{A_1C_1R_1}$ for Alice  $\ket{\varphi}^{BDS}$ for Bob, and where $C_0, C_1$ and $D$ represent side information held by the receiver Charlie. The systems $R_0, R_1$ and $S$ represent reference systems which remain untouched by the protocol. \\
\vspace{1mm} \\
We will prove the following lemma:
\begin{proposition}
\label{prop:QMACdecoupling}
Consider the quantum multiple access channel 
$\mathcal{N}^{A'B'\rightarrow C}$. Consider a pure `control state' 
$\ket{\sigma}^{A''B''A'B'} \coloneqq \ket{\Omega}^{A''A'}\ket{\Delta}^{B''B'}$. 
Let 
$\ket{\psi}^{A_1 C_1 R_1} \otimes \ket{\eta}^{A_0 C_0 R_0}$ 
and $\ket{\phi}^{B D S}$ be the states that are to be sent to
Charlie through the channel by Alice and Bob respectively, where
$C_0$, $C_1$, $D$ model the side information about 
the respective messages $A_0$, $A_1$, $B$ that Charlie possesses and
$R_0$, $R_1$, $S$ are reference systems that are untouched by
channel and coding operators.
Let $\mathbb{I}$ denote the identity superoperator.
For $\theta\in [0,1]$, let $\brak{U_{\theta}}^{A''}$ be a splitting scheme. We define 
$
\ket{\sigma(\theta)}^{A''_0 A''_1 A' B'' B'} \coloneqq
U_{\theta}\ket{\Omega}^{A'' A'}\ket{\Delta}^{B''B'} 
$
and
\[
\sigma(\theta)^{A''_0 A''_1 B'' C} \coloneqq
(\mathcal{N}^{A'B' \to C} \otimes \mathbb{I}^{A''_0 A''_1 B''})(
\sigma(\theta)^{A''_0 A''_1 A' B'' B'}
).
\]
Then there exist encoding maps 
$\mathcal{A}^{A_0 A_1 \rightarrow A'}$, 
$\mathcal{B}^{B\rightarrow B'}$ and a decoding map 
$\mathcal{C}^{C C_0 C_1 D \rightarrow A_0 C_0 A_1 C_1 B D}$ 
such that
\[
\begin{array}{l}
\left\|
(\mathcal{C} \otimes \mathbb{I}^{R_0 R_1 S})(
(\mathcal{N} \otimes \mathbb{I}^{C_0 R_0 C_1 R_1 D S})(
\right.
(\mathcal{A} \otimes \mathcal{B} \otimes \mathbb{I}^{C_0 R_0 C_1 R_1 D S})(
(\eta \otimes \psi) \otimes \phi
))) 
\left.
{} - 
\eta \otimes \psi \otimes \phi
\right\|_1
\leq \delta,
\end{array}
\]
whenever 
\begin{align*}
        &H_{\max}^{f(\eps)}(A_1)_{\psi}\leq H_{\min}^{f(\eps)}(A''_1)_{\sigma}+2\log f(\eps), \\
    &H_{\max}^{f(\eps)}(B)_{\phi}\leq H_{\min}^{f(\eps)}(B'')_{\sigma}+2\log f(\eps),
\end{align*}
where $\delta \coloneqq \delta_{\textup{enc}}+\delta_{\textup{dec}}$ and
\begin{align*}
    2\sqrt{2\delta_{\textup{enc}}(3)}+2\sqrt{2\delta_{\textup{enc}}(2)+2\delta_{\textup{enc}}(1)} 
    \coloneqq & \delta_{\textup{enc}}, \\
    2\sqrt{2\delta_{\textup{dec}}(1)}+2\sqrt{2\delta_{\textup{dec}}(2)}+2\sqrt{2\delta_{\textup{dec}}(3)} 
\coloneqq & \delta_{\textup{dec}} ,\\
2\cdot\left(2^{-\frac{1}{2}H_{\min}^{\eps}(A_1|R_1)_{\psi}-\frac{1}{2}H_{\min}^{O(\eps^2)}(A''_1|E)_{\mathcal{U}_{\mathcal{N}}\cdot \sigma}+O(\log k)}+12k\eps\right) 
\coloneqq & 2\delta_{\textup{dec}}(3), \\
2\cdot\left(2^{-\frac{1}{2}H_{\min}^{\eps}(B|S)_{\phi}-\frac{1}{2}H_{\min}^{O(\eps^2)}(B''|A''_1E)_{\mathcal{U}_{\mathcal{N}}\cdot \sigma}+O(\log k)}+12k\eps\right) 
\coloneqq & 2\delta_{\textup{dec}}(2) ,\\
2\cdot\left(2^{-\frac{1}{2}H_{\min}^{\eps}(A_0|R_0)_{\eta}-\frac{1}{2}H_{\min}^{O(\eps^2)}(A''_0|A''_1BE)_{\mathcal{U}_{\mathcal{N}}\cdot \sigma}+O(\log k)}+12k\eps\right)
\coloneqq & 2\delta_{\textup{dec}}(1) ,\\
2\cdot \left(k\cdot 2^{\frac{1}{2}H_{\max}^{f(\eps)}(B)_{\phi}-\frac{1}{2}H_{\min}^{f(\eps)}(B'')_{\Delta}}+12kf(\eps)\right) 
\coloneqq & 2\delta_{\textup{enc}}(3) ,\\
2\cdot\left(k\cdot 2^{\frac{1}{2}H_{\max}^{f(\eps)}(A_1)_{\psi}-\frac{1}{2}H_{\min}^{f(\eps)}(A''_1)_{\Omega}}+12kf(\eps)\right) 
&\coloneqq 2\delta_{\textup{enc}}(2) ,\\
2\cdot\left(2^{\frac{1}{2}H_{\max}^{\eps}(A_0)_{\eta}-\frac{1}{2}H_{\min}^{O(\eps^2)}(A''_0|A''_1)_{\sigma}+O(\log k)}+12k\eps\right)&\coloneqq 2\delta_{\textup{enc}}(1),
\end{align*}
where $k$ is a constant integer and $f(x)=O(x^2)$.
\end{proposition}

\cref{prop:QMACdecoupling} immediately implies the following theorem, by the arguments presented in \cref{corol:achievability}.

\begin{theorem}
\label{thm:QMAConeshot}
Consider the setting of Proposition~\ref{prop:QMACdecoupling}.
Let $Q_A$, $E_A$, $Q_B$, $E_B$ be the number of message qubits and
number of available ebits of Alice and Bob respectively.
Let $\theta, \epsilon \in [0,1]$ and
$\epsilon_0 \coloneqq O(\eps^2)$, where the order hides some multiplicative constant.
Then there exist encoding and decoding maps
such that any message cum ebit rate $4$-tuple satisfying either
the following set
of constraints or the set obtained by interchanging $A''_0$ with $A''_1$,
$Q_A(0)$ with $Q_A(1)$ and $E_A(0)$ with $E_A(1)$
in the right-hand sides of the first two inequalities,
is achievable with error at most $O(\sqrt{\epsilon})$:
\begin{align*}
Q_A 
& =
Q_A(0) + Q_A(1), 
E_A 
=
E_A(0) + E_A(1), \\
Q_A(0)+E_A(0) 
& <  
H_{\min}^{\epsilon_0}(A''_0|A''_1)_{\sigma(\theta)}
+O(\log\epsilon)-O(1), \\
Q_A(1)+E_A(1) 
& <  
H_{\min}^{f(\epsilon)}(A''_1)_{\sigma(\theta)}
+O(\log\epsilon), \\
Q_A(0)-E_A(0)
& < 
I_{\min}^{\epsilon_0}(A''_0 \rangle  C)_{\sigma(\theta)}
+O(\log\epsilon)-O(1), \\
Q_A(1)-E_A(1)
& < 
I_{\min}^{\epsilon_0}(A''_1 \rangle  C A_0'' B'')_{\sigma(\theta)}
+O(\log\epsilon)-O(1), \\
Q_B+E_B 
& < 
H_{\min}^{f(\epsilon)}(B'')_{\sigma(\theta)}
+O(\log\epsilon), \\
Q_B-E_B 
& <  
I_{\min}^{\epsilon_0}(B'' \rangle  C A_0'')_{\sigma(\theta)}
+O(\log\epsilon)-O(1).
\end{align*}
The $O(1)$ in the region above hides the $O(\log k)$ factors from \cref{prop:QMACdecoupling}.
\end{theorem}

\subsubsection{Intermediate States and Almost CPTP Maps for the QMAC}
The development in this section closely follows the layout and logical flow of \cref{sec:almostcptp}. \\ \vspace{2mm} \\

We first define isometric embeddings $W_0^{A_0\to A''_0}, W_1^{A_1\to A''_1}$ and $W_3^{B\to B''}$ which map
\[
\begin{aligned}
&\ket{\eta}^{A''_0C_0R_0}\coloneqq W_0\ket{\eta}^{A_0C_0R_0}, \\
&\ket{\psi}^{A''_1C_1R_1}\coloneqq W_1\ket{\psi}^{A_1C_1R_1} ,\\
&\ket{\varphi}^{B''DS}\coloneqq W_3\ket{\varphi}^{BDS}.
\end{aligned}
\]

\noindent We define an intermediate state as follows:
\begin{definition}{\bf Intermediate State for the QMAC}
We define 
\[
\ket{\omega_{12}}^{A''_0C_1R_1DSA'B'}\coloneqq \sqrt{\abs{B''A''_1}}\op^{A''_1B''\to A''_0A'B'}(\ket{\sigma})(U^{A''_1}\otimes U^{B''})\ket{\psi}^{A''_1C_1R_1}\ket{\phi}^{B''DS}.
\]
\end{definition}
\begin{lemma}{\bf Trace of Intermediate State} Given the conditions \label{lem:qmacIntermediateStateTrace}
\begin{align*}
    &H_{\max}^{f(\eps)}(A_1)_{\psi}\leq H_{\min}^{f(\eps)}(A''_1)_{\sigma}+2\log f(\eps) ,\\
    &H_{\max}^{f(\eps)}(B)_{\phi}\leq H_{\min}^{f(\eps)}(B'')_{\sigma}+2\log f(\eps),
\end{align*}
we have that
\[
\E_{U^{A''_1},U^{B''}}\left[\abs{\Tr[\omega_{12}]-1}\right] \leq 26f(\eps).
\]
\begin{proof}
First, notice that
\begin{align*}
    \ket{\omega_{12}} &= \sqrt{\abs{B''A''_1}}\op^{A''_1B''\to A''_0A'B'}(\ket{\sigma})(U^{A''_1}\otimes U^{B''})\ket{\psi}^{A''_1C_1R_1}\ket{\phi}^{B''DS} \\
    &= \left(\sqrt{\abs{B''}}\op^{B''\to B'}(\ket{\Delta})U^{A''_1}\ket{\phi}^{B''DS}\right)\otimes\left(\sqrt{\abs{A''_1}}\op^{A''_1\to A''_0A'}(\ket{\Omega})U^{A''_1}\ket{\psi}^{A''_1C_1R_1}\right) \\
    &\coloneqq \ket{\omega_1}^{B'DS}\ket{\omega_2}^{A''_0C_1R_1A'}.
\end{align*}
Notice that the state $\omega_2$ is similar to the state $\omega$ in the last section. We essentially repeat the analysis of \cref{lem:TraceIntermediateStateLemma} for $\ket{\omega_1}$ and $\ket{\omega_2}$. Notice that this implies that if the conditions given in the hypothesis of the lemma are satisfied, then
\begin{align*}
    &\E_{U^{B''}}\left[\abs{\Tr[\omega_1]-1}\right] \leq 13f(\eps) ,\\
        &\E_{U^{A''_1}}\left[\abs{\Tr[\omega_2]-1}\right] \leq 13f(\eps).
\end{align*}
Then notice that
\begin{align*}
    \abs{\Tr[\omega_{12}]-1} &= \abs{\Tr[\omega_1]\cdot \Tr[\omega_2]-1} \\
    &\leq \Tr[\omega_1]\cdot\abs{\Tr[\omega_2]-1}+\abs{\Tr[\omega_1]-1} \\
    \intertext{Since $U^{A''_1}$ and $U^{B''}$ are sampled independently, }
    \E_{U^{A''_1},U^{B''}}[\abs{\Tr[\omega_{12}]-1}]&\leq \E_{U^{B''}}\left[\Tr[\omega_1]\right]\cdot \E_{U^{A''_1}}\left[\abs{\Tr[\omega_2]-1}\right]+\E_{U^{B''}}[\abs{\Tr[\omega_1]-1}].
\end{align*}
It is not hard to see that
\[
\E_{U^{B''}}\left[\Tr[\omega_1]\right]=1.
\]
Therefore 
\[
\E_{U^{A''_1},U^{B''}}[\abs{\Tr[\omega_{12}]-1}] \leq 26f(\eps).
\]
This concludes the proof.
\end{proof}

\end{lemma}

\begin{lemma}{\bf Rewriting the Intermediate State for Coding} \label{lem:omega12toomega3}
The intermediate state $\ket{\omega_{12}}^{A''_9C_1R_1DSA'B'}$ can be rewritten as 
\[
\ket{\omega_{12}}^{A''_0C_1R_1DSA'B'}\coloneqq \sqrt{B''}\op^{B''\to A''_0A'B'C_1R_1}(\ket{\omega}_3)U^{B''}\ket{\phi}^{B''DS},
\]
where 
\[
\ket{\omega_3}^{B''A''_0A'B'C_1R_1}\coloneqq \sqrt{\abs{A''_1}}\op^{A''_1\to B''A''_0A'B'}\left(\ket{\sigma}\right)U^{A''_1}\ket{\psi}^{A''_1C_1R_1}.
\]
\begin{proof}
\begin{align*}
    \ket{\omega_{12}}^{A''_0C_1R_1DSA'B'} &=\sqrt{\abs{B''A''_1}}\op^{A''_1B''\to A''_0A'B'}(\ket{\sigma})(U^{A''_1}\otimes U^{B''})\ket{\psi}^{A''_1C_1R_1}\ket{\phi}^{B''DS} \\
    &=\sqrt{\abs{B''A''_1}}\op^{A''_1B''\to C_1R_1DS}\left((U^{A''_1}\otimes U^{B''})\ket{\psi}^{A''_1C_1R_1}\ket{\phi}^{B''DS}\right)\ket{\sigma}^{A''_0A''_1B''A'B'} \\
    &=\left(\sqrt{\abs{B''}}\op^{B''\to DS}\left(U^{B''}\ket{\phi}^{B''DS}\right)\otimes \sqrt{\abs{A''_1}}\op^{A''_1\to C_1R_1}\left(U^{A''_1}\ket{\psi}^{A''_1C_1R_1}\right)\right)\ket{\sigma}\\
    &= \sqrt{\abs{B''}}\op^{B''\to DS}\left(U^{B''}\ket{\phi}^{B''DS}\right)\left(\sqrt{\abs{A''_1}}\op^{A''_1\to C_1R_1}\left(U^{A''_1}\ket{\psi}^{A''_1C_1R_1}\right)\ket{\sigma}\right) \\
    &= \sqrt{\abs{B''}}\op^{B''\to DS}\left(U^{B''}\ket{\phi}^{B''DS}\right)\left(\sqrt{\abs{A''_1}}\op^{A''_1\to B''A''_0A'B'}\left(\ket{\sigma}\right)U^{A''_1}\ket{\psi}^{A''_1C_1R_1}\right)\\
    &= \sqrt{\abs{B''}}\op^{B''\to DS}\left(U^{B''}\ket{\phi}^{B''DS}\right) \ket{\omega_{3}}^{B''A''_0A'B'C_1R_1} \\
    &= \sqrt{\abs{B''}}\op^{B''\to A''_0A'B'C_1R_1}(\ket{\omega_3})U^{B''}\ket{\phi}^{B''DS}
\end{align*}
where e have used the properties of the op operator as proved in \cref{subsec:opoperator}.
\end{proof}

\end{lemma}

\begin{lemma}{\bf Trace of $\ket{\omega_3}$}.
\[
\Tr[\omega_3]=\Tr[\omega_2].
\]
\end{lemma}
\begin{proof}
Recall that, from \cref{lem:qmacIntermediateStateTrace}, 
\[
\ket{\omega_2}^{A''_0C_1R_1A'}= \sqrt{\abs{A''_1}}\op^{A''_1\to A''_0A'}(\ket{\Omega})U^{A''_1}\ket{\psi}^{A''_1C_1R_1}.
\]
Now
\begin{align*}
    \ket{\omega_3}^{B''A''_0A'B'C_1R_1}&= \sqrt{\abs{A''_1}}\op^{A''_1\to B''A''_0A'B'}\left(\ket{\sigma}\right)U^{A''_1}\ket{\psi}^{A''_1C_1R_1} \\
     &=\sqrt{\abs{A''_1}}\op^{A''_1\to B''A''_0A'B'}
     \left(\ket{\Omega}^{A''_0A''_1A'}\ket{\Delta}^{B''B'}\right)U^{A''_1}\ket{\psi}^{A''_1C_1R_1} \\
         &=\ket{\Delta}^{B''B'}\otimes\sqrt{\abs{A''_1}}\op^{A''_1\to A''_0A'}\left(\ket{\Omega}^{A''_0A''_1A'}\right)U^{A''_1}\ket{\psi}^{A''_1C_1R_1} \\
        &= \ket{\Delta}^{B''B'}\ket{\omega_2}^{A''_0C_1R_1A'}.
\end{align*}
Since by definition, $\Tr[\Delta]=1$, we have that
\[
\Tr[\omega_3]=\Tr[\omega_2].
\]
This concludes the proof.
\end{proof}
\begin{lemma}{\bf Approximate DPI with $\ket{\omega_{12}}$} Given the intermediate state 
\[
\ket{\omega_{12}}^{A''_0C_1R_1DSA'B'}\coloneqq \sqrt{\abs{B''A''_1}}\op^{A''_1B''\to A''_0A'B'}(\ket{\sigma})(U^{A''_1}\otimes U^{B''})\ket{\psi}^{A''_1C_1R_1}\ket{\phi}^{B''DS},
\]
and the relations

\begin{align*}
    &H_{\max}^{f(\eps)}(A_1)_{\psi}\leq H_{\min}^{f(\eps)}(A''_1)_{\sigma}+2\log f(\eps), \\
    &H_{\max}^{f(\eps)}(B)_{\phi}\leq H_{\min}^{f(\eps)}(B'')_{\sigma}+2\log f(\eps),
\end{align*}
there exist unitaries $U^{A''_1}$ and $U^{B''}$, with constant probability, such that 
\[
H_{\min}^{\sqrt{O(f(\eps))}}(A''_0|C_1R_1DSE)_{\mathcal{U}_{\mathcal{N}}\cdot \omega_{12}}\geq H_{\min}^{f(\epsilon)}(A''_0|A''_1B''E)_{\mathcal{U}_{\mathcal{N}}\cdot \sigma}-O(1).
\]
\end{lemma}
\begin{proof}
Define the map 
\begin{align*}
       \mathcal{T}^{A''_1B''\rightarrow C_1R_1DS}(\xi)\coloneqq \abs{A''_1B''}\big(\text{op}^{A''_1B''\rightarrow C_1R_1DS}(\ket{\psi}\ket{\varphi})\cdot \xi\big).
\end{align*}
First, recall the following properties of Haar integration
\begin{enumerate}
   \item  $\int U_1^{A}\otimes U_2^{B} \cdot \rho^{AB} dU_1dU_2=\Tr[\rho^{AB}]\pi^{AB}$,
   \item $\int U_1^{A}\otimes U_2^{B}\otimes I^{C} \cdot \rho^{ABC} dU_1dU_2=\pi^{AB}\otimes \rho^C$.
\end{enumerate}
It is now easy to verify that $\mathcal{T}$ is indeed an almost CPTP. The first two properties can be shown to be true using reasoning similar to that used in \cref{almostcptp}. Finally, using property $1$ of double Haar integration above, one can immediately see that 
\begin{align*}
    \int\mathcal{T}(U^{A''_1}\otimes U^{B''}\cdot \xi^{A''_1B''})dU_1dU_2 = \Tr[\xi]\mathcal{T}(\pi^{A''_1B''}).
\end{align*}
Next suppose $\Tilde{\sigma}^{A''_1A''_0B''CE}$ be a state such that $H_{\min}^{f(\epsilon)}(A''_0|A''_1B''E)_{\mathcal{U}_{\mathcal{N}}\cdot\sigma}=H_{\min}(A''_0|A''_1B''E)_{\Tilde{\sigma}}$ where $\norm{\Tilde{\sigma}-\mathcal{U}_{\mathcal{N}}\cdot\sigma}_1\leq 2f(\epsilon)$. Let $\lambda^{A''_1B''E}$ be a positive semidefinite matrix such that
\begin{align*}
&\Tr[\lambda]=2^{-H_{\min}^{f(\epsilon)}(A''_0|A''_1B''E)_{\mathcal{U}_{\mathcal{N}}\cdot\sigma}}, \\
\intertext{ and }
    &\Tilde{\sigma}^{A''_0A''_1B''E} \leq \mathbb{I}^{A''_0}\otimes \lambda^{A''_1B''E}.
\end{align*}
Then, since $\mathcal{T}$ is CP,
\[
\mathcal{T}\left(\left(U^{A''_1}\otimes U^{B''}\right)^{T}\Tilde{\sigma}^{A''_0A''_1B''E}\right) \leq \mathbb{I}^{A''_0}\otimes \mathcal{T}\left(\left(U^{A''_1}\otimes U^{B''}\right)^{T}\lambda^{A''_1B''E}\right).
\]

Then, it holds that 
\begin{align*}
    \Tr[\int\mathcal{T}\Big((U^{A''_1}\otimes U^{B''})^T\cdot \lambda^{A''_1B''E}dU^{A''_1}dU^{B''}] &= \Tr[\mathcal{T}(\pi^{A''_1B''})\otimes\lambda^{E}] \\
    &= 2^{-H_{\min}^{f(\epsilon)}(A''_0|A''_1B''E)_{\mathcal{U}_{\mathcal{N}}\cdot\sigma}} .
\end{align*}

Note that
\begin{align*}
    \mathcal{T}\left(\left( U^{A''_1}\otimes U^{B''}\right)^T\mathcal{U}_{\mathcal{N}}\cdot\ket{\sigma}^{A''_0A''_1B''A'B'}\right)&=\abs{A''_1B''}\left(\op^{A''_1B''\to C_1R_1DS}(\ket{\psi}\ket{\phi})\left( U^{A''_1}\otimes U^{B''}\right)^T\mathcal{U}_{\mathcal{N}}^{A'B'\to CE}\cdot \sigma\right) \\
    &= \abs{A''_1B''}\left(\op^{A''_1B''\to C_1R_1DS}(U^{A''_1}\ket{\psi}\otimes U^{B''}\ket{\phi}) \mathcal{U}_{\mathcal{N}}\cdot \sigma\right) \\
    &= \abs{A''_1B''}\left(\op^{A''_1B''\to A''_0CE}(\mathcal{U}_{\mathcal{N}} \ket{\sigma}) \left(U^{A''_1}\otimes U^{B''}\right)\cdot\left(\psi\otimes \phi\right)\right) \\
    &= \left(\mathcal{U}_{\mathcal{N}}\cdot \omega_{12}\right)^{A''_0C_1R_1DSCE}, \\
    \intertext{which implies that}
\Tr_{C}\mathcal{T}\left(\left( U^{A''_1}\otimes U^{B''}\right)^T\mathcal{U}_{\mathcal{N}}\cdot\ket{\sigma}^{A''_0A''_1B''A'B'}\right)&= \left(\mathcal{U}_{\mathcal{N}}\cdot \omega_{12}\right)^{A''_0C_1R_1DSE}.
\end{align*}

Also, from the arguments used in \cref{almostcptp}, we see that there exist positive matrices $\Delta^+$ and $\Delta^-$ such that
\begin{align*}
    &\E_{U^{A''_1},U^{B''}}\norm{\mathcal{T}\left(\left(U^{A''_1}\otimes U^{B''}\right)^{T}\Tilde{\sigma}\right)-\mathcal{U}_{\mathcal{N}}\cdot\omega_{12}}_1 \\
    =&~\E_{U^{A''_1},U^{B''}}\norm{\mathcal{T}\left(\left(U^{A''_1}\otimes U^{B''}\right)^{T}\Tilde{\sigma}\right)-\mathcal{T}\left(\left(U^{A''_1}\otimes U^{B''}\right)^{T}\mathcal{U}_{\mathcal{N}}\cdot\sigma\right)}_1 \\
    =&~ \E_{U^{A''_1},U^{B''}}\norm{\mathcal{T}\left(\left(U^{A''_1}\otimes U^{B''}\right)^{T}\cdot \left(\Tilde{\sigma}-\mathcal{U}_{\mathcal{N}}\cdot\sigma\right)\right)}_1 \\
    =&~ \E_{U^{A''_1},U^{B''}}\norm{\mathcal{T}\left(\left(U^{A''_1}\otimes U^{B''}\right)^{T}\cdot \left(\Delta^+-\Delta^-\right)\right)}_1 \\
    \leq&~ \E_{U^{A''_1},U^{B''}}\norm{\mathcal{T}\left(\left(U^{A''_1}\otimes U^{B''}\right)^{T}\cdot\Delta^{+}\right)}_1+\E_{U^{A''_1},U^{B''}}\norm{\mathcal{T}\left(\left(U^{A''_1}\otimes U^{B''}\right)^{T}\cdot\Delta^{-}\right)}_1 \\
    =&~ \E_{U^{A''_1},U^{B''}}\Tr\left[\mathcal{T}\left(\left(U^{A''_1}\otimes U^{B''}\right)^{T}\cdot\Delta^{+}\right)\right]+\E_{U^{A''_1},U^{B''}}\Tr\left[\mathcal{T}\left(\left(U^{A''_1}\otimes U^{B''}\right)^{T}\cdot\Delta^{-}\right)\right] \\
    =&~ \left(\Tr[\Delta^+]+\Tr[\Delta^-]\right)\cdot \Tr\left[\mathcal{T}\left(\pi^{A''_1B''}\right)\right] \\
    \leq&~ 4f(\eps).
\end{align*}
Note also that given the entropic conditions in the hypothesis of the lemma, we see via \cref{lem:qmacIntermediateStateTrace} that
\[
\E_{U^{A''_1},U^{B''}}\left[\abs{\Tr[\omega_{12}]-1}\right]\leq 26f(\eps).
\]

Then, via the derandomisation arguments used in \cref{almostcptp}, we see that there exists, with probability at least $1-\frac{3}{k}$, for some constant integer $k$, fixed unitaries $U^{A''_1}$ and $U^{B''}$ such that
\[
\norm{\frac{\mathcal{T}\left(\left(U^{A''_1}\otimes U^{B''}\right)^{T}\Tilde{\sigma}\right)}{\Tr\left(\mathcal{T}\left(\left(U^{A''_1}\otimes U^{B''}\right)^{T}\Tilde{\sigma}\right)\right)}-\frac{\mathcal{U}_{\mathcal{N}}\cdot\omega_{12}}{\Tr[\omega_{12}]}}_1 \leq k\cdot 60 \cdot f(\epsilon),
\]
and 
\[
H_{\min}^{\sqrt{60kf(\epsilon)}}(A''_0|C_1R_1DSE)_{\mathcal{U}_{\mathcal{N}}\cdot \omega_{12}} \geq H_{\min}^{f(\epsilon)}(A''_0|A''_1B''E)_{\mathcal{U}_{\mathcal{N}}\cdot \sigma}-\log k + \log (1-30kf(\epsilon)).
\]
Setting $f(\epsilon)\coloneqq \frac{\epsilon^2}{60k}$ and plugging this into the above inequality we get that
\[
H_{\min}^{\eps}(A''_0|C_1R_1DSE)_{\mathcal{U}_{\mathcal{N}}\cdot \omega_{12}}\geq H_{\min}^{O(\eps^2)}(A''_0|A''_1B''E)_{\mathcal{U}_{\mathcal{N}}\cdot \sigma}-\log k+\log (1-O(\eps^2)).
\]
This concludes the proof.
\end{proof}
We will now prove the main theorem for the QMAC.
\subsubsection{Proof of Proposition \ref{prop:QMACdecoupling}}
\begin{proof}
We will begin with the assumptions that
\begin{align*}
    &H_{\max}^{f(\eps)}(A_1)_{\psi}\leq H_{\min}^{f(\eps)}(A''_1)_{\sigma}+2\log f(\eps), \\
    &H_{\max}^{f(\eps)}(B)_{\phi}\leq H_{\min}^{f(\eps)}(B'')_{\sigma}+2\log f(\eps).
\end{align*}
We will use the intermediate states
\begin{align*}
    &\ket{\omega_{12}}^{A''_0C_1R_1DSA'B'}\coloneqq \sqrt{\abs{B''A''_1}}\op^{A''_1B''\to A''_0A'B'}(\ket{\sigma})(U^{A''_1}\otimes U^{B''})\ket{\psi}^{A''_1C_1R_1}\ket{\phi}^{B''DS}, \\
    &\ket{\omega_3}^{B''A''_0A'B'C_1R_1}\coloneqq \sqrt{\abs{A''_1}}\op^{A''_1\to B''A''_0A'B'}\left(\ket{\sigma}\right)U^{A''_1}\ket{\psi}^{A''_1C_1R_1}\intertext{and} \\
    &\ket{\omega_2}^{A''_0C_1R_1A'}=\sqrt{\abs{A''_1}}\op^{A''_1\to A''_0A'}(\ket{\Omega})U^{A''_1}\ket{\psi}^{A''_1C_1R_1}.
\end{align*}
Recall from \cref{lem:omega12toomega3} we have that
\[
\ket{\omega_{12}}^{A''_0C_1R_1DSA'B'}\coloneqq \sqrt{\abs{B''}}\op^{B''\to A''_0A'B'C_1R_1}(\ket{\omega_3})U^{B''}\ket{\phi}^{B''DS}.
\]
We define the maps
\begin{align*}
    &\mathcal{E}^{A_0A_1\to A'}(\ket{\xi})\coloneqq\sqrt{\abs{A''_0A''_1}}\op^{A''_0A''_1\to A'}(\ket{\Omega})\left(U^{A''_0}\otimes U^{A''_1}\right)\left(W_0^{A_0\to A''_0}\otimes W_1^{A_1\to A''_1}\right)  \ket{\xi} ,\\
    &\mathcal{F}^{B\to B'}(\ket{\zeta})\coloneqq \sqrt{\abs{B''}}\op^{B''\to B'}(\ket{\Delta})U^{B''}~W^{B\to B''}\ket{\zeta}.
\end{align*}
We start with the vector
\begin{align*}
    & \left(\mathcal{E}\otimes \mathcal{F}\right)(\ket{\eta}^{A_0C_0R_0}\ket{\psi}^{A_1C_1R_1}\ket{\phi}^{BDS}) \\
   = &\sqrt{\abs{A''_0B''A''_1}}\op^{A''_0B''A''_1\to A'B'}(\ket{\sigma})U^{A''_0}\otimes U^{B''}\otimes U^{A''_1}\ket{\eta}^{A''_0C_0R_0}\ket{\phi}^{B''DS}\ket{\psi}^{A''_1C_1R_1} \\
    = & \sqrt{\abs{A''_0B''A''_1}}\op^{A''_0B''A''_1\to C_0R_0DSC_1R_1}(U^{A''_0}\otimes U^{B''}\otimes U^{A''_1}\ket{\eta}^{A''_0C_0R_0}\ket{\phi}^{B''DS}\ket{\psi}^{A''_1C_1R_1})\ket{\sigma}^{A''_0B''A''_1A'B'} \\
    = &\left(\sqrt{\abs{A''_0}}\op^{A''_0\to C_0R_0}(U^{A''_0}\ket{\eta}^{A''_0C_0R_0})\otimes \sqrt{\abs{B''A''_1}} \op^{B''A''_1\to DSC_1R_1}(U^{B''}\otimes U^{A''_1}\ket{\phi}^{B''DS}\ket{\psi}^{A''_1C_1R_1})\right)\ket{\sigma} \\
    = & \sqrt{\abs{A''_0}}\op^{A''_0\to C_0 R_0}(U^{A''_0}\ket{\eta}^{A''_0C_0R_0}) \left( \sqrt{\abs{B''A''_1}} \op^{B''A''_1\to DSC_1R_1}(U^{B''}\otimes U^{A''_1}\ket{\phi}^{B''DS}\ket{\psi}^{A''_1C_1R_1}) \ket{\sigma}\right) \\
    = & \sqrt{\abs{A''_0}}\op^{A''_0\to C_0R_0}(U^{A''_0}\ket{\eta}^{A''_0C_0R_0}) \left(\sqrt{\abs{B''A''_1}} \op^{B''A''_1\to A''_0A'B'}(\ket{\sigma}) U^{B''}\otimes U^{A''_1}\ket{\phi}^{B''DS}\ket{\psi}^{A''_1C_1R_1}\right) \\
    = & \sqrt{\abs{A''_0}}\op^{A''_0\to C_0R_0}(U^{A''_0}\ket{\eta}^{A''_0C_0R_0}) \ket{\omega_{12}}^{A''_0DSC_1R_1A'B'} \\
    = &\sqrt{\abs{A''_0}}\op^{A''_0\to DSC_1R_1A'B'}(\ket{\omega_{12}})U^{A''_0}\ket{\eta}^{A''_0C_0R_0}.
\end{align*}
Using similar reasoning one can also see that
\begin{align*}
    &\sqrt{\abs{A''_0A''_1}}\op^{A''_0A''_1\to A'}(\ket{\Omega})U^{A''_0}\otimes U^{A''_1}\ket{\eta}^{A''_0C_0R_0}\ket{\psi}^{A''_1C_1R_1} \\
    =& \sqrt{\abs{A''_0}}\op^{A''_0\to C_1R_1A'}(\ket{\omega_2}^{A''_0C_1R_1A'})U^{A''_0}\ket{\eta}^{A''_0C_0R_0}.
\end{align*}
\vspace{2mm} {\large {\bf The Decoupling Step}} \\
As before, we will first consider the relevant decoupling conditions that will ensure the existence of our encoders and decoders. \\ \vspace{2mm} \\ {\bf The Encoding Equations} \\
\[
\begin{aligned}
&\begin{aligned}
    \E_{U^{A''_0}}\left\lVert\abs{A''_0}\Tr_{A'}\op^{A''_0\to C_1R_1A'}(\ket{\Tilde{\omega}_2})U^{A''_0}\cdot \eta^{A''_0C_0R_0}\right.-&\left.\eta^{R_0C_0}\otimes \Tilde{\omega}_2^{C_1R_1}\right\rVert_1 \\
    \leq & 2^{-\frac{1}{2}H_{\min}^{\eps}(A_0|C_0R_0)_{\eta}-\frac{1}{2}H_{\min}^{\eps}(A''_0|C_1R_1)_{\Tilde{\omega}_2}}+12\eps,
\end{aligned} \\
&\begin{aligned}
\E_{U^{A''_1}}\left\lVert \abs{A''_1}\Tr_{A''_0A'}\op^{A''_1\to A''_0A'}(\ket{\Omega})U^{A''_1}\cdot \psi^{A''_1C_1R_1}\right.-&\left. \psi^{C_1R_1}\right\rVert_1 \\
\leq & 2^{-\frac{1}{2}H_{\min}^{f(\eps)}(A_1|C_1R_1)_{\psi}-\frac{1}{2}H_{\min}^{f(\eps)}(A''_1)_{\Omega}}+12f(\eps),
\end{aligned} \\
&\begin{aligned}
\E_{U^{B''}}\left\lVert \abs{B''}\Tr_{B'}\op^{B''\to B'}(\ket{\Delta})U^{B''}\cdot \phi^{B''DS}\right.-&\left. \phi^{DS}\right\rVert_1 \\
\leq & 2^{-\frac{1}{2}H_{\min}^{f(\eps)}(B|DS)_{\phi}-\frac{1}{2}H_{\min}^{f(\eps)}(B'')_{\Delta}}+12f(\eps).
\end{aligned}
\end{aligned}
\]
\vspace{2mm} \\ {\bf The Decoding Equations} \\
\[
\begin{aligned}
&\begin{aligned}
\E_{U^{A''_0}}\left\lVert \abs{A''_0}\Tr_{C}~\mathcal{U}_{\mathcal{N}}^{A'B'\to CE}~\op^{A''_0\to DSC_1R_1A'B'}(\ket{\Tilde{\omega}_{12}})U^{A''_0}\cdot \right.&\left.\eta^{A''_0R_0}-\eta^{R_0}\otimes \Tilde{\omega}_{12}^{DSC_1R_1E}\right\rVert_1 \\
\leq & 2^{-\frac{1}{2}H_{\min}^{\eps}(A_0|R_0)_{\eta}-\frac{1}{2}H_{\min}^{\eps}(A''_0|DSC_1R_1E)_{\mathcal{U}_{\mathcal{N}}\cdot \Tilde{\omega}_{12}}}+12\eps ,
\end{aligned} \\
&\begin{aligned}
\E_{U^{B''}}\left\lVert \abs{B''}\Tr_{A''_0C}~\mathcal{U}_{\mathcal{N}}^{A'B'\to CE}~\op^{B''\to A''_0C_1R_1A'B'}(\ket{\Tilde{\omega}_3})U^{B''}\cdot \phi^{B''S}\right. &- \left. \phi^{S}\otimes \Tilde{\omega}_3^{C_1R_1E}\right\rVert_1 \\
\leq & 2^{-\frac{1}{2}H_{\min}^{\eps}(B|S)_{\phi}-\frac{1}{2}H_{\min}^{\eps}(B''|C_1R_1E)_{\mathcal{U}_{\mathcal{N}}\cdot \Tilde{\omega}_3}}+12\eps ,
\end{aligned} \\
&\begin{aligned}
\E_{U^{A''_1}}\left\lVert \abs{A''_1} \Tr_{A''_0B''C}~\mathcal{U}_{\mathcal{N}}^{A'B'\to CE}~ \op^{A''_1\to B''A''_0A'B'}(\ket{\sigma})U^{A''_1}\cdot\psi^{A''_1R_1}\right.-&\left. \psi^{R_1}\otimes \sigma^{E}\right\rVert_1 \\
\leq & 2^{-\frac{1}{2}H_{\min}^{\eps}(A_1|R_1)_{\psi}-\frac{1}{2}H_{\min}^{\eps}(A''_1|E)_{\mathcal{U}_{\mathcal{N}}\cdot \sigma}}+12\eps .
\end{aligned}\end{aligned}
\]
\vspace{2mm} \\ {\bf Derandomisation} \\
Using the derandomisation arguments used previously, we see that there exists a constant integer $k$ and constant integers $n_1, n_2, n_2$ such that, there exist fixed unitaries $U^{A''_0}, U^{B''}$ and $U^{A''_1}$ with probability at least $1-\frac{m}{k}$ (where $m$ is the number of events in the intersection and $k$ is chosen to be larger than $m$) such that the encoding and decoding conditions are satisfied along with the three data processing inequalities
\begin{align*}
    &H_{\min}^{\sqrt{n_1\cdot k\cdot f(\eps)}}(A''_0|C_1R_1)_{\Tilde{\omega}_2}\geq H_{\min}^{f(\eps)}(A''_0|A''_1)-\log k +\log(1-n_1\cdot k \cdot f(\eps)) ,\\
    &H_{\min}^{\sqrt{n_2\cdot k \cdot f(\eps)}}(B''|C_1R_1E)_{\mathcal{U}_{\mathcal{N}}\cdot \Tilde{\omega}_3}\geq H_{\min}^{f(\eps)}(B''|A''_1E)_{\mathcal{U}_{\mathcal{N}}\cdot \sigma}-\log k +\log (1-n_2\cdot k\cdot f(\eps)) ,\\
    &H_{\min}^{\sqrt{n_3\cdot k\cdot f(\eps)}}(A''_0|DSC_1R_1E)_{\mathcal{U}_{\mathcal{N}}\cdot \Tilde{\omega}_{12}} \geq H_{\min}^{f(\eps)}(A''_0|A''_1BE)_{\mathcal{U}_{\mathcal{N}}\cdot \sigma}-\log k + \log (1-n_3\cdot k\cdot f(\eps)).
\end{align*}
We choose $f(\eps)\coloneqq \frac{\eps^2}{k\cdot \max\brak{n_1,n_2,n_3}}$. Then using the fact that the smooth min-entropy increases with increasing $\eps$, we see that
\begin{align*}
        &H_{\min}^{\eps}(A''_0|C_1R_1)_{\Tilde{\omega}_2}\geq H_{\min}^{O(\eps^2)}(A''_0|A''_1)_{\sigma}-\log k +\log(1-O(\eps^2)) \\
    &H_{\min}^{\eps}(B''|C_1R_1E)_{\mathcal{U}_{\mathcal{N}}\cdot \Tilde{\omega}_3}\geq H_{\min}^{O(\eps^2)}(B''|A''_1E)_{\mathcal{U}_{\mathcal{N}}\cdot \sigma}-\log k +\log (1-O(\eps^2)) \\
    &H_{\min}^{\eps}(A''_0|DSC_1R_1E)_{\mathcal{U}_{\mathcal{N}}\cdot \Tilde{\omega}_{12}} \geq H_{\min}^{O(\eps^2)}(A''_0|A''_1BE)_{\mathcal{U}_{\mathcal{N}}\cdot \sigma}-\log k + \log (1-O(\eps^2)).
\end{align*}
Then, plugging in the above bounds into the encoding and decoding equations, using the definitions of $\omega_{12}, \omega_{2}$ and $\omega_3$ and using \cref{lem:NormalisationLemma} we see that the derandomised encoding and decoding equations are equivalent to
\\ \vspace{2mm} \\ {\bf Encoding} \\ 
\[
\begin{aligned}
&\begin{aligned}
    \left\lVert \frac{\Tr_{A'}\mathcal{E}^{A_0A_1\to A'}(\eta^{A_0R_0C_0}\otimes \psi^{A_1R_1C_1})}{\Tr[\mathcal{E}(\eta\otimes \psi)]} \right.-&\left.\eta^{R_0C_0}\otimes \Tilde{\omega}_2^{C_1R_1}\right\rVert_1 \\
    \leq & 2\cdot\left(2^{\frac{1}{2}H_{\max}^{\eps}(A_0)_{\eta}-\frac{1}{2}H_{\min}^{O(\eps^2)}(A''_0|A''_1)_{\sigma}+O(\log k)}+12k\eps\right) \\
    &\coloneqq 2\delta_{\textup{enc}}(1),
\end{aligned} \\
&\begin{aligned}
\left\lVert \Tilde{\omega}_2^{C_1R_1} - \psi^{C_1R_1}\right\rVert_1\leq & 2\cdot\left(k\cdot 2^{\frac{1}{2}H_{\max}^{f(\eps)}(A_1)_{\psi}-\frac{1}{2}H_{\min}^{f(\eps)}(A''_1)_{\Omega}}+12kf(\eps)\right) \\
&\coloneqq 2\delta_{\textup{enc}}(2),
\end{aligned} \\
&\begin{aligned}
\left\lVert \frac{\Tr_{B'}\mathcal{F}^{B\to B'}(\phi^{BDS})}{\Tr[\mathcal{F}(\phi)]}\right.-&\left. \phi^{DS}\right\rVert_1 \\
\leq & 2\cdot \left(k\cdot 2^{\frac{1}{2}H_{\max}^{(f\eps)}(B)_{\phi}-\frac{1}{2}H_{\min}^{f(\eps)}(B'')_{\Delta}}+12kf(\eps)\right) \\
\coloneqq & 2\delta_{\textup{enc}}(3).
\end{aligned}
\end{aligned}
\]
\noindent {\bf Decoding} \\
\[
\begin{aligned}
&\begin{aligned}
\left\lVert \frac{\Tr_{CC_0}~\mathcal{U}_{\mathcal{N}}^{A'B'\to CE}~\left(\mathcal{E}\otimes \mathcal{F}\right)\cdot\left(\eta\otimes \psi\otimes \phi\right)}{\Tr\left[\left(\mathcal{E}\otimes \mathcal{F}\right)\cdot\left(\eta\otimes \psi\otimes \phi\right)\right]}\right.& -\left.\eta^{R_0}\otimes \Tilde{\omega}_{12}^{DSC_1R_1E}\right\rVert_1 \\
\leq & 2\cdot\left(2^{-\frac{1}{2}H_{\min}^{\eps}(A_0|R_0)_{\eta}-\frac{1}{2}H_{\min}^{\eps}(A''_0|A''_1BE)_{\mathcal{U}_{\mathcal{N}}\cdot \sigma}+O(\log k)}+12k\eps\right)\\
\coloneqq & 2\delta_{\textup{dec}}(1),
\end{aligned} \\
&\begin{aligned}
\left\lVert \Tr_{A''_0CD}\left[\Tilde{\omega}_{12}^{DSC_1R_1A''_0CE}\right]\right. &- \left. \phi^{S}\otimes \Tilde{\omega}_3^{C_1R_1E}\right\rVert_1 \\
\leq & 2\cdot\left(2^{-\frac{1}{2}H_{\min}^{\eps}(B|S)_{\phi}-\frac{1}{2}H_{\min}^{\eps}(B''|A''_1E)_{\mathcal{U}_{\mathcal{N}}\cdot \sigma}+O(\log k)}+12k\eps\right) \\
\coloneqq & 2\delta_{\textup{dec}}(2),
\end{aligned} \\
&\begin{aligned}
\left\lVert  \Tr_{A''_0B''CC_1}~\left[\Tilde{\omega}_3^{B''A''_0C_1R_1CE}\right]\right.&- \left.\psi^{R_1}\otimes \sigma^{E}\right\rVert_1 \\
\leq & 2\cdot\left(2^{-\frac{1}{2}H_{\min}^{\eps}(A_1|R_1)_{\psi}-\frac{1}{2}H_{\min}^{\eps}(A''_1|E)_{\mathcal{U}_{\mathcal{N}}\cdot \sigma}}+12k\eps\right) \\
\coloneqq & 2\delta_{\textup{dec}}(3).
\end{aligned}\end{aligned}
\]
\vspace{2mm} \\ {\large{\bf Uhlmann's Theorem}} \\
From the first two inequalities in the encoding part, by using the triangle inequality we see that
\[
\norm{\frac{\Tr_{A'}\mathcal{E}^{A_0A_1\to A'}(\eta^{A_0R_0C_0}\otimes \psi^{A_1R_1C_1})}{\Tr[\mathcal{E}(\eta\otimes \psi)]} -\eta^{R_0C_0}\otimes \psi^{R_1C_1}}_1 \leq 2\delta_{\textup{enc}}(1)+2\delta_{\textup{enc}}(2).
\]
The third equation in the encoding part gives
\[
 \norm{\frac{\Tr_{B'}\mathcal{F}^{B\to B'}(\phi^{BDS})}{\Tr[\mathcal{F}(\phi)]}- \phi^{DS}}_1 \leq 2\delta_{\textup{enc}}(3).
\]
Uhlmann's Theorem then implies that there exist isometries $V_{\textup{Alice}}^{A_0A_1\to A'}$ and $V_{\textup{Bob}}^{B\to B'}$ such that
\begin{align*}
    &\norm{\frac{\mathcal{E}^{A_0A_1\to A'}(\eta^{A_0R_0C_0}\otimes \psi^{A_1R_1C_1})}{\Tr[\mathcal{E}(\eta\otimes \psi)]} -V_{\textup{Alice}}^{A_0A_1\to A'}\cdot\left(\eta^{A_0R_0C_0}\otimes \psi^{A_1R_1C_1}\right)}_1 \leq 2\sqrt{2\delta_{\textup{enc}}(1)+2\delta_{\textup{enc}}(2)} ,\\
    & \norm{\frac{\mathcal{F}^{B\to B'}(\phi^{BDS})}{\Tr[\mathcal{F}(\phi)]}- V_{\textup{Bob}}^{B\to B'}\cdot\phi^{BDS}}_1 \leq 2\sqrt{2\delta_{\textup{enc}}(3)}.
\end{align*}
Then
\begin{align*}
    &\norm{\frac{\mathcal{E}\otimes \mathcal{F}(\eta\otimes\psi\otimes \phi)}{\Tr[\mathcal{E}\otimes \mathcal{F}(\eta\otimes\psi\otimes \phi)]}-V_{\textup{Alice}}\otimes V_{\textup{Bob}}\cdot\left(\eta\otimes \psi\otimes \phi\right)}_1 \\
    \leq & \norm{\frac{\mathcal{E}\otimes \mathcal{F}(\eta\otimes\psi\otimes \phi)}{\Tr[\mathcal{E}\otimes \mathcal{F}(\eta\otimes\psi\otimes \phi)]}-\frac{\mathcal{F}(\phi)}{\Tr[\mathcal{F}(\phi)]}\otimes V_{\textup{Alice}}\cdot(\eta\otimes \psi)}_1 \\
    +&\norm{\frac{\mathcal{F}(\phi)}{\Tr[\mathcal{F}(\phi)]}\otimes V_{\textup{Alice}}\cdot(\eta\otimes \psi)-V_{\textup{Alice}}\otimes V_{\textup{Bob}}\cdot(\eta\otimes \psi\otimes \phi)}_1 \\
    =& \norm{\frac{\mathcal{E}(\eta\otimes\psi)}{\Tr[\mathcal{E}(\eta\otimes\psi)]}- V_{\textup{Alice}}\cdot(\eta\otimes \psi)}_1+\norm{\frac{\mathcal{F}(\phi)}{\Tr[\mathcal{F}(\phi)]}- V_{\textup{Bob}}\cdot\phi}_1 \\
    \leq & 2\sqrt{2\delta_{\textup{enc}}(3)}+2\sqrt{2\delta_{\textup{enc}}(1)+2\delta_{\textup{enc}}(1)} \\
    \coloneqq & \delta_{\textup{enc}}.
\end{align*}
Again, from the decoding inequalities, by applying Uhlmann's theorem we see that there exist isometries
\begin{align*}
    &V_{\textup{dec\_1}}^{CC_0\to A''_0CC_0A_0} ,\\
    & V_{\textup{dec\_2}}^{A''_0CD\to A''_0B''CBD}\\
    \intertext{and}
    &V_{\textup{dec\_3}}^{A''_0B''CC_1\to A_1C_1F}
\end{align*}
such that
\[
\begin{aligned}
&\begin{aligned}
    \left\lVert V_{\textup{dec\_1}}^{CC_0\to A''_0CC_0A_0}\cdot\frac{\mathcal{U}_{\mathcal{N}}^{A'B'\to CE}~\left(\mathcal{E}\otimes \mathcal{F}\right)\cdot\left(\eta\otimes \psi\otimes \phi\right)}{\Tr\left[\left(\mathcal{E}\otimes \mathcal{F}\right)\cdot\left(\eta\otimes \psi\otimes \phi\right)\right]} -\eta^{A_0R_0C_0}\otimes\Tilde{\omega}_{12}^{DSC_1R_1A''_0CE} \right\rVert_1& \\ \leq & 2\sqrt{2\delta_{\textup{dec}}(1)} ,
    \end{aligned} \\
    &\begin{aligned}
    \left\lVert V_{\textup{dec\_2}}^{A''_0CD\to A''_0B''CBD}\cdot\Tilde{\omega}_{12}^{DSC_1R_1A''_0CE}- \phi^{BDS}\otimes \Tilde{\omega}_3^{B''A''_0C_1R_1CE}  \right\rVert_1 \leq 2\sqrt{2\delta_{\textup{dec}}(2)} ,
    \end{aligned} \\
    &\begin{aligned}
    \left\lVert  V_{\textup{dec\_3}}^{A''_0B''CC_1\to A_1C_1F}\cdot\Tilde{\omega}_3^{B''A''_0C_1R_1CE}-\psi^{A_1R_1C_1}\otimes \sigma^{EF}\right\rVert_1
\leq 2\sqrt{2\delta_{\textup{dec}}(3)} .
    \end{aligned}
\end{aligned}
\]
Finally, defining
\[
V_{\textup{dec}}^{CC_0DC_1\to A_0C_0BDA_1C_1F}\coloneqq V_{\textup{dec\_3}}\circ V_{\textup{dec\_2}}\circ V_{\textup{dec\_1}} ,
\]
and using the triangle inequality shows that
\[
\begin{aligned}
&\norm{V_{\textup{dec}}\cdot \frac{\mathcal{U}_{\mathcal{N}}^{A'B'\to CE}~\left(\mathcal{E}\otimes \mathcal{F}\right)\cdot\left(\eta\otimes \psi\otimes \phi\right)}{\Tr\left[\left(\mathcal{E}\otimes \mathcal{F}\right)\cdot\left(\eta\otimes \psi\otimes \phi\right)\right]}-\eta^{A_0R_0C_0}\otimes \phi^{BDS}\otimes \psi^{A_1R_1C_1}\otimes \sigma^{EF}}_1 \\
\leq & 2\sqrt{2\delta_{\textup{dec}}(1)}+2\sqrt{2\delta_{\textup{dec}}(2)}+2\sqrt{2\delta_{\textup{dec}}(3)} \\
\coloneqq & \delta_{\textup{dec}}.
\end{aligned}
\]
Tracing out the systems $EF$ from the above inequality gives us the promised decoding map 
\[
\mathcal{C}^{CC_0DC_1\to A_0C_0BDA_1C_1}.
\]
A further triangle inequality with the encoding condition shows that
\[
\norm{\mathcal{C}\circ \mathcal{N}^{A'B'\to C}\circ \left(V_{\textup{Alice}}\otimes V_{\textup{Bob}}\right)\cdot \left(\eta\otimes \psi\otimes \phi\right)-\eta\otimes \psi\otimes \phi}_1 \leq \delta_{\textup{enc}}+\delta_{\textup{dec}}.
\]
\\ {\large {\bf Successive Cancellation}} \\
The decoding algorithm is now clear:
\begin{enumerate}
    \item Alice creates a state close to $\frac{\mathcal{E}(\eta\otimes \psi)}{\Tr[\mathcal{E}(\eta\otimes \psi)]}$ by using the encoding isometry $V_{\textup{Alice}}$.
    \item Bob creates a state close to $\frac{\mathcal{F}(\phi)}{\Tr[\mathcal{F}(\phi)]}$ by using the isometric encoder $V_{\textup{Bob}}$.
    \item They then enter the $A'$ and $B'$ parts of their respective encoded states into the channel.
    \item Charlie first decodes for $\ket{\eta}^{A_0R_0C_0}$ by using the map $V_{\textup{dec\_1}}$ on the systems $CC_0$, and also locally prepares the system $A''_0$ and a copy of $C$.
    \item He then decodes for $\ket{\phi}^{BDS}$ by using the decoder $V_{\textup{dec\_2}}$ on the systems $A''_0CD$ which also locally prepares the systems $A''_0B''$ and also another copy of $C$.
    \item Finally, Charlie decodes for the state $\ket{\psi}^{A_1R_1C_1}$ by using the map $V_{\textup{dec\_3}}$ on the systems $A''_0B''C_1$.
    \item The composition of all three decoding maps and disregarding the environment $E$ and the junk system $F$ gives us the decoder $\mathcal{C}$.
\end{enumerate}
This concludes the proof of the theorem.
\end{proof}
\subsection{The QIC}

In this section, we prove inner bounds for rate-limited entanglement assisted entanglement transmission through the Quantum Interference Channel (QIC) $\mathcal{N}^{A'B'\rightarrow CD}$. We wish for Alice to send EPR pairs to Charlie and for Bob to send EPR pairs to Damru. Note that, for a fixed control state $\ket{\sigma}^{A''A'B''B'}\coloneqq \ket{\Omega}^{A''A'}\ket{\Delta}^{B''B'}$, one can consider this situation as two point-to-point channels, one from Alice to Charlie and one from Bob to Damru. In that case, the achievable region becomes a rectangle of all non negative rate pairs less than $\big(I_{\min}^{\epsilon}(A''\rangle C)_{\sigma},I_{\min}^{\epsilon}(B''\rangle D))_{ \sigma}\big)$ (suppressing the additive log terms).

\begin{figure}[H]
\begin{center}
\begin{tikzpicture}
\draw [->, line width=1 pt] (1,-2) -- (1,2);
\draw [->, line width=1 pt] (1,-2) -- (5,-2);
\draw (1,2) node[anchor=east] {Bob};
\draw (5,-2) node[anchor=west] {Alice};
\draw [line width=1 pt, dotted] (3,-2) -- (3,0);
\draw [line width=1 pt, dotted] (1,0) -- (3,0);
\draw (3,0) node[anchor=west] 
	{{\footnotesize $(I^{O(\eps^2)}_{\mathrm{min}}(A\rangle C),
		  I^{O(\eps^2)}_{\mathrm{min}}(B\rangle D))$}};
\draw [line width=1 pt] (2,-2) -- (2,1);
\draw [line width=1 pt] (1,1) -- (2,1);
\draw (2,1) node[anchor=west] 
	{{\footnotesize $(I^{O(\eps^2)}_{\mathrm{min}}(A_0\rangle C), 
		  I^{O(\eps^2)}_{\mathrm{min}}(B\rangle A_1 D))$}};
\draw [line width=1 pt, dash pattern=on 3pt off 2pt] (4,-2) -- (4,-1);
\draw [line width=1 pt, dash pattern=on 3pt off 2pt] (1,-1) -- (4,-1);
\draw (4,-1) node[anchor=west] 
	{{\footnotesize $(I^{O(\eps^2)}_{\mathrm{min}}(A\rangle B_1 C), 
		  I^{O(\eps^2)}_{\mathrm{min}}(B_0 \rangle  D))$}};
\draw (4.5,1.5) node[anchor= west] {{\footnotesize $n = 1$}};
\end{tikzpicture}
\end{center}
\vspace*{-5mm}
\caption[One-shot achievable rate region for the unassisted QIC.]{One-shot achievable rate region (for single channel use
only) for the unassisted QIC.
The trivial region is shown dotted. Alice can sacrifice her rate
in order to boost Bob's rate with respect to the trivial region, 
as shown by the solid rectangle. The dashed rectangle can be similarly
obtained by Bob sacrificing his rate in order to boost Alice's.
$O(\log \epsilon)$ additive factors have been ignored in the figure. The above region is with respect to a fixed control state. The actual achievable rate region is a union over all such regions.}
\label{fig:QIConeshot}
\end{figure}

The trivial inner bound treats the QIC as two independent unassisted point-to-point
channels from Alice to Charlie and Bob to Damru. 
Rate splitting and successive cancellation can be similarly used to obtain
non-trivial rate regions for the unassisted QIC where one party,
say Alice, sacrifices her rate in order to boost Bob's rate with
respect to the trivial inner bound. The situation is summarised in
Figure~\ref{fig:QIConeshot}.

In order to show that a larger region is achievable, we use splitting schemes and successive cancellation. Essentially, we split Alice into two senders, Alice\sub{0} and Alice\sub{1}, and we require Alice\sub{0}'s input to be decoded by Damru instead of Charlie. This allows Damru to treat Alice\sub{0}'s input as side information while decoding Bob's input, which allows us to boost Bob's rate. Alice's rate to Charlie, however, takes a hit because of this. Using a splitting scheme to do this allows us to adjust the amount of resources that Alice dedicates towards boosting Bob's rate, with the extreme cases $\theta\in\brak{0,1}$ corresponding to situations when either Alice does not help Bob at all (the case of the two point-to-point channels) to when Alice dedicates all her resources to help Bob while her own rates drops to $0$.

The precise statements can be found in \cref{prop:QICnewDecoupling} and \cref{thm:QICOneShot}.

  \begin{proposition}\label{prop:QICnewDecoupling}
  Consider the quantum interference channel 
$\mathcal{N}^{A'B'\rightarrow CD}$. Consider a pure `control state' 
$\ket{\sigma}^{A''B''A'B'} \coloneqq \ket{\Omega}^{A''A'}\ket{\Delta}^{B''B'}$. 
Let 
$\ket{\psi}^{A_1 C_1 R_1}$ and  $\ket{\eta}^{A_0 R_0}$ 
 be the states that are to be sent by Alice to Charlie and Damru respectively and let $\ket{\phi}^{B D_0 S}$ be the state to be sent from Bob to Damru, where $C_1$, $D_0$ model the side information about 
the respective messages  $A_1$, $B$ that Charlie and Damru possess and
$R_0$, $R_1$, $S$ are reference systems that are untouched by
channel and coding operators.
Let $\mathbb{I}$ denote the identity superoperator.
For $\theta\in [0,1]$, let $\brak{U_{\theta}^{A''}}$ be a splitting scheme. We define 
$
\ket{\sigma(\theta)}^{A''_0 A''_1 A' B'' B'} \coloneqq
U_{\theta}\ket{\Omega}^{A'' A'}\ket{\Delta}^{B''B'} 
$
and
\[
\sigma(\theta)^{A''_0 A''_1 B'' CD} \coloneqq
(\mathcal{N}^{A'B' \to CD} \otimes \mathbb{I}^{A''_0 A''_1 B''})(
\sigma(\theta)^{A''_0 A''_1 A' B'' B'}
).
\]
Then there exist encoding maps $\mathcal{A}^{A_0A_1\rightarrow A'}$ and $\mathcal{B}^{B\rightarrow B'}$ and decoding maps $\mathcal{C}^{CC_1\rightarrow A_1C_1}$ and $\mathcal{D}^{DD_0\rightarrow A_0BD_0}$ such that 
  \begin{align*}
      \norm{\big(\mathcal{C}\otimes \mathcal{D}\big)\circ\mathcal{N}\circ\big(\mathcal{A}\otimes \mathcal{B}\big)\cdot\big(\psi\otimes \eta\otimes\varphi\big)-\psi\otimes\eta\otimes \varphi}_1\leq \delta.
  \end{align*}

Here, $\delta=\delta_{\textup{enc}}+\delta_{\textup{dec}}$ where,
\begin{align*}
    &\delta_{\textup{enc}}=    2\sqrt{2\delta_{\textup{enc}}(3)}+2\sqrt{2\delta_{\textup{enc}}(2)+2\delta_{\textup{enc}}(1)} , \\
    &\delta_{\textup{dec}}=4\sqrt{2\delta_{\textup{dec}}(0)}+4\sqrt{2\delta_{\textup{dec}}(1)}+2\sqrt{2\delta_{\textup{dec}}(2)} ,\\
    \intertext{and}
    &\delta_{\textup{enc}}(0)=2\cdot2^{\frac{1}{2}H_{\max}^{\epsilon}(A_0)_{\eta}-\frac{1}{2}H_{\min}^{O(\eps^2)}(A''_0|A''_1)_{\sigma(\theta)}+O(1)}+O(\eps) \\
    &\delta_{\textup{enc}}(1)= 2\cdot2^{\frac{1}{2}H_{\max}^{f(\epsilon)}(A_1)_{\psi}-\frac{1}{2}H_{\min}^{f(\epsilon)}(A''_1)_{\sigma(\theta)}+O(1)}+O(\eps) ,\\
    &\delta_{\textup{enc}}(2)= 
2\cdot 2^{\frac{1}{2}H_{\max}^{f(\epsilon)}(B)_{\varphi}-\frac{1}{2}H_{\min}^{f(\epsilon)}(B'')_{\sigma(\theta)}+O(1)}+O(\eps),\\
    &\delta_{\textup{dec}}(0)=2\cdot2^{-\frac{1}{2}H_{\min}^{\epsilon}(A_0|R_0)_{\eta}-\frac{1}{2}I_{\min}^{O(\eps^2)}(A''_0\rangle D)_{ \sigma(\theta)}+O(1)}+O(\eps), \\
    &\delta_{\textup{dec}}(1)=2\cdot2^{-\frac{1}{2}H_{2}^{\epsilon}(B|S)_{\varphi}-\frac{1}{2}I_{\min}^{O(\eps^2)}(B''\rangle DA''_0)_{ \sigma(\theta)}+O(1)}+O(\eps), \\
    &\delta_{\textup{dec}}(2)=2\cdot 2^{-\frac{1}{2}H_{2}^{\epsilon}(A_1|R_1)_{\psi}-\frac{1}{2}I_{\min}^{O(\eps^2)}(A''_1\rangle C)_{\sigma(\theta)}+O(1)}+O(\eps) ,
\end{align*}
where $f(\eps)=O(\eps^2)$.
\end{proposition}

\begin{remark}
  The $O(1)$ in the bounds in \cref{prop:QICnewDecoupling} hide the $\log k$ terms, as in \cref{prop:QMACdecoupling}, where $k$ is some constant integer.
\end{remark}

  We are now ready to state our main one-shot coding theorem. In this case, we denote by $Q_0$ the number of qubits available to Alice for sending to Damru, to use as side information to boost Bob's rate. The quantities of interest however are $(Q_A,E_A,Q_B,E_B)$ which denote, in order, the number of message qubits and ebits available to Alice, and the analogous quantities for Bob.
  
  \begin{theorem}\label{thm:QICOneShot}
Consider the setting of \cref{prop:QICnewDecoupling}.
Let $Q_A$, $E_A$, $Q_B$, $E_B$ be the number of message qubits and
number of available ebits of Alice and Bob respectively. Additionally, let $Q_0$ denote the number of message qubits available to Alice for transmission to Damru.
Let $\theta, \epsilon \in [0,1]$ and
$\epsilon_0 \coloneqq O(\epsilon^2)$.
Then there exist encoding and decoding maps
such that any message cum ebit rate $4$-tuple satisfying the following inequalities,
is achievable with error at most $O(\sqrt{\epsilon})$ achievable for partial entanglement assisted entanglement transmission
  
  \begin{align*}
      &Q_0 < I_{\min}^{\epsilon_0}(A''_0\rangle D)_{ \sigma(\theta)}+O(\log\epsilon)-O(1) ,\\
      &Q_0 < H_{\min}^{\epsilon_0}(A''_0|A''_1)_{\sigma(\theta)}+O(\log\epsilon)-O(1) ,\\
      &Q_A+E_A < H_{\min}^{f(\epsilon)}(A''_1)_{\sigma(\theta)}+O(\log\epsilon)\\
      &Q_A-E_A < I_{\min}^{\epsilon_0}(A''_1\rangle C)_{\sigma(\theta)}+O(\log\epsilon)-O(1) ,\\
      & Q_B+E_B < H_{\min}^{f(\epsilon)}(B'')_{\sigma(\theta)} +O(\log\epsilon) ,\\
      &Q_B-E_B < I_{\min}^{\epsilon_0}(B''\rangle A''_0D)_{\sigma(\theta)}+O(\log\epsilon)-O(1).
  \end{align*}
  \end{theorem}
  
  The proof of \cref{thm:QICOneShot} follows from \cref{prop:QICnewDecoupling} using the arguments in \cref{chap:RateSplit1}. We present the proof of \cref{prop:QICnewDecoupling} below.
  
\begin{proof}
As mentioned before, the idea is for Alice to use some part of her input to boost Bob's rate to Damru. to do this we split Alice into Alice\sub{0} and Alice\sub{1}. We can then treat the interference channel as a QMAC from Alice\sub{0} and Bob to Damru, and as a point-to-point quantum channel from Alice\sub{1} to Charlie. We will use the techniques used to prove \cref{prop:QMACdecoupling} to derive achievable rates for entanglement transmission from Bob to Charlie. Note that in this case, we assume that Alice\sub{0} shares \emph{no} pre-shared entanglement with Damru. The inner bound for entanglement transmission from Alice\sub{1} to Charlie can be derived by considering the coding scheme for entanglement transmission over a point-to-point channel. 

Note that the above analysis will give us two separate $1$-norm expressions, one for the QMAC among Alice\sub{0}, Bob and Damru and the other for the point-to-point channel from Alice\sub{1} to Charlie. To combine these two expressions we will need the following fact, whose proof can be found in the appendix. This fact appears in \cite[Lemma 5.1]{Dupuis_thesis}: \\

\begin{fact}\label{fact:DupuisBroadcastFact}
Given density operators $\rho^{ABC},\sigma^A,\omega^{BC}, \tau^{AB},\eta^C$ such that
\begin{align*}
    &\norm{\rho^{ABC}-\sigma^A\otimes \omega^{BC}}_1\leq \eps_1 ,\\
    &\norm{\rho^{ABC}-\tau^{AB}\otimes \eta^C}_1 \leq \eps_2,
\end{align*}
then
\[
\norm{\rho^{ABC}-\sigma^A\otimes \tau^B\otimes \eta^C}_1 \leq 2\eps_1+\eps_2.
\]
\end{fact}
\vspace{2mm} 
 We denote by $\mathcal{U}_{\mathcal{N}}^{A'B'\to CDE}$ the Stinespring dilation of the interference channel $\mathcal{N}^{A'B'\to CD}$. We define our encoding maps $\mathcal{E}^{A_0A_1\to A'}$ and $\mathcal{F}^{B\to B'}$ as in the proof of \cref{prop:QMACdecoupling}.
First, we repeat the decoding protocol for the QMAC, where Damru decodes Alice\sub{0} first and then Bob. The intermediate states used for this part of the protocol are as follows:
\begin{align*}
    &\ket{\omega_{12}}^{A''_0A'B'C_1R_1D_0S}\coloneqq\sqrt{\abs{B''A''_1}} \op^{A''_1B''\to A''_0A'B'}(\ket{\sigma})\left(U^{A''_1}\otimes U^{B''}\right)W_1^{A_1\to A''_1}\ket{\psi}^{A_1R_1C_1}W_2^{B\to B''}\ket{\varphi}^{B''_1D_0S} ,\\
    &\ket{\omega_3}^{B''A''_0A'B'C_1R_1}\coloneqq \sqrt{A''_1}\op^{A''_1\to B''A''_0A'B'}(\ket{\sigma})U^{A''_1}W_1^{A_1\to A''_1}\ket{\psi}^{A_1R_1C_1},
\end{align*}
where $W_1$ and $W_2$ are isometric embeddings, as before. Damru decodes Alice\sub{0}'s message and then Bob's. Damru is not required to decode Alice\sub{1}'s message. In fact, for the protocol to work, we will treat Damru as part of the environment so that Charlie can decode Alice\sub{1}'s message, effectively making it impossible for Damru to decode Alice\sub{1}'s message. We can then show the existence of a decoding isometry
\[
V^{DD_0\to A''_0B''DD_0A_0B}_{\textsc{Bob}}
\]
 such that
\[
\begin{aligned}
&\norm{V_{\textsc{Bob}}^{DD_0\to A''_0B''DD_0A_0B}\frac{\mathcal{U}_{\mathcal{N}}^{A'B'\to CDE}(\mathcal{E}\otimes \mathcal{F})\cdot (\eta\otimes \phi\otimes \psi))}{\Tr[(\mathcal{E}\otimes \mathcal{F})\cdot (\eta\otimes \phi\otimes \psi)]}-\eta^{A_0R_0}\otimes \phi^{BD_0S}\otimes \Tilde{\omega}_3^{B''A''_0C_1R_1DCE}}_1 \\
\leq & 2\sqrt{2\delta_{\textup{dec}}(0)}+2\sqrt{2\delta_{\textup{dec}}(1)}.
\end{aligned}
\]
We now consider the channel from Alice\sub{1} to Charlie. We will need the following intermediate state:
\[
\ket{\omega_4}^{A''_1A'B'R_0D_0S}\coloneqq \sqrt{\abs{A''_0B''}}\op^{A''_0B''\to A''_1A'B'}(\ket{\sigma})\left(U^{A''_0}\otimes U^{B''}\right)W^{A_0\to A''_0}_0 \ket{\eta}^{A_0R_0}W_2^{B\to B''}\ket{\varphi}^{B''_1D_0S}.
\]
We can then show the existence of an isometric decoder
\[
V_{\textsc{Alice}}^{CC_1\to CC_1A''_1A_1}
\]
such that
\[
\begin{aligned}
&\norm{V_{\textsc{Alice}}^{CC_1\to CC_1A''_1A_1}\frac{\mathcal{U}_{\mathcal{N}}^{A'B'\to CDE}(\mathcal{E}\otimes \mathcal{F})\cdot (\eta\otimes \phi\otimes \psi))}{\Tr[(\mathcal{E}\otimes \mathcal{F})\cdot (\eta\otimes \phi\otimes \psi)]}-\psi^{A_1C_1R_1}\otimes \Tilde{\omega}_{4}^{DD_0ESR_0A''_1C}}_1 \\
\leq & 2\sqrt{2\delta(2)}.
\end{aligned}
\]
Next, through some standard algebraic manipulation, we see that the two inequalities above are equivalent to
\[
\begin{aligned}
&\norm{V_{\textsc{Alice}}\otimes V_{\textsc{Bob}}\frac{\mathcal{U}_{\mathcal{N}}^{A'B'\to CDE}(\mathcal{E}\otimes \mathcal{F})\cdot (\eta\otimes \phi\otimes \psi))}{\Tr[(\mathcal{E}\otimes \mathcal{F})\cdot (\eta\otimes \phi\otimes \psi)]}-\eta^{A_0R_0}\otimes \phi^{BD_0S}\otimes \zeta_1^{B''A''_0A''_1A_1CC_1R_1DE}}_1 \\
\leq & 2\sqrt{2\delta_{\textup{dec}}(0)}+2\sqrt{2\delta_{\textup{dec}}(1)} ,\\
&\norm{V_{\textsc{Bob}}\otimes V_{\textsc{Alice}}\frac{\mathcal{U}_{\mathcal{N}}^{A'B'\to CDE}(\mathcal{E}\otimes \mathcal{F})\cdot (\eta\otimes \phi\otimes \psi))}{\Tr[(\mathcal{E}\otimes \mathcal{F})\cdot (\eta\otimes \phi\otimes \psi)]}-\psi^{A_1C_1R_1}\otimes \zeta_2^{DD_0A''_0B''A_0BESR_0A''_1C}}_1 \\
\leq & 2\sqrt{2\delta(2)},
\end{aligned}
\]
where 
\begin{align*}
    &\ket{\zeta_1}^{B''A''_0A''_1A_1CC_1R_1DE}\coloneqq V_{\textsc{Alice}}\ket{\Tilde{\omega}_3}^{B''A''_0C_1R_1DCE} ,\\
    &\ket{\zeta_2}^{DD_0A''_0B''A_0BESR_0A''_1C}\coloneqq V_{\textsc{Bob}}\ket{\Tilde{\omega}_4}^{A''_1DCER_0D_0S}.
\end{align*}
We can now use \cref{fact:DupuisBroadcastFact} to conclude that:
\[
\begin{aligned}
&\norm{V_{\textsc{Alice}}\otimes V_{\textsc{Bob}}\frac{\mathcal{U}_{\mathcal{N}}^{A'B'\to CDE}(\mathcal{E}\otimes \mathcal{F})\cdot (\eta\otimes \phi\otimes \psi))}{\Tr[(\mathcal{E}\otimes \mathcal{F})\cdot (\eta\otimes \phi\otimes \psi)]}-\eta^{A_0R_0}\otimes \phi^{BD_0S}\otimes \zeta_2^{A''_0B''A''_1DCE}\otimes \psi^{A_1R_1C_1}}_1 \\
\leq & 4\sqrt{2\delta_{\textup{dec}}(0)}+4\sqrt{2\delta_{\textup{dec}}(1)}+2\sqrt{2\delta_{\textup{dec}}(2)}.
\end{aligned}
\]
The rest of the proof is identical to the analysis of the encoding error in the proof of \cref{prop:QMACdecoupling}. This concludes the proof.
\end{proof}  
\section{Asymptotic IID Analysis}\label{sec:IID}

In  this section we present the asymptotic iid versions of the one-shot achievability results presented in the previous section. Based on the discussion so far, we have seen that we can achieve the following rate point for Alice\sub{0}, Bob and Alice\sub{1}:
\[
\left(H_{\min}^{\eps}(A''_0|BA_1''E), H_{\min}^{\eps}(B''|A''_1E), H_{\min}^{\eps}(A''_1|E)) \right)
\]
 for all values of the parameter $\theta$.  It is tempting to conclude using the Quantum Asymptotic Equipartition Property \cite{QAEP} to the above rate point and conclude the achievability of the following rate point in the asymptotic iid setting:
 \[
 \left(H(A''_0|BA_1''E), H(B''|A''_1E), H(A''_1|E)), \right).
 \]
 
 Things are not so simple however, due to the fact that after rate splitting, one of the terms in this rate point could be negative. For example, if Alice\sub{0}'s rate is negative, then the protocol no longer works for decoding Bob and Alice\sub{1}. 
 
 First, note that we only have to worry about Alice\sub{0}'s being negative. This is because Bob's rate could never be negative due to the following data processing inequality:
 \[
 H_{\min}^{\eps}(B''|A''E)=H_{\min}^{\eps}(B''|A''_0A''_1E) \leq H_{\min}^{\eps}(B''|A''_1E) \leq H_{\min}^{\eps}(B''|E).
 \]
This also implies that Alice\sub{0} and Alice\sub{1}'s combined rate cannot be negative, since the sum rate of Alice and Bob is invariant. Thus, if Alice\sub{1}'s rate is negative, that implies that Alice\sub{0}'s rate is \emph{more} than the combined rate given of Alice\sub{0} and Alice\sub{1}. This is a good situation since we could then simply perform the protocol for only Alice\sub{0} and Bob.

Thus, the difficulty lies in the case when Alice\sub{0}'s rate is negative. We will show that we can achieve the desired rate region in the asymptotic iid limit, with a small amount of pre-shared entanglement between Alice\sub{0} and Charlie. These pre-shared EPR pairs will be used catalytically with the added advantage that the rate at which we require these pre-shared EPR pairs go to $0$ in the asymptotic iid limit. 

We will divide the $n$ channels into $\sqrt{n}$ blocks, where each block is of size $\sqrt{n}$. To avoid cumbersome notation, we use the following convention:
\[
H_{\min}^{\eps}(A|B)_{\rho}(n)\coloneqq H_{\min}^{\eps}(A^{\otimes n}|B^{\otimes n})_{{\rho^{AB}}^{\otimes n}}.
\]

We will consider two situations:
\begin{enumerate}
    \item $H_{\min}^{\eps}(A_0|A_1)(\sqrt{n})\geq 0$,
    \item $H_{\min}^{\eps}(A_0|A_1)(\sqrt{n})<0$.
\end{enumerate}
\vspace{2mm}  {\bf Case I: $H_{\min}^{\eps}(A_0|A_1)(\sqrt{n})\geq 0$.} \\ \vspace{2mm} \\
For each block of size $\sqrt{n}$ set:
\begin{align*}
    Q_{A_0^{\sqrt{n}}} =& 0 ,\\
    E_{A_0^{\sqrt{n}}} =& \abs{H_{\min}^{\eps}(A''_0|A''_1BE)(\sqrt{n})}.
\end{align*}
Since $H_{\min}^{\eps}(A_0|A_1)(\sqrt{n})$ is positive, this implies that there exists an isometric encoder for Alice and the protocol can start. Note that at the end of the protocol, Alice\sub{1} shares a maximally entangled state with Charlie of rank
\[
2^{H_{\min}^{\eps}(A''_1|E)(\sqrt{n})}.
\]
 Alice can now keep aside $2^{\abs{H(A''_0|A''_1BE)(\sqrt{n})}}$ EPR pairs and use them as the seed pre-shared EPR states for the next block of $\sqrt{n}$ channels. Repeating this argument for each block, we see that Alice's rate for entanglement transmission is
\[
\frac{1}{n}\sqrt{n} \left(H_{\min}^{\eps}(A''_1|E)(\sqrt{n})+H(A''_0|A''_1BE)(\sqrt{n})) \right).
\]
In the asymptotic iid limit, the above quantity is equal to
\[
H(A''_1|E)+H(A''_0|A''_1B''E),
\]
which is Alice's desired rate. Note that we only needed to use $2^{H_{\min}^{\eps}(A''_1|E)(\sqrt{n})}$ EPR pairs for the very first block, and thereafter these EPR pairs were regenerated by the protocol. This implies that the rate of seed EPR pairs is given by
\[
\frac{1}{n} H_{\min}^{\eps}(A''_1|E)(\sqrt{n}),
\]
which is $0$ in the asymptotic iid limit. Thus in this case, we can prove the following theorem:
\begin{theorem}\label{iid}
   Given a quantum multiple access channel $\mathcal{N}^{A'B'\rightarrow C}$ all rate points in the closure of the following region are achievable for partial entanglement assisted entanglement generation:
   \begin{align*}
       \bigcup\limits_{k=1}^{\infty}\frac{1}{k}\mathcal{Q}(\mathcal{N}^{\otimes k}),
   \end{align*}
   where $\mathcal{Q}(\mathcal{N}^{\otimes k})$ is the set of non negative rate tuples $(Q_A,E_A,Q_B,E_B)$ in the set
   \begin{align*}
   Q_A+E_A &< H(A^{''k})_{\sigma_k}    ,\\
       Q_A-E_A &< I(A^{''k}\rangle  {B}^{''k}C^k)_{\mathcal{U}_{\mathcal{N}}\cdot\sigma_k} ,\\
       Q_B+E_B &< H(B^{''k})_{\sigma_k} ,\\
       Q_B-E_B &< I({B}^{''k}\rangle  A^{''k}C^k)_{\mathcal{U}_{\mathcal{N}}\cdot\sigma_k} ,\\
       Q_A-E_A+Q_B-E_B &< I(A^{''k}{B}^{''k}\rangle  C^k)_{\mathcal{U}_{\mathcal{N}}\cdot\sigma_k},
   \end{align*}
   where $\ket{\sigma_k}^{A^{''k}{B}^{''k}A^{'k}B^{'k}}\coloneqq   \ket{\Omega}^{A^{''k}A^{'k}} \ket{\Delta}^{{B}^{''k}B^{'k}}$.
   \end{theorem}
 \vspace{2mm}  {\bf Case II: $H_{\min}^{\eps}(A_0|A_1)(\sqrt{n})< 0$.} \\ \vspace{2mm} \\
 In this case we cannot prove the existence of an isometric encoder for Alice. We get around this by using the state merging protocol. To be precise, for the value for $\theta$ for which Case II occurs, Alice and Bob simply send the bipartite pure state $\left(\ket{\Omega}^{A
''_0A''_1A'}\ket{\Delta}^{B''B'}\right)^{\otimes n}$ through $n$ copies of the channel. This results in the following state:
\[
\left(\ket{\sigma}^{A''_0A''_1B''CE}\right)^{\otimes n}.
\]
We divide this state into $\sqrt{n}$ blocks, where each block corresponds to the state 
\[
\left(\ket{\sigma}^{A''_0A''_1B''CE}\right)^{\otimes \sqrt{n}}.
\]
The parties then do a multi-party state merging protocol for this state \cite{dutil}, with parties Alice\sub{0}, Bob and Alice\sub{1}. Since the expression $H(A''_0|A''_1BE)(\sqrt{n})$ is assumed to be negative, Alice\sub{0} must use $2^{H(A''_0|A''_1BE)(\sqrt{n})}$ pre-shared EPR pairs with Charlie to merge her share of the state. The EPR pairs are regenerated when the protocol ends by Alice\sub{1} merging her state with Charlie. Then, the same arguments as in Case I show that, with a vanishing amount of pre-shared seed, the following rate point is achievable for \emph{unassisted} state merging, in the asymptotic iid limit
\begin{align*}
    Q_{A} <& H(A''_1|E)+H(A''_0|A''_1B''E) ,\\
    Q_{B} <& H(B''|A''_1E).
\end{align*}
It is known that entanglement generation and entanglement transmission are equivalent \cite{Tema}. This implies that there exists an \emph{unassisted} entanglement transmission protocol which also achieves the above rates. Finally, one can obtain the rates for the case when rate-limited entanglement assistance is available by simply time sharing between the unassisted protocol above and the completely assisted protocol of Bennet et al. \cite{EntanglementAssistedBennett}. This implies that \cref{iid} is true for all cases.

The arguments above can now be used to prove a similar theorem for the QIC as well:
\begin{theorem}
  Given a quantum interference channel $\mathcal{N}^{\otimes k}$, the control state $\ket{\sigma}^{A''A'B''B'}$ the following regularised rate region is achievable for partial entanglement assisted entanglement transmission:
  \begin{align*}
             \bigcup\limits_{k=1}^{\infty}\frac{1}{k}\mathcal{Q}(\mathcal{N}^{\otimes k}).
  \end{align*}
  
  For each $k\in \mathbb{N}$, 
  \begin{align*}
\mathcal{Q}(\mathcal{N}^{\otimes k})= \bigcup \mathcal{A}_{\theta}^k{\large \bigcup} ~\bigcup \mathcal{B}_{\theta}^k,
  \end{align*}
  where, for a fixed $\theta\in [0,1]$, $\mathcal{A}_{\theta}^k$ is the set of all non-negative tuples $(Q_A,E_A,Q_B,E_B)$ such that
  \begin{align*}
      &Q_A+E_A < H(A^{''k}_1)_{\sigma_k(\theta)},\\
      &Q_A-E_A < I(A^{''k}_1\rangle  C^k_1)_{\mathcal{U}_{\mathcal{N}}\cdot\sigma_k(\theta)},\\
      & Q_B+E_B < H(B^{''k})_{\sigma_k(\theta)},\\
      &Q_B-E_B < I(B^{''k}\rangle  A^{''k}_0C^k_2)_{\mathcal{U}_{\mathcal{N}}\cdot\sigma_k(\theta)},
      \end{align*}
      where $\ket{\sigma_k}^{A^{''k}{B}^{''k}A^{'k}B^{'k}}\coloneqq   \ket{\Omega}^{A^{''k}A^{'k}} \ket{\Delta}^{{B}^{''k}B^{'k}}$ and $\ket{\sigma_k}\coloneqq U^{A^{''k}\rightarrow A^{''k}_0A^{''k}_1}\ket{\sigma_k}$. We assume that $\brak{U_\theta}$ is a splitting scheme. Analogously, $\mathcal{B}_{\theta}^k$ is the set of those points which are obtained when the splitting isometry acts on the system $B^{''k}$.
      \end{theorem}
      
      \section{Conclusion}
      
      In this paper we use the technique of quantum rate splitting and successive cancellation decoding of entanglement transmission codes to design entanglement transmission codes for the QMAC and QIC. We recover a non-trivial rate region for the QMAC in the one-shot setting. Suitable adaptations of our techniques also achieve the ideal pentagonal
      rate region in the asymptotic iid setting.
      
      For the QIC, we show the existence of a non-trivial rate region both in the asymptotic iid and one-shot setting, which is larger than the region one would obtain by considering the QIC as two point-to-point channels.

\section{Acknowledgements}
    The work done in this paper was completed while SC and AN were graduate students at the School of Technology and Computer Science, TIFR, Mumbai. We would like to acknowledge the support of the
Department of Atomic Energy, Government of India,
under project no. RTI4001.

\bibliography{ref_collated}
\bibliographystyle{unsrt}

\appendix
\section{Appendix}

\begin{lemma}\label{contomega}
Given $\theta,\theta'\in [0,1]$ such that $\abs{\theta-\theta'}\leq \delta$, we have that 
\[ 
P\big(\Omega'(\theta)^{A''_0A''_1BE},\Omega'(\theta')^{A''_0A''_1BE}\big)\leq O(\sqrt{\delta}).
\]
\end{lemma}
   \begin{proof}[Proof of \cref{contomega}]
For the course of the proof we will neglect to mention the registers in the superscript to ease the notation, unless necessary. Since both $\Omega'(\theta)$ and $\Omega'(\theta')$ are pure, we will use the identity :
\[
P\big(\Omega'(\theta),\Omega'(\theta')\big)=\sqrt{1-\abs{\braket{\Omega'(\theta)|\Omega'(\theta')}}^2}.
\]
Recall that since
\[
\ket{\Omega'}^{A''_0A''_0BE}(\theta)=U^{A'}_{\mathcal{N}}\circ U^{A''_0A''_1A'}_{f} \Big(\sum\limits_{u\in\mathcal{A}}\sqrt{P^{\theta}_{U}(u)}\ket{u}^{A''_0}\Big)\otimes \Big( \sum\limits_{v\in\mathcal{A}}\sqrt{P^{\theta}_{V}(v)}\ket{v}^{A''_0}\Big)\ket{0}^{A'}
\]
 and similarly for $\ket{\Omega'}(\theta')$, 
 \[
 \braket{\Omega'(\theta)|\Omega'(\theta')}=F(P^{\theta}_U,P^{\theta'}_U)F(P^{\theta}_V,P^{\theta'}_V).
 \]
 It is thus sufficient to show that the distributions $P^{\theta}_U$ and $P_V^{\theta}$ are close to $P^{\theta'}_U$ and $P^{\theta'}_V$ respectively. Then, recalling the explicit form of $P_U^{\theta}$ observe that :
 \begin{align*}
     \norm{P^{\theta'}_U-P^{\theta}_U}_1 
     &=\abs{(1-\theta+\theta P_A(0))-(1-\theta'+\theta' P_A(0))}+\sum\limits_{i\neq 0}~\abs{\theta P_A(i)-\theta' P_A(i)} \\
     &\leq \abs{\theta-\theta'}+\abs{\theta-\theta'}\sum\limits_{i\in \mathcal{A}}P_A(i) \\
                                         &\leq 2\delta.
 \end{align*}
 Next, observe that , for any $i\in \mathcal{A}$
 \begin{align*}
     P^{\theta}_{V}(i) &=\frac{F_A(i)}{F^{\theta}_U(i)}-\frac{F_A(i-1)}{F^{\theta}_U(i-1)}.
 \end{align*}
 It holds that 
 \begin{align*}
     F_A(i)F^{\theta}_U(i-1)-F_A(i-1)F^{\theta}_U(i) &=(P_A(i)+F_A(i-1))F^{\theta}_U(i-1)-F_A(i-1)(\theta F_A(i)+1-\theta) \\
     &=(P_A(i)+F_A(i-1))F^{\theta}_U(i-1)-F_A(i-1)(F^{\theta}_U(i-1)+\theta P_A(i)) \\
     &=P_A(i)(\theta F_A(i-1)+1-\theta)-\theta F_A(i-1)P_A(i) \\
     &=(1-\theta)P_A(i).
 \end{align*}
Denote $F_U^{\theta}(i)F^{\theta}_U(i-1)\coloneqq g(\theta)$. Then,
\begin{align*}
&\abs{g(\theta)-g(\theta')} \\
    &=\abs{F_U^{\theta}(i)F^{\theta}_U(i-1)-F^{\theta'}_U(i)F^{\theta'}_U(i-1)} \\
    &\leq \abs{F_U^{\theta}(i)F^{\theta}_U(i-1)-F_U^{\theta'}(i)F^{\theta}_U(i-1)}+\abs{F_U^{\theta'}(i)F^{\theta}_U(i-1)-F^{\theta'}_U(i)F^{\theta'}_U(i-1)} \\
    &\leq 4\delta.
\end{align*}
 Let $p^*=\min\limits_{i\in \mathcal{A}}P_A(i)$. Then, 
 \begin{align*}
     g(\theta)&\geq (1-\theta+\theta p^*)^2 \\
     &\geq {p^*}^2
 \end{align*}
\begin{align*}
    \abs{P_V^{\theta}(i)-P_V^{\theta'}(i)} &= P_A(i)~\Big|\frac{1-\theta}{g(\theta)}-\frac{1-\theta'}{g(\theta')}\Big| \\
    &=\frac{P_A(i)}{\abs{g(\theta)g(\theta')}}\cdot \abs{(1-\theta)g(\theta')-(1-\theta')g(\theta)} \\
    &\overset{(a)}{\leq} \frac{P_A(i)}{{p^*}^4}\cdot c \cdot \delta,
\end{align*}
where $c$ is some constant and we have used the triangle inequality and the lower bound for $g(\theta)$ in $(a)$.

 Then, the above bound implies that:
\[
\norm{P^{\theta}_V-P^{\theta'}_V}_1\leq O(\delta).
\]
Using the property that $F(P,Q)\geq 1-\norm{P-Q}_1$ for any two distributions $P$ and $Q$, we see that 
\[
\abs{\braket{\Omega'(\theta)|\Omega'(\theta')}}^2\geq 1-O(\delta).
\]
This concludes the proof.
\end{proof}

\begin{lemma}{\bf [Normalisation Lemma]}\label{lem:NormalisationLemma}
Given a state $\rho$ and any positive matrix $\sigma$, not necessarily of trace $1$, given
\[
\norm{\sigma-\rho}_1 \leq \delta,
\]
then
\[
\norm{\frac{\sigma}{\Tr[\sigma]}-\rho}_1\leq 2\delta.
\]
\end{lemma}
\begin{proof}
First note that for any positive matrix $\Omega$, 
\[
\norm{\sigma}_1=\Tr[\sigma].
\]
Then
\begin{align*}
    \norm{\frac{\sigma}{\Tr[\sigma]}-\rho}_1 &\leq \norm{\frac{\sigma}{\Tr[\sigma]}-\sigma}_1+\norm{\sigma-\rho}_1 \\
    &= \abs{\frac{1}{\Tr[\sigma]}-1}\cdot \norm{\sigma}_1+\norm{\sigma-\rho}_1\\
    &=\abs{\Tr[\sigma]-1}+\norm{\sigma-\rho}_1.
\end{align*}
Now, by the given condition,
\begin{align*}
    &\norm{\sigma-\rho}_1 \leq \delta \\
    \implies &\abs{\Tr[\sigma]-1}\leq \delta
\end{align*}
by the monotonicity of $1$-norm under trace. Thus, we see that
\[
\norm{\frac{\sigma}{\Tr[\sigma]}-\rho}_1\leq 2\delta.
\]
This completes the proof.
\end{proof}

\begin{corollary}{\bf [Purification Lemma]}\label{cor:purificationLemma}
Given the setting of \cref{onehaar}, define
\[
\kappa^{R_0BR_1E}\coloneqq\abs{A''_0}~\Tr_C~\mathcal{U}_{\mathcal{N}} ~\textup{op}^{A''_0\rightarrow A'BR_1}(\omega)~U^{A''_0}\cdot \eta^{A''_0R_0}
\]
and 
\[
\delta\coloneqq 2^{-\frac{1}{2}H_{\min}^{\frac{\epsilon^2}{26k}}(A''_0|A''_1E)_{ \mathcal{U}_{\mathcal{N}}\cdot\Omega}-\frac{1}{2}H_{\min}^{\epsilon}(A_0|R_0)_{\eta}+O(\log k) }+12k\epsilon.
\]
Then
\[
\norm{\frac{\kappa^{R_0BR_1E}}{\Tr[\kappa]}-\eta^{R_0}\otimes \Tilde{\omega}^{BR_1E}}_1\leq 2\delta.
\]
\end{corollary}
\begin{proof}
\cref{onehaar} tells us that
\[
\norm{\frac{\kappa^{R_0BR_1E}}{\Tr[\omega]}-\eta^{R_0}\otimes \Tilde{\omega}^{BR_1E}}_1 \leq \delta.
\]
Then by \cref{lem:NormalisationLemma}, we see that
\[
\norm{\frac{1}{\Tr\left[\frac{\kappa}{\Tr[\omega]}\right]}\frac{\kappa^{R_0BR_1E}}{\Tr[\omega]}-\eta^{R_0}\otimes \Tilde{\omega}^{BR_1E}} \leq 2\delta,
\]
which implies that
\[
\norm{\frac{\kappa^{R_0BR_1E}}{\Tr[\kappa]}-\eta^{R_0}\otimes \Tilde{\omega}^{BR_1E}}_1\leq 2\delta.
\]
This concludes the proof. 
\end{proof}

\end{document}